\documentclass[a4paper,11pt]{article}
\pdfoutput=1 

\usepackage{jheppub} 

\usepackage[T1]{fontenc} 

\title{\boldmath General Lagrangian formulations for  mixed-antisymmetric  tensor fields  on flat backgrounds}

\author[a,b,c]{A.A. Reshetnyak,\note{Corresponding author.}}

\affiliation[a]{Center for Theoretical Physics,
    Tomsk State Pedagogical University, \\Tomsk, 634061, Russia}
\affiliation[b]{National Research Tomsk  State   University,\\Tomsk 634050, Russia}
\affiliation[c]{National Research Tomsk Polytechnic   University,\\ Tomsk 634050, Russia}

\emailAdd{reshetnyak@tpu.ru}

\abstract{We present, for the first time, local Lagrangian formulations     for (ir)reducible  integer
higher-spin massless and massive Poincare group representations subject to
Young tableau with $k$ columns  $Y[\hat{s}_1,\hat{s}_2,...,\hat{s}_k]$ in  $d$-dimensional Minkowski space-time. The particles are described   in a metric-like formulation by  tensor fields carrying $k$ groups of antisymmetric Lorentz indices $\Phi_{\mu^1[{\hat{s}_1}],\mu^2[{\hat{s}_2}],..., \mu^k[{\hat{s}_k}]}$  using the BRST procedure with both the complete operator $Q$  and incomplete $Q_c$ BRST operators. Starting from a description of
bosonic mixed-antisymmetric higher-spin fields  in terms of an auxiliary Fock space associated with a
special Poincare module, we realize a conversion of the initial
operator constraint system
into a system of first-class operator constraints. For  this purpose, we find,
 for the first time, by means of  Verma module the auxiliary representations of
 the constraint
 subalgebra, to be isomorphic due to Howe duality to $so(k,k)$
 algebra, and
containing the subsystem of second-class operators  in terms of
new  oscillator variables forming the Fock module.  An unconstrained (with  $Q$) and constrained (with $Q_c$ and BRST invariant algebraic constraints) gauge Lagrangian formulations with equivalent dynamics, but different configuration spaces are found. 
We outline within $Q_c$-approach the concept of consistent local cubic interactions and evaluate the candidate structure of cubic vertices involving $p\geq 3$ copies of interacting massless mixed-antisymmetric fields.}

\begin{document}
\maketitle
\flushbottom

\section{Introduction}
\label{sec:intro}

Higher-spin (HS) field theory continues to attract growing interest, motivated by the hope of constructing new kinds  of Lagrangian
models in classical field theory (FT) and to formulate on this
ground the new approaches to the unification of the fundamental
interactions jointly with resolution the  quantum gravity problem (see, for review,   e.g. \cite{reviews}, \cite{reviewsV}, \cite{Ponomarev}, \cite{Snowmass} and the recent paper \cite{2606.25063}  in view of the Supersymmetry activity at the Joint Institute for
Nuclear Research). HSFT
 is strongly related to
(super)string theory \cite{SFT}, \cite{SFT1}, \cite{Siegel}, \cite{tensionlessl}  operating with an infinite tower of fields  with integer and half-integer spins in $10$-dimensional space-time with compactified extra dimensions thus generating theirs masses, e.g. via the dimensional reduction mechanism. Since interacting (super)strings appear to provide the only consistent way  to describe evolution of the early Universe and
 involve the interaction of particles with different values of spin --  with lower spins (Higgs particle for $s=0$; quarks  and leptons for $s=\frac{1}{2}$; photon, gluons, $W^{\pm}, Z$ bosons for $\lambda=1$ and till undiscovered graviton with $\lambda=2$) -- the unknown particles with $s>2$  and with  generalized spins $s_{(k)}=(s_1,s_2,...,s_k)$ may also exist. From this perspective such particles  may be considered  as the probable candidates to describe the Dark Matter problem outside the SM, see for a review\cite{DM1}, \cite{DM2} (always surrounding electromagnetically observable particles) beyond the extensions of SM with sterile neutrino  \cite{Dubinin} and dark massive "photon" \cite{DMphoton}. While the problem of constructing a Lagrangian formulation (LF) for HS fields in constant-curvature spaces has been addressed within different approaches  (see for review, \cite{reviews3}, \cite{Bengtsson}) both in metric and frame-like formalism for totally, e.g. in  \cite{SinghHagen}, \cite{Fronsdint}, \cite{FronsdintA}, \cite{Franciatr2}, \cite{0702161},
  and mixed-symmetric  irreducible (spin)-tensors fields on $\mathbb{R}^{1,d-1}$ (or AdS space), for instance, in \cite{Labastida}, \cite{Labastida1}, \cite{0311164}, \cite{0501108}, \cite{08070903}, \cite{08081778}, \cite{franciamixbos},
 \cite{1110.5044}, \cite{1211.1273}, \cite{2305.00142}  (for $d=4$ as well as for irreducible representations (irreps) with continuous spin, see \cite{ShusterToro}, \cite{ShusterToro1}, \cite{Bengtsson2}, \cite{FedorukBuch}, \cite{BuchKrychcont}, \cite{BurdikPResh}, \cite{2403.14446}, \cite{2503.14290}),  the
  same problem for another basis of tensor fields, namely mixed-antisymmetric (MAS) HS fields, has been solved only for special cases of Poincaré group irreps with Young tableau   $Y[\hat{s}_1,\hat{s}_2,...,\hat{s}_k]$. This was achieved for the totally antisymmetric case ($k =1$)  in \cite{ashsb1}, \cite{ashsf2},     \cite{zinoviev} and for massless HS fields with antisymmetric spin ${s}{[2]}=(\hat{s}_1, \hat{s}_2)$ in \cite{Medeiros}, \cite{0311212}, \cite{0402180}, \cite{Boulanger2},  \cite{zinoviev1}, in BRST (Becchi-Rouet-Stora-Tyutin) approaches with complete and incomplete BRST operators for massive and massless cases in \cite{resh0}, \cite{resh1} and recently for   the fields  $\Phi_{\mu^1[{\hat{s}_1}],\mu^2[{\hat{s}_2}],\mu^3[{\hat{s}_3}]}$ in \cite{RBP}. A gauge-invariant LF with a non-local action for massless tensor fields transforming in arbitrary irreducible representations of the general linear group in $d$-dimensional constant-curvature space-time, constructed with the use of Levi-Civita tensors, was obtained in     \cite{Boulanger3}.  The non-local nature of the Lagrangians \cite{Boulanger3} makes all  calculations with them within classical and quantum diagrammatic Feynman techniques  (see, e.g. \cite{SlavnovFaddeev},  \cite{Weiberg1}, \cite{Weiberg1}, \cite{GitmanTyutin}, \cite{HenneauxTeitelboim}) extremely non-standard and, so far, unexplored. 
  
The reformulation of the local LF \cite{1110.5044} , given in the symmetric basis, in the antisymmetric basis for the same Poincaré group irrep is a non-trivial task in its own right and constitutes a separate problem.
We intend to fill this gap by studying the properties of the local Lagrangian dynamics and a unified gauge-algebra description of the general irreducible MAS HS field subject to $Y[\hat{s}_1, \ldots, \hat{s}_k]$ on the basis of the BRST cohomological technique.

The universal BRST approach originates both from  (super)string theory and from the special supersymmetry known as Becchi–Rouet–Stora–Tyutin symmetry  \cite{BRST1}, \cite{BRST2} within the Faddeev-Popov rules \cite{FP1}for the Lagrangian quantization of FT with a gauge group, and also from the generalized canonical quantization known as the BFV (Batalin-Fradkin-Vilkovisky) method \cite{BFV}, \cite{BFV1}, \cite{BFV2} for quantizing constrained dynamical systems. Here, however, it serves to solve the inverse problem of finding an LF, starting from non-Lagrangian partial differential equations that describe field (ir)reps of the  $ISO(1,d-1)$ or $SO(2,d-1)$ group. BRST approach, being also the realization of the AKSZ model \cite{AKSZ} (see, for its development \cite{Grig1}, \cite{Grig2} and references therein),
  deals with the superalgebra of operators $\{o_I\}$ containing together with isometry subsuperalgebra, $\{o_A\}$ involving only space-time derivatives, the subalgebra of holonomic second-class operator constraints  $\{o_a, o^+_a\}$ \cite{Dirac11, Dirac2, resh2}. If the nilpotent Grassmann-odd BRST operator is constructed with respect to the subalgebra $\{o_A\}$ alone, it is called \emph{incomplete}, $Q_c$ following from string field theory (SFT) \cite{tensionlessl}, \cite{BRST-first} in the presence of algebraic constraints 
\cite{resh2}, \cite{0406192}; this is to be compared with the  \emph{complete}  operator, $Q$ initiated by  A.~Pashnev, M.~Tsulaia, I.~Buchbinder  in  \cite{Pashnev1}, \cite{BurdikPashnev1}, \cite{BurdikPashnev2}, \cite{symint-adsmassless}
  \begin{equation}\label{QQc}
 Q_c = C^Ao_A+ \frac{1}{2}
    {C}^A{C}^Bf^D_{BA}{\overline{\mathcal{P}}}_D (-1)^{\epsilon({o}_D) + \epsilon({o}_A)}, \ \ Q = C^Io_I+ \frac{1}{2}
    {C}^I{C}^Jf^K_{JI}{\overline{\mathcal{P}}}_K (-1)^{\epsilon({o}_K) + \epsilon({o}_I)}
\end{equation}
with structural constants $f^K_{JI}(f^D_{BA})$ from the involution relations: $[o^I, o_J\} = f^K_{IJ}o_K$, with operators of ghost coordinates $C^I$, momenta $\overline{\mathcal{P}}_J$ of Grassmann parity $\epsilon$ opposite to that of  $o_I$, and with supercommutators $[C^I, {\overline{\mathcal{P}}}_J\} =\delta^I_J$. The respective LFs are determined by inner products in the corresponding Hilbert spaces $\mathcal{H}_{tot}$ and $\mathcal{H}_{c|tot}$ ($\mathcal{H}_{c|tot}
\subset \mathcal{H}_{tot}$) with field vectors $|\chi\rangle$,  $|\chi_c\rangle$  ($|\chi_c\rangle \subset |\chi\rangle$) containing the initial tensor field (e.g. $\Phi_{\mu^1[{\hat{s}_1}],\mu^2[{\hat{s}_2}],\mu^3[{\hat{s}_3}]}$ as in \cite{RBP}) with the actions
\begin{equation}\label{SSc}
   \mathcal{S}_{s[k]}[| \chi \rangle ] \sim \langle \chi
|K Q | \chi \rangle, \quad \texttt{and} \quad \mathcal{S}_{c|s[k]}[| \chi_c \rangle ] \sim \langle \chi_c
|Q_c | \chi_c \rangle
\end{equation}
where the operator $K$ provides Hermiticity  of the operator  $Q$; in the LF with the incomplete BRST operator, finite sequences of gauge transformations together with additional holonomic constraints are involved.

The purposes of the paper are twofold. First one is to  construct  general local
gauge LFs for both massless (and massive)
MAS tensor fields of rank $\sum_{l=1}^ks_l $, with
any integer numbers $\hat{s}_1 \geq \hat{s}_2\geq ...\geq \hat{s}_k \geq 1$ for $\hat{s}_1
\leq [d/2-1]$ in  $d$-dimensional Minkowski space as
elements of Poincare-group $ISO(1,d-1)$ irreps with a Young
tableau having $k$ columns. Second  purpose consists in study of consistent cubic interactions for   MAS HS fields.

  The application of the BRST construction  to
free HS field theory  consists of three steps. First, the
conditions that determine the representations with a given spin
are regarded as an operator dynamical
system of first- and second-class operator constraints  in an
auxiliary Fock space with one or two Hamiltonians (second and first orders for fermionic, whereas only  second order for bosonic fields). Second, the subsystem of the initial operator constraints containing only second-class constraints is converted, preserving the initial algebraic structure, into a system of first-class operators alone in an enlarged Fock space 
 (see \cite{conversion}, \cite{conversion1}, \cite{conversion2}, \cite{conversion3} for the conversion
methods); with respect to these, jointly with the first-class subsystem, one constructs the complete BRST charge  $Q$.
 Third, the Lagrangian for an HS field is constructed in terms of the BRST charge in such a way that the corresponding equations of motion reproduce the initial constraints. We emphasize that this approach automatically yields a gauge-invariant Lagrangian description with all appropriate auxiliary and Stuckelberg fields. The BRST approach to the Lagrangian formulation of HS field theories has been developed for arbitrary massless and massive, bosonic and fermionic fields in Minkowski and AdS spaces, e.g. in   \cite{symint-ads}, \cite{symint-ads-b},\cite{0505092}, \cite{symferm-flat}, \cite{symferm-flat_m}, \cite{0703049}, \cite{BuchKrycRysTak}, \cite{0706.0386}, \cite{0902BKRT}.

The inclusion of consistent interactions for HS fields requires a deformation of certain known free Lagrangian formulations; the corresponding consideration was initiated within the approach with the complete BRST operator in 
\cite{2105.12030},  \cite{2205.00488}, \cite{BKStwis} for massless HS bosonic fields, was  discussed for massive totally-symmetric and mixed-symmetric HS fields in \cite{2212.07097}, \cite{2304.10358}, \cite{2403.16164}, \cite{2505.02190} ], and was also studied within the BRST–BV approach in \cite{2303.02870} with the construction of a minimal Batalin–Vilkovisky action within the Lagrangian quantization method of gauge theories  \cite{BV}, \cite{BV1}. The cubic and quartic interaction vertices obtained in different approaches show that the construction of interacting LFs with HS fields is still far from complete, in spite of the results (see, e.g., the cited papers and references therein) for cubic vertices  \cite{Manvelyan1}, \cite{frame-like1}, \cite{Metsaev0512}, \cite{BRST-BV3},  \cite{frame-like2},  among them those obtained with the use of $N=2$ harmonic superspace  \cite{BuchIvZaig1}, \cite{BuchIvZaig2}, as well as those within the BRST approach with an incomplete BRST operator for reducible integer-spin representations \cite{brstinter1}, \cite{brstinter2}, \cite{brstinter3}, \cite{brstinter4}, \cite{BKTW}, for quartic vertices \cite{DT}, \cite{Taronna} and for the related problems of locality  \cite{Kessel}, \cite{Taronna1}, \cite{Taronna2}, \cite{Ponomarev3}, \cite{Vasil}, \cite{Didenko1}, \cite{Didenko2}.

In case of consideration in the system of operator constraints the only isometry algebra  constraints (to be by the first-class operator subsystem for the massless HS fields) as the  dynamical ones the respective incomplete BRST operator,  $Q_c$, should form closed superalgebra  with BRST extended holonomic operator constraints and spin operator as the condition to get an  LF with consistent off-shell holonomic constraints  imposed on the whole set of field and gauge parameters vectors. The configuration space then contains a smaller number of auxiliary fields compared with the Lagrangian obtained within the approach with the complete BRST charge $Q$. 
The dynamics of the Lagrangian formulations with the complete and incomplete BRST operators for the same HS field in flat space-time are equivalent for any field belonging to a Poincaré-group irrep with finite (integer or half-integer) spin, as was shown in \cite{resh2}  for mixed-symmetry HS fields; this equivalence will now be shown to hold for MAS HS fields as well, at least for free fields. Extending the Lagrangian dynamics of MAS HS fields from the free to the interacting case appears to be a major challenge; it must be addressed both within the framework of interactions among MAS HS fields themselves and in the description of interactions between MAS HS fields and mixed-symmetric HS fields.

The paper is organized as follows. In Section~\ref{Symmalgebra},
 we formulate a closed superalgebra of operators based on the constraints in an auxiliary Fock space that determines a massless irrep of the Poincaré group in  $\mathbb{R}^{1,d-1}$ with  generalized spin $\mathbf{s}
= (k,k,...,k,k-1,k-1,...,k-1,...,2,2,2, ...,2,1,...,1)$, where the number of entries equal to $k$ being equal to $\hat{s}_k$ and the number of entries equal to  $1$  is $(\hat{s}_1-\hat{s}_2)$.

In Section~\ref{Vermamodule}, we construct an
auxiliary representation for the subalgebra of the second-class operator constraints  in
terms of new (additional) creation and annihilation operators in
Fock space\footnote{Note that a similar construction for bosonic
HS fields subject to Young tableaux with $k$ rows in a flat space
has been presented in \cite{1110.5044} for the massless and  massive cases in the symmetric basis.}
 in order to arrive at the additively deformed superalgebra of converted constraints  $O_I=o_I+o'_I$. The complete BRST operator for the superalgebra of converted operator constraints, together with 
 the respective spectral problem involving the Q-complex, is found in Section~\ref{BRSToperator}. The
derivation  of an unconstrained and a component  BRST  actions, alone with a sequence of the respective reducible gauge transformations describing the propagation of a MAS field of arbitrary integer spin, is carried out in
Section~\ref{LagrFormulation}.  In  Section~\ref{constrlagrform} we solve the same problem but within suggested  BRST approach with incomplete BRST operator with BRST invariant off-shell holonomic constraints.  We find that the component LF with the incomplete BRST operators and the appropriate set of holonomic constraints is expressed, for the case for the case $k=2$ (Section~\ref{constrBRST}), in terms of only the physical massless field and a set of reducible gauge transformations.  In  Section~\ref{massiveLF} it is shown that the Lagrangian
description for a theory of a massive integer MAS HS
field in a $d$-dimensional Minkowski space is deduced by
dimensional reduction from a massless HS field theory of the same
type in  $(d+1)$-dimensional flat space. Finally, local (and non-local) Lagrangian covariant cubic vertices for interacting massless MAS (ir)reducible fields are derived in Section~\ref{cubic}.

In Appendices~\ref{ap1} and~\ref{oscrealsl2kdet}, we present the results of the Verma module construction for the algebra of second-class constraints —- which is isomorphic to algebra $so(k,k$) —- and its scalar oscillator realization in terms of a polynomial Fock module, for irreducible massless and massive MAS HS fields with  $k$ group indices. Appendix~\ref{holconstrres} contains the resolution of the mixed-antisymmetry and tracelessness constraints used to obtain the component Lagrangian formulation.

Throughout the paper, we use the conventions of
  \cite{resh1},   \cite{RBP}, \cite{2105.12030}, \cite{2212.07097}. in particular, $\eta_{\mu\nu}$ is the  metric tensor; $(\epsilon, gh_H)(F)$ denotes the Grassmann parity and ghost number of a quantity $F$;  $[\ ,\ \}$ is the supercommutator;  $[z]$ is the integer part of a real number $z$;
${s[k]}$ denotes the antisymmetric spin with components $
(\hat{s}_1, \hat{s}_2,...,\hat{s}_k)$;  $A^{\{i}B^{j\}l} =
A^{i}B^{jl}+
A^{j}B^{il}$ (and  $A^{[i}B^{j]l} =
A^{i}B^{jl}-
A^{j}B^{il}$) denotes (anti)symmetrization; $(\chi_c)_p= (\chi^{(1)}_c, \chi^{(2)}_c, ..., \chi^{(p)}_c)$ is a multicomponent vector;  $(s[k])_p=(s[k_1], s[k_2], ... , s[k_p])$ is a multicomponent antisymmetric spin; and $\theta^e{}_{i}$ (or $\theta_{e{}i}$)is the Heaviside $\theta$-symbol, equal to 
 $1(0)$ when $e>i (e\leq i)$.

\section{Integer
HS Symmetry Superalgebras subject to $Y[s_1,...,s_k]$ }\label{Symmalgebra}

In this section, we remind \cite{resh0, resh1} that a massless integer spin  Poincare group irrep in the  Minkowski space
$\mathbb{R}^{1,d-1}$  is described by a real-valued tensor field
$\Phi_{\mu[{\hat{s}_1}],..., \mu[{\hat{s}_k}]}
\hspace{-0.2em}$ $\equiv $ $\hspace{-0.2em}
\Phi_{\mu^1_1\ldots\mu^1_{\hat{s}_1},...,\mu^k_1\ldots\mu^k_{\hat{s}_k}}$
 of rank $\sum_i^k\hat{s}_i$ and generalized spin
 $\mathbf{s} = (k,k,...,k;...;2, 2,
 ... , 2;1, ..., 1)$, ($\hat{s}_1\geq... \geq s_k>0, \hat{s}_1
\leq [d/2]-1$) to be corresponding to a Young tableaux with $k$ columns
of height  $\hat{s}_1, \hat{s}_2,...,\hat{s}_k$, respectively (with omitting later the symbol "$\hat{\phantom{s}}$" under $\hat{s}_i$)
\begin{equation}\label{Young k}
\Phi_{\mu^1[{s_1}],...,\mu^k[{s_k}]}
\hspace{-0.3em}\longleftrightarrow \hspace{-0.3em}
\begin{array}{|c|c|c|c|c|}\hline
  \!\mu^1_1 \!&\! \mu^2_1\! &\! ... \!&\! \mu^{k-1}_1\!&\! \mu^k_1\! \\
   \hline
    \! \mu^1_2\! &\!\mu^2_2\! &\! ... \!&\! \mu^{k-1}_2\!&\! \mu^k_2\!  \\
  \cline{1-5} \!\cdot\! &   \!\cdot\!&   \!\cdot\!&   \!\cdot\!&   \!\cdot\!   \\
   \cline{1-5}
     \! \mu^1_{s_k}\! &\!\mu^2_{s_k}\! &\! \cdot \!&\! \mu^{k-1}_{s_k}\!&\! \mu^k_{s_k}\!   \\
     \cline{1-5}
     \! \mu^1_{s_k+1}\! &\!\mu^2_{s_k+1}\! &\! \cdot \!&\! \mu^{k-1}_{s_k+1}\!\\
       \cline{1-4} \!\cdot\! &   \!\cdot\!&   \!\cdot\!&   \!\cdot\! \\
   \cline{1-4}
     \! \mu^1_{s_{k-1}}\! &\!\mu^2_{s_{k-1}}\! &\! \cdot \!&\! \mu^{k-1}_{s_{k-1}}\!\\
  \cline{1-4} \!\cdot\! &   \!\cdot\!&   \!\cdot\! \\
          \cline{1-3} \!\mu^1_{s_2+1}\!   \\
  \cline{1-1} \!\cdots\!\\
  \cline{1-1} \!\mu^1_{s_1}\! \\
  \cline{1-1}
\end{array}\ ,
\end{equation}
This field is antisymmetric with respect to the permutations of each
type of Lorentz indices
 $\mu^i$,
  and
obeys to the wave (\ref{Eq-0b}) and independent $k$ divergentless
(\ref{Eq-1b}), $\frac{1}{2}k(k-1)$ traceless (\ref{Eq-2b}), also $\frac{1}{2}k(k-1)$ mixed-antisymmetry (Young)
equations (\ref{Eq-3b}):
\begin{eqnarray}
\label{Eq-0b} \hspace{-0.3em}&\hspace{-0.3em}&\hspace{-0.3em}
\partial^\mu\partial_\mu\Phi_{\mu^1{[s_1]},\ldots,\mu^k[{s_k}]}
 =0, \\
\hspace{-0.3em}&\hspace{-0.3em}&\hspace{-0.3em}   (\partial^{i} \Phi)_{
{\mu}^1[{s_1-\delta_{i1}}],...,{\mu}^k[{s_k-\delta_{ik}}]} \equiv  \partial^{\mu^i_{l_i}}\Phi_{\mu^1{[s_1]},\ldots,\mu^k[{s_k}]} =0,\texttt{ for } 1 \leq l_i \leq s_i,\ i=1,...,k, \label{Eq-1b}
\\
\hspace{-0.3em}&\hspace{-0.3em}&\hspace{-0.3em} (\texttt{Tr}^{ij}\Phi)_{\mu^1{[s_1-\delta_{i1}]},\ldots,\mu^k[{s_k}-\delta_{jk}]}\equiv  \eta^{\mu^i_{l_i}\mu^j_{l_j}}\Phi_{\mu^1{[s_1]},\ldots,\mu^k[{s_k}]}=0,  \ \ 1 \leq i<j\leq k,\label{Eq-2b}\\
\hspace{-0.3em}&\hspace{-0.3em}&\hspace{-0.3em}  (Y^{ij} \Phi)_{
\mu^1[{s_1}],...,[\mu^i[{s_i}],...,\mu^j_{l_j}],\hat{\mu}^j[{s_j-1}],...,\mu^k[{s_k}]}\hspace{-0.15em}   \label{Eq-3b} \\
&& \phantom{(\texttt{Tr}^{ij}\Phi)_{\mu^1{[s_1-\delta_{i1}]},\ldots,\mu^k[{s_k}-\delta_{jk}]}} \equiv (-1)^{l_j-1} \Phi_{
...,[\mu^i[{s_i}],...,\mu^j_{l_j}]\underbrace{\mu^j_1...\mu^j_{l_j-1}\mu^j_{l_j+1}}...\mu^j_{s_j},...,\mu^k[{s_k}]}\hspace{-0.2em}=\hspace{-0.1em}0, \nonumber
\end{eqnarray}
where we introduce the notations for the operations  of divergences $(\partial^{i} \Phi)$, mixed traces $(\texttt{Tr}^{ij}\Phi)$, Young-antisymmetrizations
$(Y^{ij} \Phi)$ for $l_i = 1,...,s_i$ and $l_j = 1,...,s_j$.
The  bracket below in (\ref{Eq-3b}) denote that the indices  in it do not
include in  antisymmetrization, i.e. the antisymmetrization  contains ${s_i}+1$ terms and concerns only
indices $\mu^i[{s_i}], \mu^j_{l_j} $, whereas the sign "$\hat{\mu}^j[{s_j-1}]$" means the absence of index $\mu^j_{l_j}$ in the set, ${\mu^j}[{s_j}]$.

To describe all irreps
simultaneously, one introduces  an auxiliary Fock
space $\mathcal{H}^f$, generated by $k$ pairs of Grassmann-odd (antisymmetric
basis) creation $\hat{a}^{i+}_{\mu^i}(x)$ and annihilation
$\hat{a}^{j}_{\nu^j}(x)$ operators, $\mu^i,\nu^j
=0,1...,d-1$: with anticommutation relations,
  (we enlarge the  procedure below following to the
lines of Ref. \cite{RBP} for MAS tensors for
$k>3$).
\begin{eqnarray}\label{comrels}
\{\hat{a}^i_{\mu^i},
\hat{a}_{\nu^j}^{j+}\}=-\eta_{\mu^i\nu^i}\delta^{ij}\,,
\qquad \delta^{ij} = diag(1,...,1)\,,
\end{eqnarray}
 and a set of constraints for an arbitrary string-like vector
$|\Phi\rangle \in \mathcal{H}^f$,
\begin{eqnarray}
\label{PhysState}  \hspace{-2ex}&& \hspace{-2ex} |\Phi\rangle  =
\sum_{s_1=0}^{[d/2]}\sum_{s_2=0}^{s_1}\ldots \sum_{s_k=0}^{s_{k-1}}
\frac{\imath^{\sum_{p=1}^ks_p}}{{s_1!\ldots s_k!}}\Phi_{\mu^1{[s_1]},\ldots,\mu^k[{s_k}]}\,
\prod_{i=1}^k\prod_{l_i=1}^{s_i} \hat{a}^{+\mu^i_{l_i}}_i|0\rangle,\\
\label{lilijt} \hspace{-2ex} && \hspace{-2ex} \bigl( {l}_0,\,{l}^i,\, l^{ij},\,
t^{ij}\, \bigr)|\Phi\rangle  = \bigl(\partial^\mu\partial_\mu,\,-i \hat{a}^i_\mu \partial^\mu,\,
\textstyle\frac{1}{2}\hat{a}^{i}_\mu \hat{a}^{j\mu}, \,\hat{a}^{i+}_\mu
\hat{a}^{j\mu}\bigr) |\Phi\rangle=0 .
\end{eqnarray}
 The set of $[k(k-1)+1]$ even and $k$ odd, ${l}^i$,
primary constraints  (\ref{lilijt}): $\{o_\alpha\}$
= $\bigl\{{{l}}_0, {l}^i, l^{ij}, t^{ij} \bigr\}$,  because of  translational invariance of the vacuum,
$\partial_\mu |0\rangle = 0$, are equivalent to equations
(\ref{Eq-0b})--(\ref{Eq-3b}) for all possible heights $s_1,..., s_k$ for $s_1\geq ...\geq s_k$\footnote{In case of the tensor field $\Phi_{\mu^1[{s_1}],...,\mu^k[{s_k}]}$ being subject to non-standard Young tableaux $Y[s_1,...,s_k]$ for  $s_i<s_j$ when $1\leq i<j\leq k$,  the  corresponding vector $|\Phi\rangle \in \mathcal{H}^f$ constructed by the rule (\ref{PhysState}) where $s_i \leftrightarrow s_j$ in the limits of sums,   can not already be by an element of  $\ker t_{ij} $:  $t_{ij}|\Phi\rangle \ne 0$. Now, the hermitian conjugated operator $t^+_{ij}$ should describe the property of belonging to $Y[s_1,...,s_k]$ of such  vector: $t^+_{ij}|\Phi\rangle = 0$. Indeed, for such tensor field there exist $(s_j-s_i)$ indices in the second set $[\mu^j]_{s_j}$ which are not symmetrized with any from the first group indices due to its absence.}.
The additional to (\ref{lilijt})
constraints with number particles operators, $g_0^i$,
\begin{eqnarray}\label{g0iphys}
g_0^i|\Phi\rangle =(s_i-\frac{d}{2}) |\Phi\rangle, \qquad
 g_0^i = -\frac{1}{2}[\hat{a}^{i+}_\mu,  \hat{a}^{\mu{}i}]\ =\ -\hat{a}^{i+}_\mu  \hat{a}^{\mu{}i }-\frac{d}{2},
\end{eqnarray}
where the sign $[\ ,\ ]$ is  the commutator describe the single Poincare group irrep of spin $\mathbf{s}=[s_1,\ldots, s_k]$,  making equations (\ref{lilijt}), (\ref{g0iphys})  equivalent to ones
(\ref{Eq-0b})--(\ref{Eq-3b}).

 We refer to the vector (\ref{PhysState}) as the basic vector\footnote{We
 may consider a set of
 all finite string-like vectors
 which  different choice of a spin $\mathbf{s}$ as the vector space  of
 polynomials $P_k^d(\hat{a}^{+})$
 in degree $\hat{a}^{+\mu^i}_i$. The Lorentz algebra on $P_k^d(\hat{a}^{+})$
 is realized by means of action on it
 the Lorentz transformations, $M^{\mu\nu}= \sum_{i\geq 1}^k \hat{a}^{+[\mu}_i \hat{a}^{
 \nu]i}$ with a standard rule
 $A^{[\mu}B^{\nu]}\equiv A^{\mu}B^{\nu}-A^{\nu}B^{\mu}$, thus endowing $P_k^d(\hat{a}^{+})$
 by the structure of Lorentz-module.}.

The procedure of Lagrangian formulation (LF) implies  the property of
BRST-BFV operator both complete $Q$, $Q = C^\alpha o_\alpha + more$ and incomplete $Q_c$ without involving algebraic constraints $l^{ij}, t^{ij}$,  to be
Hermitian, that is equivalent to the requirements: $\{o_\alpha\}^+
= \{o_\alpha\}$ and closedness for $\{o_\alpha\}$ with respect to
the supercommutator multiplication $[\ ,\ \}$.  To provide this we consider  an inner product on
$\mathcal{H}^f$ (with sign "$\star$" meaning complex conjugation),
\begin{eqnarray}
\label{sproduct} \langle{\Psi}|\Phi\rangle & =  & \int
d^dx\sum_{s_1=0}^{[d/2]}\sum_{s_2=0}^{s_1}\ldots\sum_{s_k=0}^{s_{k-1}} \frac{(-1)^{\sum_ps_p}}{s_1!\ldots s_k!}\Psi^{\star}_{\mu^1[{s_1}],\ldots, \mu^k[{s_k}]}(x)
\Phi^{\mu^1[{s_1}],\ldots, \mu^k[{s_k}]}(x) .
\end{eqnarray}
As the result, the set of $\{o_\alpha\}$ extended by means of the
operators,
\begin{eqnarray} \label{lilijt+} \hspace{-2ex} && \hspace{-2ex} \bigl({l}^{i+},\
l^{ij+},\ t^{ij+} \bigr)  = \bigl(-i \hat{a}^{i+}_\mu
\partial^\mu,\ \textstyle\frac{1}{2}\hat{a}^{j+}_\mu \hat{a}^{i\mu+},\
\hat{a}^{j+}_\mu \hat{a}^{i\mu}\bigr) ,\
\end{eqnarray}
is closed with respect to  Hermitian conjugation, with taken into account of self-conjugated
operators, $(l_0^+,\ {g_0^i}^+) = (l_0,\ {g_0^i})$. It is rather
simple exercise  to see the second requirement is fulfilled as
well if the number particles operators $g_0^i$ will be included
into set of all constraints $o_I$ having therefore the structure,
\begin{eqnarray}
\{o_I\} = \{o_\alpha, o_\alpha^+;\ g_0^i\}\equiv \{o_a, o_a^+ ;\
l_0,\ l^i,\ l^{i+};\ g_0^i\}. \label{inconstraints}
\end{eqnarray}
Together the set $\{o_a, o_a^+\}$ in the Eq.
(\ref{inconstraints}), for $\{o_a\} = \{l^{ij}, t^{ij}\}$ and
the one $\{o_A\}= \{l_0,\ l^i,\ l^{i+}\}$, may be considered from
the Hamiltonian analysis of the dynamical systems
 as the operator respective $2k(k-1)$ second-class and $2k+1$ first-class  constraints subsystems among
$\{o_I\}$ for  gauge system (i.e. with operator Hamiltonian $l_0$) because of,
\begin{eqnarray}
[o_a,\; o_b^+\} = f^c_{ab} o_c +\Delta_{ab}(g_0^i),\ [o_A,\;o_B\} =
f^C_{AB}o_C, \  [o_a,\; o_B\} = f^C_{aB}o_C .
\label{inconstraintsd}
\end{eqnarray}
Here $f^c_{ab}, f^C_{AB}, f^C_{aB}$ are  antisymmetric with
respect to permutations of lower indices constant quantities and
quantities $\Delta_{ab}(g_0^i)$ form the non-degenerate $2k(k-1)\times
2k(k-1)$ matrix $\|\Delta_{ab}\|$ in  Fock space  $\mathcal{H}^f$ on
the surface $\Sigma \subset \mathcal{H}^f$:
$\|\Delta_{ab}\|_{|\Sigma} \ne 0 $, which is determined by the
equations, $(o_a, l_0,\ l^i)|\Phi\rangle = 0$. The set of $o_I$
contains the operators $g_0^i$ are not being by the constraints in
$\mathcal{H}^f$.

Explicitly, operators $o_I$ satisfy to the Lie-algebra commutation
relations,
\begin{equation}\label{geninalg}
    [o_I,\ o_J]= f^K_{IJ}o_K, \  f^K_{IJ}= - (-1)^{\varepsilon(o_I)\varepsilon(o_J)} f^K_{JI},
\end{equation}
where the structure constants $f^K_{IJ}$ are used in the
Eq.(\ref{inconstraintsd}),
and determined from the
multiplication table~\ref{table in}. \hspace{-1ex}{\begin{table}
{{\footnotesize
\begin{center}
\begin{tabular}{||c||c|c|c|c|c|c|c||c||}\hline\hline
$\hspace{-0.2em}[\; \downarrow, \rightarrow
\}\hspace{-0.5em}$\hspace{-0.7em}&
 $t^{ij}$ & $t^+_{ij}$ &
$l_0$ & $l^i$ &$l^{i{}+}$ & $l^{ij}$ &$l^{ij{}+}$ &
$g^i_0$ \\
\hline\hline $t^{ep}$
    & $A^{ep,ij}$ & $B^{ep}{}_{ij}$
   & $0$&\hspace{-0.3em}
    $\hspace{-0.2em}l^{p}\delta^{e{}i}$\hspace{-0.5em} &
    \hspace{-0.3em}
    $-l^{e+}\delta^{p{}i}$\hspace{-0.3em}
    & \hspace{-0.3em}
    $-l^{p[i}\delta^{j]e}$\hspace{-0.3em} & \hspace{-0.7em} $\hspace{-0.7em}l^{e[i+}\delta^{j]p}
    \hspace{-0.9em}$ \hspace{-1.2em}& $+F^{ep,i}$ \\
\hline $t^+_{ep}$
    & -$B^{ij}{}_{ep}$ & $A^+_{ij,ep}$
&$0$   & \hspace{-0.3em}
    $\hspace{-0.2em} l_{e}\delta^{i{}}_p$\hspace{-0.5em} &
    \hspace{-0.3em}
    $-l^+_{p}\delta^{i{}}_e$\hspace{-0.3em}
    & $\delta^{[{i}}_p l_{e}{}^{j]}$ & $\delta^{[{i}}_e l^{j]}{}^+_p$ & $-F^{ep,i+}$\\
\hline $l_0$
    & $0$ & $0$
& $0$   &
    $0$\hspace{-0.5em} & \hspace{-0.3em}
    $0$\hspace{-0.3em}
    & $0$ & $0$ & $0$ \\
\hline $l^e$
   & \hspace{-0.5em}$ -\delta^{e{}i} l^{j}$ \hspace{-0.5em} &
   \hspace{-0.5em}$
  -\delta^{e{}j} l_{i}$ \hspace{-0.9em}  & \hspace{-0.3em}$0$ \hspace{-0.3em} & $0$&
   \hspace{-0.3em}
   $\delta^{ei} l_0$\hspace{-0.3em}
    & $0$ & \hspace{-0.5em}$ \textstyle\frac{1}{2}\delta^{e[i}l^{j+]}$
    \hspace{-0.9em}&$\delta^{ei}l^i$  \\
\hline $l^{e+}$ & \hspace{-0.5em}$l^{i+}
   \delta^{e{}j}$\hspace{-0.7em} & \hspace{-0.7em}
   $l_{j}^+\delta^{e}_{i}$ \hspace{-1.0em} &
   $0$&\hspace{-0.3em}
      \hspace{-0.3em}
   $\delta^{ei}l_0$\hspace{-0.3em}
    \hspace{-0.3em}
   &\hspace{-0.5em} $0$\hspace{-0.5em}
    &\hspace{-0.7em} $ -\textstyle\frac{1}{2}\delta^{e[i}l^{j]}
    $\hspace{-0.7em} & $0$ &$-l^{i+}\delta^{ei}$  \\
\hline $l^{ep}$
    & \hspace{-0.3em}$-\delta^{i[p}l^{e]j}$
    \hspace{-0.5em}  &\hspace{-0.5em} $-\delta_{j}^{[p}l^{e]}{}_{i}$\hspace{-0.3em}
   & $0$&\hspace{-0.3em}
    $0$\hspace{-0.5em} & \hspace{-0.3em}
    $ \hspace{-0.7em}\textstyle-\frac{1}{2}l^{[e}\delta^{p]i}
    \hspace{-0.5em}$\hspace{-0.3em}
    & $0$ & \hspace{-0.7em}$ -L^{ij,ep}$\hspace{-0.7em}& $\hspace{-0.7em}  {\delta^{i[p}l^{e]i}}\hspace{-0.7em}$\hspace{-0.7em} \\
\hline $l^{ep+}$
    & $ {l^{i[p+}\delta^{{e}]j}}$ & $ -{l_j{}^{[e+}\delta^{p]}_i}$
   & $0$&\hspace{-0.3em}
    $\hspace{-0.2em} \textstyle\frac{1}{2}l^{[e+}\delta^{p]i}$\hspace{-0.5em} & \hspace{-0.3em}
    $0$\hspace{-0.3em}
    & $\textstyle L^{ep,ij}$ & $0$ &$\hspace{-0.5em} {{\delta^{i[e}l^{p]i+}}}\hspace{-0.3em}$\hspace{-0.2em} \\
\hline\hline $g^e_0$
    & $- F^{ij,e}$ & $ + F_{ij}{}^{e+}$
   &$0$& \hspace{-0.3em}
    $\hspace{-0.2em}-\delta^{ie}l^e$\hspace{-0.5em} & \hspace{-0.3em}
    $l^{e+}\delta^{ei}$\hspace{-0.3em}
    & \hspace{-0.7em}$ {\delta^{e[i}l^{j]e}}$\hspace{-0.7em} & ${\delta^{e[j}l^{i]e+}}$& $0$ \\
   \hline\hline
\end{tabular}
\end{center}}} \vspace{-2ex}\caption{HS symmetry  superalgebra  $\mathcal{A}(Y[k],
\mathbb{R}^{1,d-1})$.\label{table in} }\end{table}

First note that,  in the table~\ref{table in},  the squared brackets for the indices $i$,
$j$ in the quantity $A^{[i}B^{j]k}$
mean the antisymmetrization
 $A^{[i}B^{j]k}$ =
$A^{i}B^{jk}-
A^{j}B^{ik}$ as well as these indices are
raising and lowering by means of Euclidian metric tensors
$\delta^{ij}$, $\delta_{ij}$, $\delta^{i}_{j}$}.
Second,  the products
$A^{ep,ij}$ ($A^+_{ij,ep}$), $B^{ep}_{ij}$ for $e<p$, $i<j$ and $L^{ep,ij}$ are determined as
\begin{eqnarray}\label{teptij}
 \hspace{-1em} A^{ep,ij} & = & t^{ip}\delta^{ej}-t^{ej}\delta^{ip}, \quad \big(A^{+ep,ij} \ =\ t^{+ip}\delta^{ej}-t^{+ej}\delta^{ip} \big),   \\
   \label{teptijt+}
 \hspace{-1em}  B^{ep}{}_{ij}& = & \delta^{p}_j\Big(\delta^{e}_i g_{0{}i} - \theta^e{}_{i}
  t^{+}_i{}^e - \theta_{i}{}^e
  t^{e}{}_i \Big) - \delta^{e}_i\Big(\delta^{p}_j g_{0{}j} - \theta_{j}{}^{p}
  t^{+p}{}_j - \theta^{p}{}_{j}
  t_{j}{}^p \Big), \\
  \hspace{-1em} L^{ep,ij} &=&   \textstyle\frac{1}{4}\Bigl\{\delta^{e[i}
\delta^{j]p}\Bigl[ g_0^{p} +
g_0^{e}\Bigr]  - \delta^{e[i}\Bigl[t^{pj]}\theta^{j]p} +t^{j]p+}\theta^{pj]}\Bigr] + \delta^{p[i}\Bigl[t^{j]e+}\theta^{ej]}
+t^{ej]}\theta^{j]e}\Bigr] \Bigr\}
 \,.\label{Lklij}
\end{eqnarray}
The quantities $F^{ep,i} = [t^{ep}, g_0^i\}$ ($F^{ep,i+}=[ g_0^i, t^{ep+}\}$) are described according to
\begin{equation}\label{teptij1}
  F^{ep,i} =  t^{ei}\delta^{pi}-t^{ip}\delta^{ei},\quad F^{ep,i+} =
   t^{ei+}\delta^{pi}-t^{ip+}\delta^{ei}.
\end{equation}

Traditionally the algebra of these operators is called the \emph{integer
higher-spin symmetry algebra in Minkowski space with a Young
tableaux having $k$ columns}\footnote{one should not identify the
term "\emph{higher-spin symmetry algebra}" using here for free HS
formulation starting from the paper \cite{0505092} with the
algebraic structure known as "\emph{higher-spin algebra}" (see,
for instance Ref.\cite{Vasiliev_inter}, \cite{Vasiliev_inter1}, \cite{Vasiliev_inter2}) arising to enlarge Poincare transformations by spin generator and to  describe the HS gravity} and denote it as $\mathcal{A}(Y[k],
\mathbb{R}^{1,d-1})$.

From the table~\ref{table in} it is obvious that D'alambertian
$l_0$ being by the Casimir element of the Poincare algebra
$iso(1,d-1)$  belongs to the center of superalgebra $\mathcal{A}(Y[k],
\mathbb{R}^{1,d-1})$ as well. The  elements $o_A$  of the algebra
$\mathcal{A}(Y[k], \mathbb{R}^{1,d-1})$  forms the subsuperalgebra
which describes the isometries of Minkowski space $R^{1,d-1}$. It
may be realized as direct sum (Cartan decomposition) of $k$-dimensional commutative
algebra $T^{[k]} = \{l_i\}$ and its dual $T^{[k]*}=\{l^{i+}\}$,
\begin{equation}\label{TkTk}
    \{l^i, l^{i+}, l_0\} = (T^{[k]} \oplus T^{{[k]}*}\oplus [T^{[k]},
    T^{{[k]}*}]),\quad [T^{[k]},
    T^{{[k]}*}] \sim l_0.
\end{equation}
Now, we may to describe shortly the structure of the
Lorentz-module $P^d_k(a^+)$ of all finite string-like
vectors of the form  (\ref{PhysState})   on a base of  Howe duality
\cite{Howe1} applied to the case of integer spin representations of
Lorentz group $SO(1,d-1)$. The Howe dual algebra to
$so(1,d-1)$ is real form of $so(2k,\mathbb{C})$, i.e.
$so(k,k)$ if $k=\left[\frac{d-1}{2}\right]$
with the following basis elements \cite{Howe1} for arbitrary $i,j
= 1,...,k$,
\begin{equation}\label{basissp2n}
     \hat{l}_{ij} = a^{\mu+}_j a_{i{}\mu}^+,\qquad  \hat{t}_{i}{}^j = \frac{1}{2}[a_{i{}\mu}^+,\;a^{j{}\mu }],\qquad \hat{l}^{ij} =
    a^{i\mu} a_{{}\mu}^j,
\end{equation}
which is distinguished from the elements of $\mathcal{A}^f(Y[k],
\mathcal{R}^{1,d-1})$ by the sign "hat". Their non-vanishing
supercommutator's relations have the form
\begin{align}
 & [\hat{t}_{i_1}{}^{j_1},\; \hat{t}_{i_2}{}^{j_2}] \ =\  \hat{t}_{i_1}{}^{j_2}
  \delta_{i_2}^{j_1} - \hat{t}_{i_2}{}^{j_1}
  \delta_{i_1}^{j_2},&&  [\hat{l}^{i_2{}j_2},\; \hat{l}_{i_1{}j_1}]\ =\ \delta^{[i_2}_{[i_1}\hat{t}_{j_1]}{}^{j_2]},  \nonumber \\
 & [\hat{t}_{i_1}{}^{j_1},\; \hat{l}_{i_2j_2}]  =  \hat{l}_{i_1[i_2}\delta_{j_2]}^{j_1} ,
 &&  [\hat{t}_{i_1}{}^{j_1},\; \hat{l}^{i_2j_2}]\ =\
 \hat{l}^{j_1[j_2}\delta^{i_2]}_{i_1} .
  \label{comrelsp}
\end{align}
The elements  $l^{ij}, l^{ij+}, t^{i_1j_1},
t^+_{i_1j_1}, g_0^i$  from  HS symmetry algebra
$\mathcal{A}^f(Y[k], \mathbb{R}^{1,d-1})$ are derived from the
basis elements of $so(k,k)$ by the rules (for $k=2$ when $so(2,2) \sim sl(2,\mathbb{R})\oplus sl(2,\mathbb{R})$ case see
 \cite{resh1}),
\begin{equation}\label{osp2nhssa}
   l_{ij}^+ = \frac{1}{2}\hat{l}_{ij}, \quad {l}^{ij} =
    \frac{1}{2}\hat{l}^{ij},\quad
     {t}_{i}{}^j = \hat{t}_{i}{}^j\theta^{ji},\quad  {{t}^{j}{}_{i}}^+{} = \hat{t}_{i}{}^j\theta^{ij},\quad
     g_0^i=-
     \hat{t}_{i}{}^i.
\end{equation}
so that  $\mathcal{A}(Y{[k]},
\mathbb{R}^{1,d-1})$ represents the semidirect sum of Lie
 algebra generated by $\{o_a, o_a^+;\hspace{-0.15em} g_0^i\}$ [as an algebra of internal derivations
of $(T^{[k]} \oplus T^{{[k]}*})]$ with $(T^{[k]} \oplus T^{{[k]}*}\oplus [T^{[k]},
    T^{{[k]}*}])$ analogously to interpretation of the algebra  $\mathcal{A}(Y(k), \mathbb{R}^{1,d-1})$ \cite{1110.5044},
\begin{equation}\label{identalg}
    \mathcal{A}(Y[k], \mathbb{R}^{1,d-1}) = \left(T^{[k]} \oplus T^{{[k]}*}\oplus [T^{[k]},
    T^{{[k]}*}]\right) + \hspace{-1em} \supset   so(k,k).
\end{equation}
 To be complete let us stress  that the Lie subalgebra of $\{ o_a, o_a^+ ;\ g_0^i\}$ is isomorphic for $k=3$ to $sl(4,\mathbb{R}) \sim so(3,3)$  \cite{RBP}.

 Explicitly, any element $X_{2k}$ of $so(k,k)$, see e.g. \cite{BarutRonchka}, has the matrix form composed from generators (for all parameters equal to $1$)
\begin{equation}\label{Howeso}
  X_{2k} = \begin{pmatrix}
               X_{1|k} & X_{2|k} \\
               X^T_{2|k} & X_{3|k} \\
             \end{pmatrix}, \ (X_{1|k})_{ij}=\hat{l}^{ij}, \ (X_{2|k})_{ij}=\hat{t}_{i}{}^{j}, \ (X_{3|k})_{ij}=\hat{l}_{ij}
   \end{equation}
   with arbitrary $X_{2|k}$  and skew-symmetric $X_{1|k}, X_{3|k}$ of orders $k$.

Having constructed the HS symmetry superalgebra, we  can  not still
construct complete BRST operator $Q$, for the unconstrained    LF\footnote{However, the superalgebra $\mathcal{A}(Y[k], \mathbb{R}^{1,d-1})$ is sufficient (see Sec.~\ref{constrlagrform} below) for constrained LF with incomplete BRST operator, $Q_c=\eta_0l_0+q_il_i^+ +l_iq_i^+ + \imath q_i q_i^+\mathcal{P}_0$, with half of operator   constraints $o_a$.}  with respect to the elements $o_I$
from  $\mathcal{A}(Y[k], \mathbb{R}^{1,d-1})$ because of a
presence of the non-degenerate in the Fock space $\mathcal{H}^f$
operators $g_0^i$ determining following to the relations
(\ref{inconstraints}) the system of $o_I$ as  one containing the
second-class operator constraints subsystem.   Due to general property
of  BRST-BFV method \cite{BFV} considered in details in \cite{resh2}  a such complete BRST operator $Q$ would not
reproduce the right set of initial constraints (\ref{lilijt}) in the zero ghost $Q$-cohomology subspace of total
Hilbert space, $\mathcal{H}_{tot}$ ($\mathcal{H}^f \subset
\mathcal{H}_{tot}$). To resolve the problem, we as usual consider the
procedure of conversion the set of initial operators $o_I$ into one of deformed $O_I$ which
would be by the first-class operator constraints $\{O_I\}\setminus \{G_0^i\}$ only  on the subspaces with
except for extended number particles operators $G_0^i$.

\section{Deformed  HS symmetry superalgebra}\label{Vermamodule}
Here  we describe the method of auxiliary representation
construction for the algebra  with second-class operator
constraints alone, in terms of finite polynomials in powers of new creation and annihilation
operators from auxiliary Fock space.

\subsection{Oscillator realization of the
additional parts to constraints}\label{addconvers}

Within a standard additive conversion procedure
developed in the approach with complete BRST operator, see for instance,
\cite{BurdikPashnev1},  \cite{1110.5044}, \cite{resh1}, \cite{BuchKrycRysTak}, \cite{0001195}
  which
implies the enlarging of $o_I$ to $O_I = o_I + o'_I$, where
additional parts $o'_I$ are given on a new Fock space
$\mathcal{H}'$  independent on $\mathcal{H}^f$:
$\mathcal{H}'\bigcap \mathcal{H}^f = \emptyset$. In this case the
elements $O_I$ are given on tensor product $\mathcal{H}^f\otimes
\mathcal{H}'$ so that the requirement for $O_I$ to be in
involution, i.e. $[O_I,\ O_J] \sim O_K$,  leads to the series of
the same algebraic relations,
\begin{align}\label{addrel}
&[o'_I,\ o'_J]= f^K_{IJ}o'_K, && [O_I,\ O_J]= f^K_{IJ}O_K
\end{align}
as ones for $o_I$ with the  structure constants $f^K_{IJ}$ determined by (\ref{geninalg}) for the original
set of $o_I$.

Because of only the generators, which do not contain space-time derivatives, $\partial_{\mu}$, are the second-class operator
constraints  in $\mathcal{A}(Y[k],\mathbb{R}^{1,d-1})$ to be
converted then instead of all $o'_I$ in (\ref{addrel}) one should
be used only part of them, namely $\{o'_a, {o'}^+_a\}$ and $g_0^{i \prime }$. Therefore,
one should to get new operator realization of  the  subalgebra
$o'_I$ isomorphic to $so(k,k)$.

 The conversion problem is solved with help of special procedure
known in the mathematical literature as Verma module construction
\cite{Verma1}, \cite{Dixmier} for the latter algebra admitting the Cartan  decomposition (\ref{Cartandecomp}) which results explicitly
derived in the appendix~\ref{ap1}.

A scalar oscillator realization of converted subalgebra (known also as Fock module) is found  in the  appendix~\ref{oscrealsl2kdet}.
The result has the form
\begin{eqnarray}
 l^{\prime+}_{ij} & = & b_{ij}^+\,,
 \label{l'+ijFaa}\\
g_0^{\prime i}& = &  \sum_{l<i}b^+_{li}b_{li} + \sum_{l>i} b^+_{il}b_{il}  - \sum_{s>i}d^+_{is}d_{is}+\sum_{s<i}d^+_{si}d_{si}+ h^i
 \,,\label{g'0iFaa} \\
  t^{\prime+}_{lm}   & = & d^+_{lm} - \sum_{n=1}^{l-1}d_{nl}d^+_{nm}-\sum_{n=1}^{l-1} b^+_{nm}b_{nl}+
  \sum_{n=l+1}^{m-1} b^+_{nm}b_{ln}-\sum_{n=m+1}^{k} b^+_{mn}b_{ln}
  \,,
 \label{t'+lmaa}
 \end{eqnarray}
 for the elements $l^{\prime }_{lm}$  when $l<m$
\begin{eqnarray}
 l^{\prime }_{lm}&=&-
 \frac{1}{4}\sum\limits_{n=1}^{l-1} \Bigl[\sum_{p=n+1}^{m-1}
 b^+_{np}b_{pm }- \sum_{p=m+1}^{k}
 b^+_{np}b_{mp }+ C^{nm}(d^+,d)- \sum_{n'=1}^{n-1}d^+_{n'n}d_{n' m}
 \label{l'lmboseaa}\\
 \hspace{-1em}&\hspace{-1em}+&\hspace{-1em}
\sum_{p=1}^{m-n-1}\Big(\sum_{k_1=n+1}^{m-1}\ldots \sum_{k_p=k_{p-1}+1}^{m-1}\Big\{
 C^{k_{p}m}(d^+,d)- \sum_{n'=k_{p-1}}^{k_p-1}d^+_{n'k_p}d_{n' m} \Big\}\prod_{j=1}^{p}d_{k_{j-1}k_{j}}\Big)\Bigr]b_{nl}
 \nonumber\\
\hspace{-1.5em}&\hspace{-1em}+& \hspace{-0.5em}
\frac{1}{4}\textstyle\sum\limits_{n=l+1}^{m-1} \Bigl[-\sum\limits_{p=l}^{n-1}
 b^+_{pn}b_{pm }+ \sum\limits_{p=n+1}^{m-1}
 b^+_{np}b_{pm }-\sum\limits_{p=m+1}^{k}
 b^+_{np}b_{mp }  + C^{nm}(d^+,d)- \displaystyle\sum_{n'=1}^{n-1}d^+_{n'n}d_{n' m}\nonumber
 \\
 \hspace{-1.5em}&\hspace{-1em}+&\hspace{-0.5em}
  \sum_{p=1}^{m-n-1}\Big(\sum_{k_1=n+1}^{m-1}\ldots \sum_{k_p=k_{p-1}+1}^{m-1}\Big\{
 C^{k_{p}m}(d^+,d)- \sum_{n'=k'_{p-1}}^{k_p-1}d^+_{n'k_p}d_{n' m} \Big\}\prod_{j=1}^{p}d_{k_{j-1}k_{j}}\Big)\Bigr]b_{ln}
 \nonumber
  \\
\hspace{-1.5em}&\hspace{-1em}+& \hspace{-0.5em} \frac{1}{4}\textstyle\sum\limits_{n=1}^{l-1} \Bigl[\sum\limits_{p=n+1}^{l-1}
 b^+_{np}b_{pl }- \sum\limits_{p=l+1}^{k}
 b^+_{np}b_{lp }+C^{nl}(d^+,d)- \displaystyle\sum_{n'=1}^{n-1}d^+_{n'n}d_{n' l}
\nonumber
\end{eqnarray}
 \vspace{-1.5ex}
\begin{eqnarray}
\hspace{-1.5em}&\hspace{-1em}+& \hspace{-0.5em}
\sum_{p=1}^{l-n-1} \Big(\sum_{k_1=n+1}^{l-1}\ldots \sum_{k_p=k_{p-1}+1}^{l-1}\Big\{
 C^{k_{p}l}(d^+,d)- \sum_{n'=k'_{p-1}}^{k_p-1}d^+_{n'k_p}d_{n' l} \Big\}\prod_{j=1}^{p}d_{k_{j-1}k_{j}} \Big)\Bigr]b_{nm}
 \nonumber\\
\hspace{-1.5em}&\hspace{-1em}-& \hspace{-0.5em}\frac{1}{4}\Bigl(b^+_{lm}b_{lm}+\sum_{n=m+1}^k (b^+_{ln}b_{ln}+b^+_{mn}b_{mn}) +  \sum_{n=
l+1}^{m-1}b^+_{nm} b_{nm}  - \sum_{s>l}d^+_{ls}d_{ls} -
\sum_{s>m}d^+_{ms}d_{ms} \nonumber\\
\hspace{-1em}&\hspace{-1em}+& \sum_{r<l}d^+_{rl}d_{rl} +\sum_{r<m}d^+_{rm}d_{rm} + h^{l}+
h^{m}\Bigr)b_{lm} \nonumber\\
 \hspace{-1.5em}&\hspace{-1.2em}+& \hspace{-0.9em} \frac{1}{4}\textstyle\sum\limits_{n=m+1}^{k} \Bigl[ d^+_{mn}-
\sum\limits_{n'=1}^{m-1} d^+_{n'n}d_{n'm} -
\sum\limits_{n'=l+1}^{m-1}b^+_{n'n}b_{n'm}+\sum\limits_{n'=m+1}^{n-1}b^+_{n'n}b_{mn'} - \sum\limits_{n'=n+1}^{k}b^+_{n n'}b_{mn'}   \Bigr]{b}_{ln}\nonumber\\
\hspace{-1.5em}&\hspace{-1em}+& \hspace{-0.7em}  \frac{1}{4}\textstyle\sum\limits_{n=l+1}^{m-1}
\Bigl[ d^+_{ln} - \sum\limits_{n'=1}^{l-1}d^+_{n'n}d_{n'l}
\Bigr]{b}_{nm} - \displaystyle\frac{1}{4} \textstyle\sum\limits_{n=m+1}^{k}
\Bigl[ d^+_{ln} - \sum\limits_{n'=1}^{l-1}d^+_{n'n}d_{n'l}
\Bigr]{b}_{mn}\nonumber,
 \end{eqnarray}
 and for  $t^{\prime }_{lm}$ (with notation $k_0\equiv l$),
 \begin{eqnarray}
t^{\prime }_{lm} &=&C^{lm}(d^+,d)- \sum_{n'=1}^{l-1}d^+_{n'l}d_{n'm} +
\sum_{p=1}^{m-l-1}\bigg[\sum_{k_1=k_0+1}^{m-1}\ldots \sum_{k_p=k_{p-1}+1}^{m-1}
 \Big\{C^{k_{p}m}(d^+,d)\label{t'lmFaa} \\
 && - \sum_{n'=k'_{p-1}}^{k_p-1}d^+_{n'k_p}d_{n'm} \Big\}\prod_{j=1}^{p}d_{k_{j-1}k_{j}}\bigg] -\sum_{n=1}^{l-1}b^+_{nl}b_{nm}
 + \sum_{n=l+1}^{m-1}b^+_{ln}
b_{nm}-\sum_{n=m+1}^{k}b^+_{ln}
b_{mn}\nonumber
\end{eqnarray}
where the  operators $ C^{k_{p}m}(d^+,d)$, $l<m$ given in (\ref{Clm}) and coincide with ones obtained for mixed-symmetric HS fields \cite{1110.5044} (in symmetric basis) for symplectic algebra $sp(2k)$.
The operators $t^{+\prime}_{rs}$ and $t^{\prime}_{rs}$; $l^{+\prime}_{ij}$ and $l^{\prime}_{ij}$ are respectively Hermitian conjugated to each other, as well as  the number particles operators $g_{0}^{i\prime}$ is Hermitian with help of Grassmann-even operator  $K'$ determined from the system (\ref{systemK}) - (\ref{explicit K}) in appendix~\ref{oscrealsl2kdet}.

One can directly check that these operators form the same algebra as one for $\{{o}_a, {o}^+_a\}$  in the table \ref{table in}.

Now, as it was considered previously in \cite{1110.5044} for mixed-symmetric massive HS fields subject to $Y(s_1,...,s_k)$,  we turn to the case of the massive  MAS  HS fields whose
system of second-class operator constraints contains additionally to
elements of $so(k,k)$ algebra  the operator constraints  of isometry
subalgebra of  Minkowski space  $l^i, l^+_i $.

Note, for $k=1$ (totally antisymmetric tensor fields) there is no any second-class constraints, and for $k=2,3$ group of antisymmetric indices the oscillator realizations coincide with ones obtained in \cite{resh0}, \cite{resh1} and \cite{RBP}.

\subsection{Auxiliary representations of the superalgebra
$\mathcal{A}_m(Y[k],\mathbb{R}^{1,d-1})$ for massive  HS
fields}\label{auxtmasrep}

 An  oscillator representations for the HS
symmetry superalgebra of massive bosonic HS fields with mass $m$ implies the change of
the wave equation given by (\ref{Eq-0b})   on
Klein-Gordon equation corresponding to the operator constraint $\tilde{l}_0$
($\tilde{l}_0=\partial^\mu\partial_\mu +m^2$), which acts on the same
string-vector $|\Phi\rangle$ (\ref{PhysState})
\begin{eqnarray}
\label{Eq-0bm} && (\partial^\mu\partial_\mu+
m^2)\Phi_{\mu^1[{s_1}],\ldots , \mu^k[{s_k}]}
 =0 \Longleftrightarrow \tilde{l}_0|\Phi\rangle =0.
\end{eqnarray}
  We may to consider the procedure described above in
section~\ref{addconvers}  and in details realized in the
Appendices~\ref{ap1},~\ref{oscrealsl2kdet}  for maximal
subalgebra without derivatives (see  comments in the Appendix~\ref{addalgebram} for
the massive case).
    However,
we may used a procedure of the dimensional reduction of the
initial algebra $\mathcal{A}(Y[k],\mathbb{R}^{1,d})$ for massless
HS fields in $(d+1)$-dimensional flat background to the superalgebra $\mathcal{A}_m(Y[k],\mathbb{R}^{1,d-1})$ with
dimension $d$, $\mathbb{R}^{1,d-1}$.

To realize it  we write down the rules of the dimensional reduction
from  $\mathbb{R}^{1,d}$  space-time  coordinated by numbers $x^M=(x^\mu, x^d)$  to $\mathbb{R}^{1,d-1}$  with $(2^k -1)$ auxiliary independent tensor fields of lesser ranks than $\sum_{l=1}^ks_l$ down to the field $\Phi_{{\mu^1[{s_1-1}]d,\ldots,\mu^{l}[s_{l}-1]d,\ldots , \mu^k[{s_k-1}]d}} \equiv$ $\tilde{\Phi}_{{\mu^1[{s_1-1}],\ldots,\mu^k[{s_k-1}]}} $:
\begin{eqnarray}
\label{dimrf} \hspace{-0.7em} &\hspace{-0.7em}&\hspace{-0.7em} \Phi_{{M^1[{s_1}],\ldots , M^k[{s_k}]}}(x^M) = \exp\{\imath m x^d\}\Big(\Phi_{{\mu^1[{s_1}],\ldots , \mu^k[{s_k}]}}, \ldots , \tilde{\Phi}_{{\mu^1[{s_1-1}],\ldots,\mu^k[{s_k-1}]}} \Big)(x^{\mu}),\nonumber
\end{eqnarray}
\vspace{-2.9ex}
\begin{align} \label{reduction}
   &\partial^{M} = (\partial^{\mu}, -\imath m)\,, &&\hat{a}^{M}_i = (\hat{a}^{\mu}_i, f_i)\,, &&
   \hat{a}^{M{}+}_i = (\hat{a}^{\mu{}+}_i, f_i^+)\,,  \\
   &M=0,1,\ldots ,d\,, && \mu=0,1,\ldots ,d-1\,, && \eta^{MN} =
   diag (1,-1,\ldots,-1,-1)\,,\label{reduction1}
\end{align}
Following to  (\ref{reduction}), (\ref{reduction1}) we get for the set of the original elements $o_I$ from
the massless HS symmetry superalgebra
$\mathcal{A}(Y[k],\mathbb{R}^{1,d})$ the ones $\tilde{o}_I$ in
massive HS symmetry superalgebra $\mathcal{A}_m(Y[k],\mathbb{R}^{1,d-1})$
 as follows,
\begin{align}
    &\tilde{l}_0 = \partial^{M}\partial_{M}=
l_0+ m^2, && \tilde{g}^i_{0} = - \frac{1}{2}[\hat{a}^+_{Mi}\hat{a}^{M}_i] =
g_0^i -\frac{1}{2} + f^+_if_{i} ,\label{l0tilde} \\
\label{litilde}& \tilde{l}_i = -i\hat{a}^M_i\partial_{M}= {l}_i +mf_i,
&& \tilde{l}_i^+
= -i\hat{a}^{+M}_i\partial_{M}= {l}^+_i + mf^+_i,\\
& \tilde{l}_{ij} = \frac{1}{2}\hat{a}^M_i\hat{a}_{Mj} = {l}_{ij} -
\frac{1}{2}f_if_{j}, && \tilde{l}^+_{ij} =
\frac{1}{2}\hat{a}^{M+}_j\hat{a}^+_{Mi} = {l}^+_{ij} -
\frac{1}{2}f^+_jf^+_{i}, \\
& \tilde{t}_{ij} = \hat{a}^{M+}_i\hat{a}_{Mj} = {t}_{ij} -
f^+_{i}f_j , && \tilde{t}^+_{ij} =
\hat{a}^{M+}_j\hat{a}_{Mi} = {t}^+_{ij} - f^+_{j}f_i.
\label{exprnew}
\end{align}
The generators $(\tilde{l}_0, {l}^+_i, {l}_i, {l}_{ij},
{l}^+_{ij}, {t}_{ij}, {t}^+_{ij}, g_0^i)$ obey the same
commutation relations as in the table~\ref{table in} for massless HS
symmetry superalgebra with except for the pair of the anticommutators,
\begin{equation}\label{ll+}
    \{l_i, l^+_i\} = (\tilde{l}_0 - m^2), \ \mathrm{for} \ i=1,...,k.
\end{equation}
 Relations  (\ref{l0tilde}), (\ref{ll+}) indicate
the presence of $2k$ additional second-class operator constraints, $l_i,
l_i^+$, with corresponding fermionic oscillators $f_i, f_i^+$,
$\{f_i, f_j^+\} = \delta_{ij}$, as compared to the massless
case.

Interestingly, that  the elements with "tilde" in the
 equations (\ref{l0tilde})--(\ref{exprnew}) satisfy the algebraic
relations for massless HS symmetry superalgebra
$\mathcal{A}(Y[k],\mathbb{R}^{1,d-1})$ now without central charge
(i.e. those quantities $\tilde{o}_I$ contains the same
second-class operators as ${o}_I$ in massless case). As the consequence,
the converted operator constraints $O_I$, $O_I = o_I + o'_I$, in massive
case are given by the relations,
\begin{equation}\label{conv}
O_I = \tilde{o}_I + o'_I, \qquad M^2 = m^2+{m'}^2=0,
\end{equation}
where additional parts $o'_I = o'_I(b_{i},b^+_{i})$  are determined by the relations
(\ref{g'0iFa})--(\ref{t'lmFa}).

Resuming, the auxiliary representation (Verma module) for the maximal semi-simple subalgebra from  $\mathcal{A}'(Y[k],\mathbb{R}^{1,d})$
 determines with use of the dimensional reduction procedure
the oscillator realization (Fock module) for the additional parts of massive HS
symmetry superalgebra $\mathcal{A}'_m(Y[k],\mathbb{R}^{1,d-1})$
completely.

\section{BRST-BFV operator and spectral problem}\label{BRSToperator}
\setcounter{equation}{0}

Because of the converted HS symmetry superalgebra $\mathcal{A}_c(Y[k],\mathbb{R}^{1,d-1})$ is a Lie superalgebra, the BRST operator, $Q'$, for it can be constructed according to the standard prescription \cite{BFV, BFV1, BFV2} for unconstrained case \cite{resh2}  by means of introduction to the each operator $O_I$  of a pair of ghost
coordinate  $\mathcal{C}^I$ and  momentum $\overline{\mathcal{P}}_I$ operators: $gh_H(\mathcal{C}^I)$ = $- gh_H(\overline{\mathcal{P}}_I)$= $1$,    with non-trivial supercommutation relations, $[\mathcal{C}^I, \overline{\mathcal{P}}_J\} = \delta^I_J$, and whose Grassmann parity is opposite to one of  $O_I$.
The Grassman-odd nilpotent operator $Q'$, $gh_H(Q')=1$, is given in the form (\ref{QQc}) for converted $O_I$
with structural constants, $f^K_{JI}$, determined by the Eq. (\ref{addrel}) and table~\ref{table in}.
Following to this receipt, we introduce Grassman-even ghost coordinates  $q_i^+$, $q_i$ and  ghost momenta  $p_j$, $p_j^{+}$ for Grassman-odd basis elements $l_i$,
$l_i^+$  of the superalgebra $\mathcal{A}_C(Y[k],\mathbb{R}^{1,d-1})$,   respectively, and Grassman-odd ghost coordinates $\eta_0$,  $\eta_{ij}^+$, $\eta_{ij}$, $\vartheta_{rs}^+$, $\vartheta_{rs}$, $\eta_{i}^G$
and  momenta $\mathcal{P}_0$,  $\mathcal{P}_{ij}$, $\mathcal{P}_{ij}^+$, $\lambda_{rs}$, $\lambda_{rs}^+$, $\mathcal{P}^G_{i}$ for Grassman-even basis elements $l_0$,  $L_{ij}$,
$L_{ij}^+$, $T_{rs}$,  $T_{rs}^+$,  $G_0^{i}$ of $\mathcal{A}_C(Y[k],\mathbb{R}^{1,d-1})$,  respectively, which is subject to non-trivial commutation relations,
\begin{gather}
  [q_i^+,p_j]=[p_i^+,q_j]=\delta_{ij},
 \qquad \{\eta_0, \mathcal{P}_0\}=\imath, \qquad \qquad  \{\eta^G_{i},\mathcal{P}^G_{j}\}=\imath \delta_{ij}, \label{ghost_def}\\
\{\eta_{ij}^+,\mathcal{P}_{i'j'}\}=\{\eta_{ij},\mathcal{P}_{i'j'}^+\}=\delta_{i i',jj'}, \quad
\{\vartheta_{rs},\lambda_{r's'}^+\}=\{\vartheta_{rs}^+,\lambda_{r's'}\}=\delta_{r r',ss'}. \nonumber
\end{gather}
 The zero-mode ghosts satisfy to the following hermitian conjugation rules,  $(\eta_0, \mathcal{P}_0)^+ = (\eta_0, - \mathcal{P}_0)$,
$ (\eta^G_{i},\mathcal{P}^G_{j})^+ = (\eta^G_{i},-\mathcal{P}^G_{j})$, whereas for $\eta_{ij}^{(+)},\mathcal{P}_{ij}^{(+)}$ the properties hold: $\big(\eta_{ij}^{(+)},\mathcal{P}_{ij}^{(+)}\big)$ = $-\big(\eta_{ji}^{(+)},\mathcal{P}_{ji}^{(+)}\big)$.

After application of the formula (\ref{QQc}) we can write down nilpotent BRST-BFV operator $Q'$ (with natural Wick-ordering for $\mathcal{C},\overline{\mathcal{P}}$ in question)  as
\begin{eqnarray}
Q'& = &    \eta_0l_0 +l_i q^+_i +  q_il^+_i  +iq_iq^+_i \mathcal{P}_0 + \sum_{i<j}\Big(\eta_{ij}{L}{}^+_{ij} + {L}_{ij}\eta^+_{ij}+  T_{ij}\vartheta_{ij}^+ +\vartheta_{ij}T^+_{ij}
\Big)  \label{Q'} \\
&&  +
\frac{1}{2}\sum_{i<j}\Big(\eta_{ij}q_{[i}^+ p_{j]}^+
+\eta^+_{ij}q_{[i} p_{j]}\Big)   +  \sum_{i<j}\Big((q_j^+p_i+q_ip_j^+ )\vartheta_{ij}+ \vartheta_{ij}^+(q_jp_i^++q_i^+p_j)\Big)\nonumber\\
&& + \sum_{i<j<p}[\vartheta^+_{jp}\lambda_{ip}\vartheta^+_{ij}+ \vartheta_{ij}\lambda^+_{ip}\vartheta_{jp}]  - \sum_{i<j<p}\vartheta^+_{ip}[\vartheta_{ij}\lambda_{jp}-\vartheta_{jp}\lambda_{ij}] \nonumber\\
&& - \sum_{i<j<p}\vartheta_{ip}[\vartheta_{jp}^+\lambda_{ij}^+ -\vartheta_{ij}^+\lambda_{jp}^+] \nonumber\\
&&  +\sum_{i<j}\eta_{ij}^+\big[\sum_{p<j}\vartheta_{ip}^+\mathcal{P}_{pj}
+\sum_{j<p} \vartheta_{jp}^+\mathcal{P}_{ip}-\sum_{j<p}\vartheta_{ip}^+\mathcal{P}_{j p} \big]\nonumber\\
&&  -\sum_{i<j}\eta_{ij}\big[\sum_{p<j}\vartheta_{ip}\mathcal{P}^+_{pj}
+\sum_{j<p} \vartheta_{jp}\mathcal{P}_{ip}^+-\sum_{j<p}\vartheta_{ip}\mathcal{P}_{j p}^+ \big] \nonumber
\\
&&  +\sum_{i<j<p}\vartheta_{ij}^+\eta_{jp}\mathcal{P}_{ip}^+ -  \sum_{i<j<p} \big[\vartheta_{ip}^+\eta_{jp} + \eta_{ip}\vartheta^+_{jp}
 \big]\mathcal{P}_{ij}^+  \nonumber\\
&&  -\sum_{i<j<p}\vartheta_{ij}\eta_{jp}^+\mathcal{P}_{ip} +  \sum_{i<j<p} \big[\vartheta_{ip}\eta_{jp}^+ + \eta_{ip}^+\vartheta_{jp}
 \big]\mathcal{P}_{ij}+\frac{1}{4}\sum_{i<j<p}\Big\{\eta_{ij}^+\big[\eta_{ip}\lambda^+_{jp}-\eta_{jp}\lambda^+_{ip}\big]
 \nonumber
 \\
&&  -\eta_{ij}\big[\eta_{ip}^+
\lambda_{jp}-\eta_{jp}^+\lambda_{ip}\big] + \eta_{ip}^+\eta_{jp}\lambda^+_{ij}+\eta_{jp}^+\eta_{ip}\lambda_{ij}
    \Big\}
+ \sum_i\eta^G_{i}\sigma^i(G)\nonumber
 \end{eqnarray}
 \vspace{-1.5ex}
\begin{eqnarray}
&&
  - i \sum_{i<j}\left[  \frac{1}{4}
\eta_{ij}^+ \eta_{ij}(\mathcal{P}_{i}^G+\mathcal{P}_{j}^G)-\vartheta_{ij}^+\vartheta_{ij}(\mathcal{P}_{i}^G - \mathcal{P}_{j}^G)\right] \nonumber.
\end{eqnarray}
with generalized spin operator
\begin{eqnarray}
 && \sigma_i(G) = G_{0i}- q_i^+ p_i  - q_i p_i^+  + \sum_{j>i}[\eta_{ij}^+ \mathcal{P}_{ij} -\eta_{ij} \mathcal{P}^+_{ij}]+
  \sum_{j<i}[\eta_{ji}^+ \mathcal{P}_{ji} -\eta_{ji} \mathcal{P}^+_{ji}] \label{gspin}\\
   &&  + \sum_{l<i}[\vartheta^+_{li}
\lambda_{li} - \vartheta_{li}\lambda^+_{li}]-
\sum_{i<l}[\vartheta^+_{il} \lambda_{il} -
\vartheta_{il}\lambda^+_{il}].\nonumber
\end{eqnarray}
The operator $Q'$ can be presented with help of incomplete  BRST operator   $Q_c$ with only differential first-class constraints $\{l_0,l_i,  l_i^+\}$ and with complete BRST-extended  converted second-class  constraints $\{\mathcal{O}_a, \mathcal{O}^+_a\}$ as
\begin{eqnarray}
Q'& = &   Q_c +  \sum_{i<j}\Big(\eta_{ij}\mathcal{L}{}^+_{ij} + \mathcal{L}_{ij}\eta^+_{ij}+  \mathcal{T}_{ij}\vartheta_{ij}^+ +\vartheta_{ij}\mathcal{T}^+_{ij}
\Big)+ \sum_i[\eta^G_{i}\sigma^i(G)+ \imath \mathcal{B}^i\mathcal{P}_{i}^G ] , \label{Q'Q}
\end{eqnarray}
with definite operators $\mathcal{B}^i$ (obtained from (\ref{Q'})) and with complete traceless and Young constraints (and theirs  hermitian conjugated ones)
\begin{eqnarray}
\mathcal{L}_{ij}&=& L_{ij} + \frac{1}{2} q_{[i} p_{j]}    +\sum_{p<j}\vartheta_{ip}^+\mathcal{P}_{pj}
+\sum_{j<p} \vartheta_{jp}^+\mathcal{P}_{ip}-\sum_{j<p}\vartheta_{ip}^+\mathcal{P}_{j p}\label{BRSTextTr}\\
&&  +\frac{1}{4}\Big\{\sum_{p>j}\big[\eta_{ip}\lambda^+_{jp}-\eta_{jp}\lambda^+_{ip}\big]  + \sum_{p<j}\eta_{pj}\lambda^+_{ip}
    \Big\} ,  \nonumber \\
 \mathcal{T}_{ij} &=& {T}_{ij}+(q_jp_i^++q_i^+p_j)+\sum_{p>j} \vartheta^+_{jp}\lambda_{ip} - \sum_{p<j}[\vartheta_{ip}\lambda_{pj}-\vartheta_{pj}\lambda_{ip}]
   \label{BRSTextYo}\\
   && +\sum_{p>j}\eta_{jp}\mathcal{P}_{ip}^+ -  \sum_{i<p<j} \eta_{pj}\mathcal{P}_{ip}^+ + \sum_{p<i}\eta_{pj}
 \mathcal{P}_{pi}^+. \nonumber
\end{eqnarray}
Sum of the operator   $Q_c$ with rest term
$$Q(\mathcal{O}_a, \mathcal{O}^+_a)=\sum_{i<j}\big( \mathcal{L}_{ij}\eta^+_{ij}+  \mathcal{T}_{ij}\vartheta_{ij}^+ +h.c,\big) $$  without  number particles ghost coordinates, $\eta^G_{i}$, and momenta, $\mathcal{P}_{i}^G$ determines complete BRST operator $Q$ for only converted
constraint system with extracted zero-mode ghosts
\begin{eqnarray}
Q& =&  Q_c + Q(\mathcal{O}_a, \mathcal{O}^+_a) = \eta_0 l_0 + i  q_iq_i^+  \mathcal{P}_0
 +\Delta Q \label{separation1}\\
\Delta Q & = & \Big\{l_i q^+_i +\sum_{i<j}\big( \mathcal{L}_{ij}\eta^+_{ij}+  \mathcal{T}_{ij}\vartheta_{ij}^+\big)\Big\} +h.c.
\label{separation4}
\end{eqnarray}
Here,  the Hermitian conjugation is understood in the sense of the rule determined by the equations,
 \begin{eqnarray}\label{HermQ}
&& Q^{\prime +}K\  =\ K Q',  \ (\mathcal{L}_{ij}, \mathcal{T}_{ij}, \sigma_i(G))^+K\  =\ K(\mathcal{L}^+_{ij}, \mathcal{T}^+_{ij}, \sigma_i(G)) \\
&& \qquad \texttt{   with    } K\ = \ 1 \otimes K' \otimes 1_{gh}.\nonumber
\end{eqnarray}
The non-degenerate Grassmann-even operator  $K$  is given in the total Hilbert space, $\mathcal{H}_{tot}$, $\mathcal{H}_{tot}$ = $\mathcal{H}^f \otimes \mathcal{H}'$ $\otimes \mathcal{H}_{gh}$ and  can be constructed
from the whole set of oscillators, whereas $K'$ (intertwining $\{|\vec{N}\rangle_V\}$ and $\{|\vec{N}\rangle\}$  and introduced in appendix~\ref{oscrealsl2kdet}  by means of the equations (\ref{systemK}) and (\ref{explicit K}))   from only  $b_{ij}, d_{rs}$ and $b_{ij}^+, d_{rs}^{+}$.

Thus, we have  constructed  Hermitian complete (and incomplete) BRST operator for the HS symmetry superalgebra $\mathcal{A}_C(Y[k],\mathbb{R}^{1,d-1})$ of converted operators $O_I$.
Including oscillators $f_i^+$ and the ghost operators, we extend our basic vector  $|\Phi\rangle$ (\ref{PhysState}) given in $\mathcal{H}^f$  to
\begin{eqnarray}
|\chi\rangle & = &
\sum_{\{n\}_b=0}^{\infty}
\sum_{\{n\}_f=0}^{1}
\eta_{0}^{n_{\eta_{0}}}\prod_{i<j, i,j=1}^{k}
\eta_{ij}^{+n_{\eta_{ij}}}\mathcal{P}_{ij}^{+n_{P_{ij}}}\prod_{r<s, r,s=1}^{k}
\vartheta_{rs}^{+n_{\vartheta_{rs}}}
\lambda_{rs}^{+n_{\lambda_{rs}}} \prod_{i=1}^{k}
(\eta_{i}^{G})^{ n_{i}}
q_i^{+n_{q_i}}
p_i^{+n_{p_i}}\nonumber
\\
\hspace{-1em}& \hspace{-1em}\times &\hspace{-1em}
\prod_{i<j, i,j=1}^{k}f_j^{+n_{f_j}}
b_{ij}^{+n_{b_{ij}}}\hspace{-0.3em}\prod_{r<s, r,s=1}^{k}d_{rs}^{+n_{d_{rs}}}
\left|\chi_{n_{\eta_{0}}n_{\eta_{ij}}n_{\vartheta_{rs}}n_{P_{ij}}n_{\lambda_{rs}}n_{i} n_{q_i}n_{p_i}n_{f_j} n_{b_{ij}}n_{d_{rs}}}(\hat{a}{}_i^+)\rangle\right. ,
\label{extState}
\end{eqnarray}
where
$\{n\}_b = n_{q_i},n_{p_i}, n_{b_{ij}}, n_{d_{rs}}$ and $\{n\}_f = n_{\eta_0},n_{\eta_{ij}},n_{P_{ij}},n_{\vartheta_{rs}},n_{\lambda_{rs}}, n_{f_j}$\footnote{for the
massless  real-valued HS field $\Phi_{[\mu^1]_{s_1},\ldots,[\mu^k]_{s_k}}$
there are no operators $f^+_j$ in the decomposition (\ref{extState}),
i.e. indices $(n)_{f_j} = (0)_{f_j}$ as well as all component functions should have special complex-conjugated properties.}.
We should restrict ourselves to the Hilbert subspace   do not depending on ghost $\eta_{i}^{G}$ (i.e. $n_i\equiv 0$, due to special character of $G_{i}$ operators),
\begin{eqnarray}
\mathcal{P}_{i}^G | \chi \rangle &=&0, \; i = 1,\ldots,  k \label{restriction1}
\end{eqnarray}
and choose the representation in the Hilbert subspace $\mathcal{H}_{gh}$   in accordance with representation (\ref{extState}) as follows,
\begin{eqnarray}
 \left( q_i,  p_i,  \eta_{ij},  \mathcal{P}_{ij}, \vartheta_{rs}, \lambda_{rs}, \mathcal{P}_0\right) | 0 \rangle &=&0, \; i,j, r,s = 1,\ldots,  k. \label{ghreps}
 \end{eqnarray}
 From nilpotency of $Q'$ it follows the relations for BRST and spin operators
  \begin{align}\label{geneq}
   & Q^2 =  2\sum{i}\mathcal{B}^i \sigma^i(G) ,\ &&  [Q,\, \sigma^i(G)\} =0.
  \end{align}
Therefore, from the  BRST-like equation, determining the
physical vector, $Q'|\chi\rangle$ = $0$, (for $|\chi\rangle =
|\chi^{0}\rangle$) and from the set of reducible gauge
transformations, $\delta|\chi\rangle$ = $Q' |\chi^{1}\rangle$,
$\delta|\chi^{1}\rangle = Q'|\chi^{2}\rangle$, $\ldots$,
$\delta|\chi^{s-1}\rangle = Q'|\chi^{s}\rangle$, for $gh(|\chi^{n}\rangle)=-n $, $n=0,...,s$, for some integer $s>0$, a finite
sequence of relations underlying the $\eta^i_G$-independence of
all of the above homogeneous in ghost number vectors:
\begin{align}
\label{Qchi} & Q|\chi\rangle=0, && \sigma^i(G)|\chi\rangle=0,
&& \left(\varepsilon, {gh}_H\right)(|\chi\rangle)=(\varepsilon_\chi,0),
\\
& \delta|\chi\rangle=Q|\chi^{1}\rangle, &&
\sigma^i(G)\chi^{1}\rangle=0, && \left(\varepsilon,
{gh}_H\right)(|\chi^{1}\rangle)=(\varepsilon_\chi+1,-1), \label{QLambda}
\\
& \delta|\chi^{1}\rangle=Q|\chi^{2}\rangle, &&
\sigma^i(G)|\chi^{2}\rangle=0, && \left(\varepsilon,
{gh}_H\right)(|\chi^{2}\rangle)=(\varepsilon_\chi,-2),\\
& \ldots\ldots && \ldots\ldots && \ldots\ldots \nonumber \\
& \delta|\chi^{n-1}\rangle=Q|\chi^{n}\rangle, &&
\sigma^i(G)|\chi^{n}\rangle=0, && \left(\varepsilon,
{gh}_H\right)(|\chi^{n}\rangle)= (\varepsilon_\chi +n \hspace{-0.5em}\mod{}2,-n). \label{QLambdai}
\end{align}
The middle set of equations in
(\ref{Qchi})--(\ref{QLambdai}) determines the possible values of
the parameters $h_i$ and the eigenvectors of the generalized spin operator
$(\sigma_1,\ldots,  \sigma_k)$.  Its resolution leads to a sequence of
eigenvectors, $|\chi^{0}\rangle_{[m_1,\ldots,  m_k]}\equiv |\chi\rangle_{[m_1, \ldots,  m_k]}$, $|\chi^{1}\rangle_{[m_1, \ldots, m_k]}$,
$\ldots$, $|\chi^{n}\rangle_{[m_1, \ldots,  m_k]}$, $m_i \geq m_j \geq  0$, for $i<j$ to a set of eigenvalues for   $(\sigma^1,\ldots, \sigma^k)$ and therefore for the
spectrum of the Cartan subalgebra constants $h_i^{[m_1,\ldots, m_k]}$ with notation $[m_1,\ldots, m_k]\equiv m[k]$,
\begin{eqnarray}
&& \sigma^i(G) |\chi\rangle_{ m[k]}
 = \left(h_i + m_i - \frac{d-6+\theta_{m0}}{2}  -2i  \right)|\chi\rangle_{ m[k]}=0 \Longleftrightarrow \nonumber\\
 &&   -  h_i^{ m[k]}\ = \  m_i- \frac{d-6+\theta_{m0}}{2}-2i  , \ (\mathrm{for}\ h_i^{ m[k]} = h_i(m_i)) \label{hgen}
\end{eqnarray}
for some integers $m_i$ for massless  and massive ($\theta_{m0}=1$) mixed-antisymmetric HS fields

Then, operator $Q$ is nilpotent on $| \chi^{n} \rangle_{ m[k]} $ due to relations (\ref{geneq})
 because one can show from $Q'^2=0$ that:
\begin{equation}\label{Q2s}
Q^2| \chi^{n} \rangle_{m[k]} =  2\sum_{i=1}^k\mathcal{B}^{i} \sigma^i(G)| \chi^{n} \rangle_{m[k]} =0 ,
\end{equation}
Restriction (\ref{restriction1}) is easily satisfied by obvious  setting $n_{i}=0$ in the decomposition (\ref{extState}) because of  the spin operator is already extracted from the BRST operator $Q'$ and only
the operator $Q$ being independent on $\eta_{i}^G$ ghosts encodes the first-class operator constraints system (in $\{O_I\}\setminus G_0^i$).

For the basic state $\left|\chi_{0_{\eta_{0}}0_{\eta_{ij}}0_{\vartheta_{rs}}0_{P_{ij}}0_{\lambda_{rs}}0_{i} 0_{q_i}0_{p_i}0_{f_j} 0_{b_{ij}}0_{d_{rs}}}(\hat{a}{}_i^+)\rangle\right.  = |\Phi\rangle_{s[k]}$ which depend on only oscillator $\hat{a}{}^+$'s, that is
$|\Phi\rangle_{{s[k]}}
= \sum_{i}\sum_{n_{a_i}=0}^{s_i}\,
\prod_{j=1}^k\prod_{l_j=1}^{s_j}a_j^{+ \mu_{l_j}}
|0\rangle \Phi_{\mu^1[{n_1}],\ldots, \mu^k[n_k]}$, the middle equation in  (\ref{Qchi}), e.g. for massless case is calculated as
\begin{eqnarray}\label{hiphysst}
(\sigma^i+h_i) |\Phi\rangle_{s[k]}
 = \left( h_i + s_i- \frac{d-6}{2}  -2i  \right) |\Phi\rangle_{s[k]}=0.
\end{eqnarray}
Therefore, due to homogeneity in spin value for our spin $[s_1,\ldots, s_k]$ model, we fix parameters $h_i^{(m)|m[k]}$ from the equations (\ref{hgen})
by the equality, $m_i=s_i$, to be
\begin{eqnarray}
(h_i^{s[k]}, h_i^{m|s[k]}) & = & \left( - s_i +  \frac{d-6}{2} + 2i \right)(1,1) +\left(0, {1}/{2}\right)  \Rightarrow  |\chi^{n}\rangle_{m[k]} = |\chi^{n}\rangle_{s[k]}   .
\label{h}
\end{eqnarray}
 Thus,  middle equations (\ref{Qchi}) is satisfied for any $|\chi^{n}\rangle_{m[k]}$, in particular, for $|\chi\rangle_{s[k]}=|\Phi\rangle_{s[k]}$.

By one of the peculiarity of BRST  approach to mixed-antisymmetric HS fields is the fact that the Grassmann parity of any of $|\chi^n\rangle_{s[k]}$  depends on the values of spin, $s[k]$, due to fermionic nature of oscillators $\hat{a}_i^{\mu}$. Therefore for the vectors with definite spin values and with definite  ghost numbers the spectrum of Grassmann gradings  is determined as
\begin{equation}\label{grasparn}
  \varepsilon\big(|\chi^{0}\rangle_{s[k]}\big)\ =\ \sum_is_i \ \mathrm{mod} \ 2,\   \Rightarrow \  \varepsilon\big(|\chi^{n}\rangle_{s[k]}\big)\ =\
  \varepsilon\big(|\chi^{0}\rangle_{s[k]}\big) +n \ =\ n+\sum_is_i \ \mathrm{mod} \ 2,
\end{equation}
so that $|\chi^{0}\rangle_{s[k]}$ appears by bosonic or fermionic respectively   for  the initial tensor fields of the even or odd rank.

 Thus, we have constructed a complete BRST
operator for the  algebra $\mathcal{A}_C(Y[k],
\mathbb{R}^{1,d-1})$ of $O_I$ and specified the spectrum of the field and gauge parameters together with generalized spin values problem for them.
 Below, this operator with the set of determined field equations and gauge transformations
will be used to construct a Lagrangian action for bosonic HS
fields subject to Young tableaux  $Y[s_1,...,s_k]$ in a flat space for massless and massive cases.

\section{Lagrangians with complete BRST operator}\label{LagrFormulation}
\setcounter{equation}{0}
Here, we will follow to the receipt for the construction of  unconstrained BRST LFs and then will derive the component Lagrangians with use of the partial gauge-fixing procedure within BRST complex resolution.

\subsection{Unconstrained BRST Lagrangian Action}\label{unBRSTL}
The construction of Lagrangians for bosonic  HS fields in
a $d$-dimensional Minkowski  space can be developed by partially
following the algorithm of \cite{BurdikPashnev1}, \cite{1110.5044}, \cite{resh2}.

In order to
construct a Lagrangian for the massless  field corresponding to a definite
Young tableau (\ref{Young k}), the numbers $m_i$ must be equal to
the numbers of the cells in the $i$-th row of the corresponding
Young tableau, i.e., $m_i=s_i$. Therefore, the state
$|\chi\rangle_{s[k]}$ contains the physical field
(\ref{PhysState}) and all of its auxiliary fields. Let us fix some
values of $m_i=s_i$. Then, one has to substitute $h_i$
corresponding to the chosen $s_i$ (\ref{h}) into
(\ref{Qchi})--(\ref{QLambdai}). Thus, the equation of
motion (\ref{Qchi}) corresponding to the field with a given Young tableaux
$Y[s_1,\ldots, s_k]$ has the form
\begin{eqnarray}
Q_{s[k]}|\chi^0\rangle_{s[k]}=0, \label{QIJ}
\end{eqnarray}
where the ordered value  $s_1\geq s_2\geq \ldots \geq s_k $ for the vector
$|\chi^l\rangle_{s[k]}$ should be composed
from the set of integers, for $i<j$,  $r<s$: $n_{\eta_{0}},n_{\eta_{ij}},n_{\vartheta_{rs}},n_{P_{ij}},n_{\lambda_{rs}}, n_{f_j}$ taking two values, $0,1$, and ones $  n_{i}, n_{q_i}, n_{p_i}, n_{b_{ij}},  n_{d_{rs}}$, for $i,j,r,s
=1,...,k$  in (\ref{extState})
and (\ref{PhysState}) {running from $0$ to infinity}. In the decomposition (\ref{extState}) the
coefficients are to be restricted
for all the vectors $| \chi^l \rangle_{s[k]}$, $l=0,\ldots,
\sum_{i=1}^k s_i+\frac{1}{2}k(k-1)$, in view of the solution for the
spectral problem (\ref{hgen}) by the formulae on spin and ghost number homogeneity values
\begin{eqnarray}
s_i &=& n_{a_i}+\theta_{m0}n_{f_i}+ n_{q_{i}}+n_{p_{i}}+  \sum_{j\ne i} (n_{\eta_{ij}}+n_{P_{ij}}+n_{b_{ij}})\nonumber \\
&{}& +\sum_{r<i}(n_{\vartheta_{ri}}+n_{\lambda_{ri}}+n_{d_{ri}})-\sum_{i<r}(n_{\vartheta_{ir}}+n_{\lambda_{ir}}+n_{d_{ir}})
 ,\label{spin-number}\\
{gh (|\chi^l\rangle_{s[k]})} &=& n_{\eta_0}+\sum_{i}(n_{q_{i}}-n_{p_{i}})
+ \hspace{-0.2em}
\sum_{i< j}\hspace{-0.1em}\bigl(n_{\eta_{ij}}-n_{P_{ij}} \bigr) + \hspace{-0.2em}\sum_{r<s}\hspace{-0.1em} \bigl(
n_{\vartheta{}rs}- n_{\lambda{}rs}\bigr) = -l
 \label{ghnumf}.
\end{eqnarray}
for massless and
massive HS fields.

Because of the complete  BRST operator ${Q}'$ is nilpotent (\ref{Q'}) for
any values of $h_i$, and, due to the proportionality of $Q^2$
(\ref{Q2s}) to the generalized spin operator, and because of a
joint solution of the spectral problem
(\ref{Qchi})--(\ref{QLambdai}), we have a sequence of reducible
gauge transformations (for notation $|\Lambda^{l}\rangle \equiv |\chi^{l+1}\rangle$):
\begin{align}
\label{dx0} &\delta|\chi^0 \rangle_{s[k]}
=Q_{s[k]}|\Lambda\rangle_{s[k]} \,, &&
\delta|\Lambda^0\rangle_{s[k]} =Q_{s[k]}|\Lambda^{1}\rangle_{s[k]}
\,,
\\
&\ldots &&\ldots
\nonumber \\
\label{dxs} & \delta|\Lambda^{l-1} \rangle_{s[k]}
=Q_{s[k]}|\Lambda^{l}\rangle_{s[k]} \,, &&
\delta|\Lambda^{l} \rangle_{s[k]} =0,\ l=\sum_{i=1}^k s_i+\frac{1}{2}k(k-1)-1\,,
\end{align}
with a nilpotent $Q_{s[k]}$ in its action on proper
eigenfunctions $|\chi\rangle_{s[k]}$ of the operator $\sigma^i$:
\begin{eqnarray}
Q_{s[k]}^2\Bigl(|\chi \rangle_{s[k]}, |\Lambda^0 \rangle_{s[k]}, \ldots, |\Lambda^{\sum_{i=1}^k s_i+\frac{1}{2}k(k-1)-1 } \rangle_{s[k]}\Bigr) \equiv  0.
\end{eqnarray}
Summarizing  we have obtained the equations of motion
(\ref{QIJ}) for an arbitrary integer-spin HS field gauge theory
subject to $Y[s_1, \ldots, s_k]$ with a mixed antisymmetry in any
space-time dimension, as well as the tower of reducible gauge
transformations (\ref{dx0})--(\ref{dxs}).
One can show that
Lagrangian action for fixed spin $s[k]$ is defined up to an
overall factor as follows
\begin{eqnarray}
\mathcal{S}_{s[k]}& = &\int d \eta_0 \; {}_{s[k]}\langle \chi
|K_{s[k]} Q_{s[k]}| \chi \rangle_{s[k]}
\label{Scl}\\
& = &  \int d^dx \frac{(-1)^{\sum_{p=1}^ks_p}}{s_1!\ldots s_k!}\Big(\Phi_{\mu^1[{s_1}],\ldots, \mu^k[{s_k}]}(x)(\partial^2+m^2)
\Phi^{\mu^1[{s_1}],\ldots, \mu^k[{s_k}]}(x) + \texttt{more}\Big),  \nonumber
\end{eqnarray}
where the usual inner product for the creation and
annihilation operators is assumed with measure $d^dx$ over
Minkowski space with additional terms: "\emph{more}" with auxiliary fields differed for massless and massive basic HS field $\Phi^{\mu^1[{s_1}],\ldots, \mu^k[{s_k}]}(x)$. The vector $| \chi \rangle_{s[k]}\equiv | \chi^0 \rangle_{s[k]}$   and the
operator $K_{s[k]}$ in (\ref{Scl}) are  respectively the vector
$| \chi \rangle$ (\ref{extState}) subject to spin distribution
relations (\ref{hgen}), (\ref{spin-number}) for massless (when $n_{f_i}\equiv 0$) or   for  massive ($n_{f_i}\ne 0$)
HS tensor field $\Phi_{\mu^1[s_1],\ldots, \mu^k[{s_k}]}(x)$ with
vanishing value of ghost number and operator $K$ (\ref{HermQ}), (\ref{explicit K}), where
the  substitution should be done
$h_i\to-(s_i-\frac{d-6+\theta_{m0}}{2}-2i)$.
The  choice with $m>0$ corresponds to a theory of massive HS bosonic
field whereas the one for $(m)=0$ to a theory of
massless HS bosonic field. In both cases the LF appears by the gauge theory of $L=\sum_{i}^k s_i+\frac{1}{2}k(k-1)-1$ stage reducibility.

Let us consider now the  derivation of  LF with incomplete BRST operator for the same HS tensor field
$\Phi_{\mu^1[{{s}_1}],\ldots , \mu^k[{{s}_k}]}$.

\section{Lagrangians with incomplete BRST operator}\label{constrlagrform}
\setcounter{equation}{0}

Following to the  concept for mixed-symmetric HS fields \cite{resh2}, we apply it  here for the HS fields subject to $Y[s_1,\ldots, s_k]$.   Our aim in this direction to develop, first,   the constrained  BRST-BFV Lagrangians, then to establish the equivalence among the LFs for  complete and incomplete BRST operators and apply  constrained BRST complex resolution  to get  component Lagrangians  with less set of auxiliary fields.

\subsection{Constrained BRST Lagrangian Action}\label{conBRSTL}

To construct  LF with incomplete BRST operator we may follow by two equivalent ways, as it was shown in \cite{resh2} (see, as well \cite{mg}) for mixed-symmetric integer HS fields on $\mathbb{R}^{1,d-1}$ subject to $Y(s_1,...,s_k)$.  The result can be reached,  first,    from  the (unconstrained)  Lagrangian with complete BRST operator  by extracting  the BRST extended  second-class operator  constraints  subsystem, $(\widehat{O}_a, \widehat{O}^+_a)$,    from a total superalgebra of constraints $O_I$, second,  in the self-consistent way  by means of finding BRST-extended  initial off-shell algebraic operator constraints $\widehat{O}_a$.
From the former variant we have
\begin{equation}\label{constbrstop}
\widehat{O}_a \ \equiv\  \big(\widehat{L}_{ij}\,, \widehat{T}_{rs}\big) \ = \ \mathcal{O}_a\big|_{b_{ij}^{(+)}=d_{rs}^{(+)} =\eta_{ij}^{(+)}=\mathcal{P}_{ij}^{(+)}=\vartheta_{rs}^{(+)}=\lambda_{rs}^{(+)}=0}
\end{equation}
not depending on auxiliary: $b_{ij}^{(+)}, d_{rs}^{(+)}$, and ghost: $\eta_{ij}^{(+)}, \mathcal{P}_{ij}^{(+)}$, $\vartheta_{rs}^{(+)}$, $\lambda_{rs}^{(+)}$, oscillators as compared to  (\ref{BRSTextTr}), (\ref{BRSTextYo}).
We consider from the total set (\ref{Eq-0b})--(\ref{Eq-3b}) of irreducible representation conditions for the field $\Phi_{[\mu^1]_{s_1},[\mu^2]_{s_2}}$ described by Young tableaux (\ref{Young k}) the only differential relations (\ref{Eq-0b}), (\ref{Eq-1b}) as dynamical, which should be reproduced
 from minimal action principle whereas the algebraic: traceless (\ref{Eq-2b}) and mixed-antisymmetric (\ref{Eq-3b}) relations will be realized as off-shell constraints.

 Doing so, the derivation of the constrained  LF  is simplified as compared to the above described unconstrained case and repeats all the steps without
 conversion procedure but with imposing BRST-extended set of holonomic constraints.

 Corresponding incomplete\footnote{in  \cite{resh1} instead of "incomplete"  the term  "restricted"  and "constrained" have been  used}  BRST operator  $Q'_c=Q_c  + \eta^G_{i}\sigma^i_{{}c}$  (\ref{Q'Q})  is easily derived from complete  $Q'$ (\ref{Q'}),  operator by means of vanishing of the ghosts  $\eta_{ij}, \mathcal{P}_{ij}, \vartheta_{rs}$, $\lambda_{rs}$, auxiliary oscillators $b_{ij}, d_{rs}$  and theirs hermitian conjugated ones for $h^i  =0$ as follows,
 \begin{eqnarray}
Q'_c & = & Q'\big|_{\eta_{ij}^{(+)}= \mathcal{P}_{ij}^{(+)}= \vartheta_{rs}^{(+)}= \lambda_{rs}^{(+)} =b_{ij}^{(+)}=d_{rs}^{(+)}=h^i=0} \label{Q'Qr}\\
Q'_c & = & \eta_0l_0 + q^+_il_i +  q_il^+_i +iq_iq^+_i \mathcal{P}_0 + \eta^G_{i}\sigma^i_c(g) \ = \ Q_c + \eta^G_{i}\sigma^i_c (g), \label{Q'r}\\
\texttt{ where } && \sigma^i_{c}(g) \ =\ g^{i}_0 - q_i^+ p_i  - q_i p_i^+\label{s'r}
\end{eqnarray}
determines the incomplete  spin operator $\vec{\sigma}_{c}(g)=(\sigma^1_{c}(g),\ldots, \sigma^k_{c}(g))$.  These operators as well as  BRST-extended set of holonomic constraints $\widehat{L}_{ij}$, $\widehat{T}_{rs}$, (\ref{constbrstop})
 are given on the incomplete    Hilbert space $\mathcal{H}_c$:  $\mathcal{H}_c =\mathcal{H}^f\otimes H^{o_A}_{gh}$.  The incomplete BRST operator $Q_c$  is nilpotent in $\mathcal{H}_c$.

 The set of  operators $Q_c, {\sigma}{}^i_c(g)$ and BRST-extended constraints $\widehat{O}_a$ should form the closed superalgebra, i.e. they satisfy  to the \emph{generating equations}
 \begin{align}\label{eqQctot}
  & [Q_c,\, \widehat{O}_a\}    = 0, &   [Q_c,\, {\sigma}{}^i_c(g) \}= 0, &&   [\widehat{O}_a,\, \widehat{O}_b\}= f_{ab}^c\widehat{O}_c.
\end{align}
In case if  we would not know the final form of ${\sigma}{}^i_c(g)$ and $\widehat{O}_a$ the only known values are theirs boundary conditions instead: $({\sigma}{}^i_c(g)\,,\widehat{O}_a)\big|_{(\mathcal{C}=\overline{\mathcal{P}})=0}$ = $(g_0^i\,, o_a)$.

The algebra of  Grassmann-even with vanishing ghost number operators $({\sigma}{}^i_c(g), \widehat{O}_a) $  is the same as one for $(g_0^i\,, o_a)$:
\begin{eqnarray}
  \label{algrest} [\widehat{L}_{ij},\sigma^l_{c}(g)\}\ =\ \delta^l_{[j} \widehat{L}_{i]l}, \quad [\widehat{T}_{rs},\sigma^i_{c}(g)\}\ =\ \widehat{T}_{rs}(\delta_{si}-\delta_{ri}).
\end{eqnarray}
Thus, we have the realization of the same Howe dual algebra $so(k,k)$ enlarged from $\mathcal{H}^f$ to Hilbert space $\mathcal{H}_c$ with elements $({\sigma}{}^i_c(g)\,,\widehat{O}_a\,, \widehat{O}{}^+_a)$.

The operator $Q_c$  encodes the first-class constraints systems from $2k$ Grassmann-odd differential operators, $l_i, l_i^+$ and operator of Hamiltonian $l_0$ providing the solution of the problem of LF construction without conversion procedure.

The algebra of the operators  $\{Q_c$, $\sigma^i_{c}(g)$, $\widehat{O}_a\}$
permits to find joint set of  proper eigen-functions, which for the  first  variant of derivation from   approach with complete  BRST operator, follows from the $Q'_c$-BRST equations and constraints,
\begin{eqnarray}
Q'_c|\chi_c\rangle &=& 0,\ \delta|\chi_c\rangle\ =\ Q'_c|\chi_c^{1}\rangle,\
\delta|\chi_c^{1}\rangle\ =\ Q'_c|\chi_c^{2}\rangle, \ \ldots \ ,
\delta|\chi_c^{n-1}\rangle = Q'_c|\chi_c^{n}\rangle, \label{Q'rbrst} \\
\widehat{O}_a|\chi_c\rangle & =& 0, \ \ldots \ , \widehat{O}_a|\chi_c^{n}\rangle \ =\ 0,\label{constbrst}\\
\texttt{for }|\chi_c\rangle  &=& \sum_{\{n\}_b=0}^{\infty}
\sum_{\{n\}_f=0}^{1}
\eta_{0}^{n_{\eta_{0}}}
 \prod_{i=1}^{k}
(\eta_{i}^{G})^{ n_{i}}
q_i^{+n_{q_i}}
p_i^{+n_{p_i}}\left|\chi(\hat{a}{}_i^+)_{c|n_{\eta_{0}}n_{i} n_{q_i}n_{p_i}}\rangle\right. .
\label{extStater}
 \end{eqnarray}
The constrained vector is related to unconstrained one $|\chi \rangle$: $|\chi_c\rangle = \hspace{-0.1em}|\chi \rangle\big|_{(\eta_{ij}^{+}, \mathcal{P}_{ij}^{+}, \vartheta_{rs}^{+}, \hspace{-0.1em}\lambda_{rs}^{+},b_{ij}^{+},d_{rs}^{+})=\hspace{-0.1em}\mathbf{0}}$.
Here, first, we have chosen the same representation for constrained  vector, $|\chi^0_c\rangle \equiv |\chi_c\rangle$, as for $|\chi\rangle$ in (\ref{extState}), second,
$\{n'\}_b = n_{q_i},n_{p_i}$,  $\{n'\}_f = n_{\eta_0},{ n_{i}}$, third,
 restrict ourselves to the   $\eta_{i}^{G}$-independent vector   as for unconstrained case  (\ref{restriction1}).

 The analogous to the equations (\ref{Qchi})--(\ref{QLambdai}) system which maybe derived from (\ref{Q'rbrst}), (\ref{constbrst}), [if we follow to the first way of  LF  with incomplete BRST operator from LF with complete BRST operator, which appears by imposing of special gauge on the latter LF] has the form
 \begin{align}
\label{Qchir} & Q_c|\chi_c\rangle=0, && \sigma^i_{c}(g)|\chi_c\rangle=(s_i-\textstyle\frac{d-2}{2})|\chi_c\rangle,
&& \left(\varepsilon, {gh}_H\right)(|\chi_c\rangle)=(\varepsilon_{\chi}, 0),
\\
& \delta|\chi_c\rangle=Q_c|\chi^{1}_c\rangle, &&
\sigma^i_{c}(g)|\chi^{1}_c\rangle=(s_i-\textstyle\frac{d-2}{2})|\chi^{1}_c\rangle, && \left(\varepsilon,
{gh}_H\right)(|\chi^{1}_c\rangle)=(\varepsilon_{\chi}+1,-1), \label{QLambdar}
\\
& \delta|\chi^{1}_c\rangle=Q_c|\chi^{2}_c\rangle, &&
\sigma^i_{c}(g)|\chi^{2}_c\rangle=(s_i-\textstyle\frac{d-2}{2})|\chi^{2}_c\rangle, && \left(\varepsilon,
{gh}_H\right)(|\chi^{2}_c\rangle)=(\varepsilon_{\chi},-2),\\
& \ldots\ldots &&  \ldots\ldots&& \ldots\ldots \nonumber\\
& \delta|\chi^{n-1}_c\rangle=Q_c|\chi^{n}_c\rangle, &&
\sigma^i_{c}(g)|\chi^{n}_c\rangle=(s_i-\textstyle\frac{d-2}{2})|\chi^{n}_c\rangle, && \left(\varepsilon,
{gh}_H\right)(|\chi^{n}_c\rangle)= (\varepsilon_{\chi}+n \hspace{-0.5em}\mod{}\hspace{-0.1em}2,-n), \label{QLambdair}
\end{align}
for $\delta|\chi^{n}_c\rangle=0$ and $n = \sum_{i=1}^ks_i$.  All vectors $|\chi^{l}_c\rangle$   are subject to the  off-shell holonomic constraints,
\begin{eqnarray}
\hspace{-0.7em}&\hspace{-0.7em}&\hspace{-0.7em} \Big(\widehat{L}_{ij}, \widehat{T}_{rs}\Big) | \chi^{l}_c \rangle  = \Big(l_{ij} + \frac{1}{2} q_{[i} p_{j]}, {t}_{rs}+(q_sp_r^++q_r^+p_s)\Big)| \chi^{l}_c \rangle\ =\ (0,0), \quad l=0,1,...,n . \label{offshellcon}
\end{eqnarray}
Let us stress, the  system (\ref{Qchir})--(\ref{QLambdair}) for the second way of    Lagrangian derivation with incomplete BRST operator should be imposed independently from the equations (\ref{constbrst}).
The middle  set of the relations (\ref{Qchir})--(\ref{QLambdair})   determines the  proper eigen vectors $|\chi^{l}_c\rangle \equiv |\chi^{l}_c\rangle_{s[k]}$, so that due to the commutation relations (\ref{algrest})  they satisfy to (\ref{offshellcon}),  and therefore   the constrained gauge invariant LF of $L=(\sum_is_i-1)$ stage reducibility with the action $\mathcal{S}_{c|s[k]}$  for HS tensor field subject to Young tableaux $Y[s_1,\ldots,s_k]$  reads as,
\begin{eqnarray}
 \mathcal{S}_{c|s[k]}(\chi_c)& =& \int d \eta_0 \; {}_{s[k]}\langle\chi_c
|Q_c| \chi_c \rangle_{s[k]} =    \label{Sclred}
\\
& = &  \int d^dx \frac{(-1)^{\sum_{p=1}^ks_p}}{s_1!\ldots s_k!}\Big(\Phi_{\mu^1[{s_1}],\ldots, \mu^k[{s_k}]}(x)\partial^2
\Phi^{\mu^1[{s_1}],\ldots, \mu^k[{s_k}]}(x) + \texttt{more}\Big), \nonumber \\
\hspace{-1.5em}&\hspace{-1.5em}&\hspace{-1.5em}\Big(\delta;   \widehat{L}_{ij}, 
   \widehat{T}_{rs} \Big)| \chi^l_c \rangle_{s[k]} =
\Big(Q_c| \chi^{l+1}_c \rangle_{s[k]}\theta_{\sum_is_i,l};  {0}, 0\Big), \ l=0,1,...,\sum_is_i .\label{Sr}
\end{eqnarray}
For $\sum_is_i=0$ (which corresponds to the scalar field) the LF  appears by non-gauge  one.
In the $\eta_0$-independent form with use of the decomposition:  $| \chi^l_c\rangle=| S^l_c \rangle+ \eta_0| B^l_c \rangle$, the action and reducible gauge transformations read:
\begin{eqnarray}&& \hspace{-0.9em}\mathcal{S}_{c|s[k]} =  (-1)^{\sum_is_i} {}_{s[k]}\Big(\langle S^{0}_{c}\big| ,\langle B^{0}_{c}\big|\Big)\left(\begin{array}{cc}
      l_0 & - \Delta Q_c  \\
                   -\Delta Q_c  &  q_iq_i^+
       \end{array}\right)
       \left(\begin{array}{c}
                             \big|S^{0}_{c} \rangle_{s[k]} \\
                             \big|B^{0}_{c} \rangle_{s[k]}
            \end{array}\right)
,\label{Sclfvectf}\\
&& \hspace{-0.9em} \delta\left(\begin{array}{c}
                             \big|S^{l}_{c} \rangle_{s[k]} \\
                             \big|B^{l}_{c} \rangle_{s[k]}
            \end{array}\right) = \left(\begin{array}{cc}
    -\Delta Q_c  &  q_iq_i^+    \\
                   l_0 & - \Delta Q_c
       \end{array}\right)
       \left(\begin{array}{c}
                             \big|S^{l+1}_{c} \rangle_{s[k]} \\
                             \big|B^{l+1}_{c} \rangle_{s[k]}
            \end{array}\right)\theta_{\sum_is_i,l}\label{gaugecvectf}
\end{eqnarray}
where $\big|S^{\sum_is_i}_{c} \rangle$ is the gauge independent lowest level gauge parameter with $\big|B^{\sum_is_i}_{c} \rangle \equiv 0$
 due to spin and ghost number: $gh_H(\big|B^{\sum_is_i}_c \rangle)=-1- \sum_is_i$,  distributions.
Without off-shell BRST-extended constraints, we have obtained from  (\ref{Sclred}), (\ref{Sr})  or (\ref{Sclfvectf}) a generalized triplet formulation, which describes reducible Poincare group representations  with different spins  being (lexicographically) less than $[s_1,\ldots, s_k]$. In turn, the generalized triplet formulation may serve by an initial point to construct a generalized quartet formulation, following to \cite{0702161}, by incorporating whole set of holonomic constraints (\ref{Sr}) into sequence of reducible gauge transformations
and field equations by means of compensator mechanism.

More general, the LFs with complete (unconstrained) and incomplete (constrained) BRST operators  are equivalent in the sense that they both reproduce  the same irreducible representation relations given by the equations (\ref{Eq-0b})--(\ref{Eq-3b}), but latter one contains less auxiliary HS fields as compared to unconstrained formulation. This fact is analogous to one for mixed-symmetric integer HS fields on the flat space-time subject to $Y(s_1,...,s_k)$ \cite{resh2} (see Subsection 5.2).

Thus, repeating the arguments from \cite{resh2} we come to  validity of the

\vspace{1ex}

\noindent
\textbf{Theorem}: The set of solutions, $H_{(m,s[k])}$,    for the equations (\ref{Eq-0b})--(\ref{Eq-3b}) [or in the form (\ref{lilijt}, (\ref{g0iphys})] extracting the Poincare group massless ($m=0$) irreducibile  representation  of  spin $[s_1.,...,s_k]$ in terms of tensor  field, $\Phi_{\mu^1[s_1],...,\mu^k[s_k]}$  is equivalent to the solutions of the Lagrangian equations of motion, for $l=-1$  in (\ref{Sr})   subject to the reducible gauge transformations (\ref{Sr}) for $l=0,...,\sum_is_i$ and off-shell holonomic constraints (\ref{offshellcon}):
\begin{eqnarray}\label{equivconstinit}
        H_{(0,s[k])} &\hspace{-0.5em} =& \big\{ |\Phi\rangle |  \, \Big(l_0,\, l_i,\, l_{ij},\, t_{rs},\, g_0^i - [s_i-d/2]\Big) |\Phi\rangle =0   \big\} 
        \\
      &\hspace{-0.5em}=&  \hspace{-0.5em} \left\{|\chi^{0}_{c}\rangle \Big|    \hspace{-0.2em}\left[Q_c ,\  \Big\{\sigma^i_c - s_i+\frac{d-2}{2}\Big\}  \right] \hspace{-0.3em} | \chi^{0}_{c}\rangle  = 0, \right. \nonumber\\
&&  \left.  \delta |\chi^{l}_{c}\rangle = Q_c|\chi^{l+1}_{c}\rangle ,  \   \delta |\chi^{\sum_is_i}_{c}\rangle =0
\right. \nonumber
 \\
\phantom{H_{(0,s[k])} }&&  \left.\Big(\widehat{L}_{ij} , \widehat{T}_{rs},  \Big\{\sigma^i_c - s_i+\frac{d-2}{2}\Big\} \Big) |\chi^{l}_{c}\rangle = 0
\right\},\label{equivconstinit2}
    \end{eqnarray}
where, $l=0,...,\sum_is_i$;
\vspace{1ex}

 One should note, first, that for massive case the same equivalence is true. Second, the same statement holds if we instead of Lagrangian dynamics with incomplete BRST operator with superalgebra of $\{Q_c, \sigma^i_c , \widehat{L}_{ij} , \widehat{T}_{rs}\}$ consider LF with complete BRST $Q$ and spin $\sigma^i$ operators.
 Indeed, in this case the space of $Q$- BRST  local cohomology with vanishing ghost number with fixed spin given by (\ref{hgen}) coincide with the set  $H_{(m,s[k])}$ as well.

It is sufficiently easy (equivalently to unconstrained case) to get a component constrained LF (for $m=0$)  in terms of the initial HS tensor field
$\Phi_{\mu^1[{{s}_1}],...,\mu^k[{{s}_k}]}$ only
from the constrained BRST  complex resolution.  Now, we consider the case of two-columns Young tableaux with the field $\Phi_{\mu^1[{{s}_1}],\mu^2[{{s}_2}]}$.

\subsection{Component Lagrangians from incomplete BRST  complex: $k=2$}\label{constrBRST}

In this case, the general result  reduced to one obtained in \cite{resh1} for the field $\Phi_{\mu^1[{{s}_1}],\mu^2[{{s}_2}]}$.  The vectors $|\chi^l_c \rangle = |S^l_c \rangle+\eta_0|B^l_c \rangle$, for $l=0,...,\sum_{i=1}^2 s_i$ and $|B^{\sum_is_i}_C \rangle \equiv 0$ due to spin and $gh_{\mathrm{H}}$-numbers distributions have the presentation
\begin{eqnarray}
&&  |S^{l}_c\rangle_{s[2]}  \ = \ \sum_{k=\max(l-s_1 , 0)}^{\min(s_2,l)}(p_1^+)^{l -k}(p_2^+)^{k}|\varphi^{l}_{c|00k}(\hat{a}^+)\rangle_{[s_1+k-l,s_2-k]} +\sum_{n=1}^{\left[s_2+s_1 -l/2\right]}\sum_{m=0}^n \label{slcstr}
  \\
 && \phantom{|S^{l}_c\rangle}  \sum_{k\geq \max(l-s_1+2n,m)}^{\min(s_2, l+n+m)}\hspace{-0.7em} (p_1^+)^{l+n+m-k}(q_1^+)^{n-m}(q_2^+)^m  (p_2^+)^{ k-m}  |\varphi^{l}_{c|nmk}(\hat{a}^+)\rangle_{[s_1+k-2n-l,s_2-k]}, \nonumber \\
 && B^{l}_c\rangle_{s[2]} \ =\  \sum_{k=\max(l-s_1+1 , 0)}^{\min(s_2,l+1)}(p_1^+)^{l-k+1}(p_2^+)^{k}|\varphi^{l}_{c|0;00k}(\hat{a}^+)\rangle_{[s_1+k-l-1,s_2-k]}
  \label{blcstr}\\
 && \phantom{|S^{l}_C}  +\sum_{n=1}^{\left[s_2+s_1-l-1/2\right]}\sum_{m=0}^n\sum_{k\geq \max(l-s_1+2n+1,m)}^{\min(s_2, l+n+m+1)}\hspace{-1em} (p_1^+)^{l+n+m-k+1}  (q_1^+)^{n-m}(q_2^+)^m  (p_2^+)^{ k-m} 
 \nonumber \\
 &&\phantom{|S^{l}_C} \times
 |\varphi^{l}_{c|0;nmk}(\hat{a}^+)\rangle_{[s_1-l-2n+k-1,s_2-k]}.
        \nonumber
\end{eqnarray} with vectors  $|\varphi^{l}_{c|00k}(\hat{a}^+)\rangle$, $|\varphi^{l}_{c|nmk}(\hat{a}^+)\rangle$, $|\varphi^{l}_{c|0;00k}(\hat{a}^+)\rangle$, $|\varphi^{l}_{c|0;nmk}(\hat{a}^+)\rangle$ given on the initial Fock space $\mathcal{H}^f$.

Resolution  of Young (mixed-antisymmetry) constraints (\ref{offshellcon}) on $|\chi^l_c \rangle$  leads to the system of linear homogeneous equations with unknowns $|\varphi^{l}_{c|nmk}(\hat{a}^+)\rangle$, starting from highest level gauge parameter value $l = \sum_i s_i$  for $n=m=0$:
\begin{eqnarray}
 & & \theta_{s_2-1,k} t_{12}\big|\varphi^{l}_{c|00k}\rangle_{(s_1+k-l,s_2-k)}  - (k+1)
  \big|\varphi^{l}_{c|00(k+1)}\rangle_{(s_1+k-l+1,s_2-k-1)}\  =\ 0 , \label{n0>=s1} \\
  && \qquad  \mathrm{for} \ \  \textstyle l = \sum_i s_i,..., s_1+1,  \ k=l-s_1,..., s_2-1, \nonumber\\
 & &  t_{12} \big|\varphi^{l}_{c|00k}\rangle_{(s_1+k-l,s_2-k)}  - \theta_{l,k}(k+1)
  \big|\varphi^{l}_{c|00(k+1)}\rangle_{(s_1+k-l+1,s_2-k-1)}\  =\ 0 ,  \label{n0<s1} \\
  && \qquad \mathrm{for} \ \ \textstyle  l = s_1,..., 0,  \ {k=0,...,s_2-1}, \nonumber
  \end{eqnarray}
  for $m=0, 1$:
\begin{eqnarray}
&& \theta_{s_2-1,k}t_{12}\big|\varphi^{l}_{c|n0k}\rangle_{(s_1+k-l-2n,s_2-k)}-  (k+1) \big|\varphi^{l}_{c|n0(k+1)}\rangle_{(s_1+k-2n-l+1,s_2-k-1)} \label{n10>=s1}
  \\
  && \qquad\qquad   - \big|\varphi^{l}_{c|n 1 (k+1)}\rangle_{(s_1+k-l-2n+1,s_2-k-1)} = 0, \nonumber\\
  && \quad \mathrm{for} \ \  \textstyle l = \sum_is_i-2,..., s_1+1,  \  n= 1, ..,\left[(\sum_is_i-l)/2\right],  \ k=l-s_1+2n,..., s_2-1, \nonumber\\
&&t_{12} \big|\varphi^{l}_{c|n0k}\rangle_{(s_1+k-l-2n,s_2-k)}-  (k+1) \big|\varphi^{l}_{c|n0(k+1)}\rangle_{(s_1+k-l-2n+1,s_2-k-1)}
 \label{n10<s1}\\
  && \qquad\qquad - \big|\varphi^{l}_{c|n 1  (k+1)}\rangle_{(s_1+k-l-2n+1,s_2-k-1)} = 0, \nonumber\\
  && \quad \mathrm{for} \ \  \textstyle l =  s_1-1,...,0 , \  n= 1, ..,\left[(\sum_is_i-l)/2\right],  \ k=0,..., s_2-1,
 \nonumber \end{eqnarray}
  for $m \geq 1$:
\begin{eqnarray}
   & &  t_{12}\big|\varphi^{l}_{c|nmk}\rangle_{(s_1+k-l-2n,s_2-k)} - (k-m+1)  \big|\varphi^{l}_{c|nm(k+1)}\rangle_{(s_1+k-l-2n+1,s_2-k-1)} \label{n11>=s1} \\
  && \qquad\qquad - \theta_{n,m}(m+1)\big|\varphi^{l}_{c|n(m+1)(k+1)}\rangle_{(s_1+k-l-2n+1,s_2-k-1)}= 0,
    \nonumber\\
    && \quad \mathrm{for} \ \textstyle l = \sum_is_i-2,..., s_1, \  n= 1, ..,\left[(\sum_is_i-l)/2\right], \  m=1,...,n,\    k=n-s_1,..., s_2-1, \nonumber\\
   & & t_{12} \big|\varphi^{l}_{c|nmk}\rangle_{(s_1+k-l-2n,s_2-k)} - (k-m+1)  \big|\varphi^{l}_{c|nm(k+1)}\rangle_{(s_1+k-l-2n+1,s_2-k-1)}\label{n11<s1}\\
  && \qquad \qquad - \theta_{n,m}(m+1)\big|\varphi^{l}_{c|n(m+1)(k+1)}\rangle_{(s_1+k-l-2n+1,s_2-k-1)}= 0,\nonumber\\
    && \quad  \mathrm{for}  \ \textstyle l = s_1-1,..., 0,  \  n= 1, ..,\left[(\sum_is_i-l)/2\right], \  m=1,...,n,\   k=1,..., s_2-1. \nonumber
\end{eqnarray}
For the unknown vectors $|\varphi^{l}_{c|0;nmk}(\hat{a}^+)\rangle$ from  $|B^{(l)}_{c}\rangle$ the analogous  systems take place when repeating the systems for
$|S^{(l-1)}_{c}\rangle$.

The resolution of the equations (\ref{n0>=s1}), (\ref{n10>=s1}), (\ref{n11>=s1}) above, leads to vanishing of $| \chi^{l}_c  \rangle$  for $l=s_1+1, ..., \sum_is_i$ and to $| B^{s_1}_c  \rangle =0$. Then for $l=s_1$  from the rest equations (\ref{n0<s1}), (\ref{n10<s1}), (\ref{n11<s1}) the solution for parameter $|{\chi}^{s_1}_{c|g}  \rangle$ maybe presented according to auxiliary Lemmas form the appendix~\ref{holconstrres} as
\begin{eqnarray} \label{chis1r}
|{\chi}^{s_1}_{c|g}  \rangle_{s[2]} & = & |S^{s_1}_{c|g}  \rangle_{s[2]} \ = \ \sum_{k=0}^{s_2}(p_1^+)^{s_1-k}(p_2^+)^{k}\frac{t^{k}_{12}}{k!} \big|\varphi^{s_1}_{c|000}\rangle_{[0,s_2]},
\end{eqnarray}
with trivially satisfied traceless  constraint   $\widehat{L}_{12}| \chi^{s_1}_{c|g}  \rangle = l_{12}| \chi^{s_1}_{c|g}  \rangle  \equiv 0$.
Then, for $l=s_1-1$ we, first, get  that $|B^{s_1-1}_{c|g}  \rangle = |{\chi}^{s_1}_{c|g}  \rangle\big|_{(\varphi^{s_1}_{c|000}\to \varphi^{s_1-1}_{c|0;000})} $. Second,
 \begin{eqnarray}
|{S}^{s_1-1}_{c|g}\rangle  &=&    \sum_{k=0}^{\min(s_2,s_1-1)}(p_1^+)^{s_1-1-k}(p_2^+)^{k}
 \frac{t^{k}_{12}}{k!} \sum_{i=0}^1\big|\varphi^{s_1-1}_{c|000}\rangle_{[1-i+\{i,i\}+s_2-i]}\theta_{s_1+i,s_2} \label{chis1-1rf}\\
  &&  + \sum_{k=1}^{s_2} (p_1^+)^{s_1-k}q_1^+(p_2^+)^{ k-1} \frac{t^{k-1}_{12}}{(k-1)!}\Big(q_1^+p_2^+- q_2^+p_1^+\Big) \big|\varphi^{s_1-1}_{c|101}\rangle_{[0,s_2-1]}   \nonumber
\end{eqnarray}
with expressing a vector $\big|\hspace{-0.1em}\varphi^{s_1-1}_{c|101}\hspace{-0.1em}\rangle_{[0,\hspace{-0.05em}s_2-1]}$ from traceless constraint in terms of  $ \big|\hspace{-0.1em}\varphi^{s_1-1}_{c|000}\hspace{-0.1em}\rangle_{[\{1,1\}\hspace{-0.05em}+s_2-1]}$ being subject to $Y[1,s_2]$ as
\begin{equation}\label{chis1-1simr}
\big|  \varphi^{s_1-1}_{c|101}\rangle_{[0,s_2-1]} = \frac{2}{(s_1+1)} l_{12}\big|  \varphi^{s_1-1}_{c|000}\rangle_{[\{1,1\}+s_2-1]}.
\end{equation}
Here, it was used the decomposition of a vector $ \big|\varphi^{s_1}_{00k}\rangle_{(k,s_2-k)}$ (tensor $\varphi^{s_1}_{00k|\mu^1[{k}], \mu^2[{s_2-k}]}$) in sum of Young antisymmetry irreducible vectors (tensors) according to (\ref{decompYir01}).

From the detailed resolution of the incomplete BRST complex  it follows the representation for the gauge-fixed field and gauge parameters $\big|S^{s_1-e}_{c|g}\rangle_{s[2]}$,  $\big|B^{s_1-e}_{c|g}\rangle_{s[2]}$, $e=0,1,...,s_1$
 having the form (\ref{ggsi-s1-eshggf})--(\ref{bl=s1-eap22})
  \begin{eqnarray}
&& \hspace{-1em}
|S^{s_1-e}_{c|g}\rangle_{s[2]}  \ =\ \sum_{r=0}^{\min(e,s_2)}\frac{2^{r}(s_1-e+1)!}{r! (s_1-e+1+r)!}\sum_{m=0}^{r} \frac{(-1)^m\, r!}{(r-m)!m!} \sum_{k=r}^{\min(s_2,s_1-e+r)}(p_1^+)^{s_1-e-k+m+r}\nonumber\\
 && \quad \times (q_1^+)^{r-m}(q_2^+)^m
 (p_2^+)^{ k-m}\frac{t^{k-r}_{12}}{(k-r)!}l_{12}^r
\sum_{i=r}^{\min(s_2,e)}\big|\varphi^{s_1-e|i}_{000}\rangle_{[e-i+\{i,i\}+s_2-i]}\theta_{s_1-e+1+i,s_2},\\
     \label{ggsi-s1-eshggf}
  && \hspace{-1em}|B^{s_1-e}_{c|g}\rangle_{s[2]}  =  |S^{s_1-e+1}_{c|g}\rangle\big|_{\big(|\varphi^{s_1-e+1|i}_{000}\rangle_{[e-1,s_2]}  \to
|\varphi^{s_1-e|i}_{0;000}(\varphi^{s_1-e}_{000})\rangle_{[e-1,s_2]}\big)} , 
\\
&&\hspace{-1em} \texttt{ with }  \big|\varphi^{s_1-e}_{0;000}(\varphi^{s_1-e}_{000})\rangle_{[e-1,s_2]}\  = \  \frac{1}{s_1-e+1}\big[l_1- \frac{2}{s_1-e+2} l_2^+ l_{12}\big] \big|\varphi^{s_1-e}_{000}\rangle_{[e,s_2]}. \label{bl=s1-eap22}
\end{eqnarray}
  with allowance made for $\big|\varphi^{s_1-e}_{c|000}\rangle \equiv \big|\varphi^{s_1-e}_{000}\rangle$.

Resolving partially equations of motion we get for the field vectors (\ref{ggsi-s1-eshggf})--(\ref{bl=s1-eap22}) (when $e=s_1$) with gauge transformations
\begin{eqnarray}
\hspace{-0.7ex}&\hspace{-0.7ex}& \hspace{-0.7ex}|S^{0}_{00|g}\rangle  =
    \sum_{r=0}^{s_2}\frac{2^{r}}{r! (r+1)!}\sum_{m=0}^{r} \frac{(-1)^m\, r!}{(r-m)!m!}(p_1^+)^{m}(q_1^+)^{r-m} (q_2^+)^m
 (p_2^+)^{ r-m}l_{12}^r
|\Phi\rangle_{[s]_2}  ,   \label{sl=s1-s1app1}
\\
\hspace{-0.7ex}&\hspace{-0.7ex}& \hspace{-0.7ex}|B^{0}_{00|g}\rangle  =  \sum_{r=0}^{\min(s_2,s_1-1)}\frac{2^{r+1}}{r! (r+2)!}\sum_{m=0}^{r} \frac{(-1)^m\, r!}{(r-m)!m!}(q_1^+)^{r-m}(q_2^+)^m(p_1^+)^{m}(p_2^+)^{ r-m}\label{bl=s1-s1app1}
\\
 && \phantom{\hspace{-0.7ex}|B^{0}_{00|g}\rangle}  \ \ \times l_{12}^r \Big(p_1^+\big[  l_2^+ l_{12}-l_1\big]-   p_2^+ \big[  l_1^+ l_{12}+l_2\big]\Big)
 \big|\Phi\rangle_{[s]_2},\nonumber \\
\hspace{-0.7ex}&\hspace{-0.7ex}&\hspace{-0.7ex}   \label{gtrl=s1-s1app}   \delta |\Phi\rangle_{[s]_2}  =   -\Big(  l_1^+    +  l_2^+t^{}_{12}\Big)   \sum_{i=s_2-1}^{s_2}\big|\varphi^{1|i}_{000}\rangle_{[s_1-1-i+\{i,\,i\}+s_2-i]}\theta_{s_1+s_2-i,s_2}.
\end{eqnarray}
  Calculating inner products for ghost operators we come to the result, first,  in the oscillator form: $\langle \Phi\big|\sum_{r\geq 0} (l_{12}^+)^r$ $K(l_0,l_i, l_i^+)  l_{12}^r\big|\Phi\rangle$ with hermitian kinetic operator $K(l_0,l_i, l_i^+)$ for the action with accuracy up to numeric factor
\begin{eqnarray}
&&    \mathcal{S}_{s[2]}(\Phi)  =  (-1)^{\sum_is_i}{}_{s[2]}\langle \Phi\big|\bigg\{\sum_{r=0}^{s_2}(-1)^{r}\frac{2^{2r}}{r!(r+1)!} (l_{12}^+)^r  \Big[l_0   - \left\{1+\textstyle\frac{r}{2}\right\}\sum_il_i^+ l_i        \label{Sclfcompghun}\\
 &&  \phantom{\mathcal{S}_{s[2]}(\Phi)} -\frac{2}{r+2} l_{12}^+\sum_il_i l_i^+l_{12} + 2l_{12}^+l_2l_1+  2l_1^+l_2^+l_{12}\Big]  l_{12}^r\bigg\} \big|\Phi\rangle_{s[2]}.\nonumber
 \end{eqnarray}
 with the sequence of gauge transformations together with Young antisymmetry condition in case of $(s_1-e)$-level   for arbitrary $e= 0,1,...,s_1-1$ in dependence on difference $r=s_1-s_2=0,1,...,s_1$:
 \begin{eqnarray}\label{gtrl=s1-eapp}   && \delta \sum_i \big|\varphi^{s_1-e|i}_{000}\rangle_{(e,s_2)} =   -\Big((s_1-e+1) l_1^+    +  l_2^+t^{}_{12}\Big) \sum_j\big|\varphi^{s_1-e+1|j}_{000}\rangle_{(e-1,s_2)},\\
\label{gtrl=s1-eapp2}   &&  \left\{\begin{array}{ll}  t_{12}^{s_2} \big|\varphi^{s_1-e}(a^+)\rangle_{(e,s_2)} \in Y[s_2+e,0],&  r=s_1-s_2>e-1; \\
t_{12}^{s_1-e} \big|\varphi^{s_1-e}(a^+)\rangle_{(e,s_2)} \in Y[s_1,s_2-s_1+e],&   r=s_1-s_2\leq e-1\end{array}\right..
\end{eqnarray}
Here we use the identification, $|\varphi^{s_1-e|i}_{000}\rangle_{(e,s_2)} \equiv |\varphi^{s_1-e|i}\rangle_{(e,s_2)}$ (see \cite{resh1} as well).

The model describes the gauge theory of free field $\Phi_{\mu^1{[s_1]},\mu^2[{s_2}]}$  of $(s_1-1)$-stage reducibility.

In the tensor form, the action (\ref{Sclfcompghun}) has the form
\begin{eqnarray}
&& \hspace{-1em} \mathcal{S}_{s[2]}(\Phi) \hspace{-0.1em} = \hspace{-0.2em} \int \hspace{-0.2em}d^dx \hspace{-0.1em}\bigg\{\hspace{-0.15em}\sum_{r=0}^{s_2}\frac{(-1)^{r}(\mathrm{Tr}^r\Phi)_{\mu^1[{s_1-r}],\mu^2[{s_2-r}]}}{(s_1-r)!(s_2-r)!r!(r+1)!}
 \hspace{-0.1em}\Big[
\partial^2  (\mathrm{Tr}^r\Phi)^{\mu^1[{s_1-r}],\mu^2[{s_2-r}]}\label{Sclfcomp2} 
  \\
 &&  - \left\{1+\textstyle\frac{r}{2}\right\} \Big( (s_1-r){\partial}^{\mu^1_1}   {\partial}_{\nu^1_1}\delta_{\nu^2_1}^{\mu^2_1}+ (s_2-r){\partial}^{\mu^2_1}   {\partial}_{\nu^2_1} \delta_{\nu^1_1}^{\mu^1_1}\Big) (\mathrm{Tr}^r\Phi)^{\nu^1_1\mu^1_2...\mu^1_{s_1-r}, \nu^2_1\mu^2_2...\mu^2_{s_2-r}}    \nonumber \\
 &&
     -\frac{(s_1-r)(s_2-r)}{2(r+2)}  \eta^{\mu^1_{s_1-r}\mu^2_1}  \Big\{2\partial^2(\mathrm{Tr}^{r+1}\Phi)^{\mu^1[{s_1-r-1}],\mu^2_2...\mu^2_{s_2-r}}
      - \Big( (s_2-r-1) {\partial}^{\mu^2_2}   {\partial}_{\rho^2_1}\delta_{\rho^1_1}^{\mu^1_1}\nonumber \\
 && \ +(s_1-r-1) {\partial}^{\mu^1_2}   {\partial}_{\rho^2_1}\delta_{\mu^2_2}^{\mu^1_1}\Big)(\mathrm{Tr}^{r+1}\Phi)^{\rho^1_1\mu^1_2...\mu^1_{s_1-r-2},\rho^2_1\mu^2_3...\mu^2_{s_2-r}}\Big\}
      \nonumber \\
 &&  + \frac{1}{2} (s_1-r)(s_2-r)\Big\{\eta^{\mu^1_{s_1-r}\mu^2_1} {\partial}_{\mu^1_{s_1-r}}   {\partial}_{\mu^2_1}(\mathrm{Tr}^{r}\Phi)^{\mu^1[{s_1-r}],\mu^2[{s_2-r}]} \nonumber \\
 &&
+  \eta_{\rho^1_{s_1-r}\rho^2_1} {\partial}^{\mu^1_{s_1-r}}   {\partial}^{\mu^2_1}(\mathrm{Tr}^{r}\Phi)^{\mu^1[{s_1-r-1}]\rho^1_{s_1-r}, \rho^2_1\mu^2_2...\mu^2_{s_2-r}}\Big\}    \Big]\bigg\}.\nonumber 
\end{eqnarray}
with the notation for a multiple trace,
\begin{equation}\label{mtr}
(\mathrm{Tr}^r \Phi)_{\mu^1[{s_1-r}],\mu^2[{s_2-r}]}\equiv \Phi_{[\mu^1]_{s_1-r}\nu_1...\nu_r,}{}^{\nu_r...\nu_1}{}_{[\mu^2]_{s_2-r}}\equiv \prod_{i=1}^r\eta^{\mu^1_{s_1+1-i} \nu^2_{i} }\Phi_{\mu^1[{s_1}],\mu^2[{s_2}]}.
\end{equation}
The gauge transformations   maybe presented as follows
 \begin{eqnarray}
\nonumber  && \hspace{-1em} \delta \varphi^{s_1-e}_{\mu^1[{e}],\mu^2[{s_2}]}  =  (s_1-e+1) \partial_{[\mu^1_1} \varphi^{s_1-e+1}_{[\mu^1_2...\mu^1_e]],\mu^2[{s_2}]} + (-1)^e\partial_{[\mu^2_1} (Y\varphi^{s_1-e+1})_{[\mu^1[{e-1}],{\mu^1_e}]\mu^2_2...\mu^2_{s_2}]}, \\
&&  \hspace{-1em} \mathrm{for} \left\{\begin{array}{l}  (Y^{s_2}\varphi^{s_1-e}) _{[\mu^1[e],\mu^1_{e+1}...\mu^1_{e+s_2}]}  \in Y[s_2+e,0],\ e=0,...,s_1-s_2 ; \\
(Y^{s_1-e}\varphi^{s_1-e}) _{[\mu^1[e],\mu^1_{e+1}...\mu^1_{s_1}] \mu^2[{s_2-s_1+e}]}   \in Y[s_1,s_2-s_1+e],\  e=s_1-s_2+1,...,s_1\end{array}\right..
\label{gtrl=s1-2c} \\
&& \hspace{-1em}\ \mathrm{where} \  (Y\varphi^{s_1-e})_{[\mu^1[e],{\mu^1}]\mu^2_2...\mu^2_{s_2}}  =  -   \varphi^{s_1-e}_{[\mu^1[e],{\mu^1}]\mu^2_2...\mu^2_{s_2}},    \nonumber \end{eqnarray}
and antisymmetrization in $\partial_{[\mu^1} \hspace{-0.1em}\varphi_{\mu^1[{e-1}]],\mu^2[{s_2}]}$, $(\hspace{-0.1em}Y\hspace{-0.1em}\varphi^{s_1-e}\hspace{-0.1em})_{[\mu^1\hspace{-0.1em}[e],{\mu^1}\hspace{-0.1em}]\mu^2_2...\mu^2_{s_2}}$ ($\hspace{-0.1em}\partial_{[\mu^2_1}\hspace{-0.1em} (\hspace{-0.1em}\varphi^{s_1-e}\hspace{-0.1em})_{\mu^1\hspace{-0.1em}[{e}],\mu^2_2...\mu^2_{s_2}]}$) does not contain  the factor $1/e$ ($1/s_2$).

The resulting LF is a gauge theory of $(s_1-1)$-th stage of reducibility, which describes the free dynamics of a massless Bose-particle of spin $[s_1,s_2]$, with the single off-shell restriction of Young antisymmetry on the field
$\Phi\equiv \varphi^{0}$ and the gauge parameters $\varphi^{1}_{(s_1-1,s_2)},..., \varphi^{s_1}_{(0,s_2)}$.


We stress, that there are no any traceless constraints on any gauge parameters and physical
tensor field. Note, the LF for the massless field $\Phi_{\mu^1{[s_1]},\mu^2[{s_2}]}$  subject to rectangular Young tableaux $Y[s_1,s_1]$ was considered in \cite{0402180}, for arbitrary case $s_1>s_2$ in \cite{Boulanger2}. In frame-like formalism it was realized by Zinoviev \cite{zinoviev1}.

Now, we intend to consider LFs for massive mixed-antisymmetric HS fields.

\subsection{Lagrangian Formulations for  Massive  Fields}
\label{massiveLF}
There are two ways to construct LFs for the massive field with mass, $m$ and generalized spin $[s_1,...,s_k]$. First, following from unconstrained or constrained  BRST  LFs  repeating the procedures developed in Section~\ref{LagrFormulation} and in Section~\ref{constrlagrform}.
Second, to apply the dimensional reduction of the massless theory in $\mathbb{R}^{1,d}$ initiated in Subsection~\ref{auxtmasrep} to derive the HS symmetry superalgebra $\mathcal{A}_m(Y[k],\mathbb{R}^{1,d-1})$ for massive
fields.
Whereas, the unconstrained BRST  LF is given by (\ref{dx0})--(\ref{Scl}) and spin values distribution (\ref{spin-number}) for field and gauge parameter vectors for $m\ne 0$ the constrained one is determined by the relations (\ref{Qchir})--(\ref{Sclred}) with replacements of initial operators $o_I$ from the massless HS symmetry superalgebra to the elements $\tilde{o}_I$ (\ref{l0tilde})--(\ref{exprnew}). This  substitution determined incomplete BRST operator and off-shell BRST extended algebraic constraints  acting on the constrained field, $|\chi^m_c\rangle$, and gauge parameter, $|\chi_c^{m(n)}\rangle$, $n=1,...,\sum_i s_i$, vectors (for $m=0$ coinciding with ones for massless case)
\begin{eqnarray}
\hspace{-1em}Q^{\prime m}_c|\chi^{m(0)}_c\rangle &=& 0,\ \delta|\chi^{m(0)}_c\rangle\ =\ Q^{\prime m}_c|\chi_c^{m(1)}\rangle, \ldots,
\delta|\chi_c^{m(n-1)}\rangle = Q^{\prime m}_c|\chi_c^{m(n)}\rangle, \label{Q'rbrstm} \\
\hspace{-1em} \widehat{L}^m_{ij}|\chi^m_c\rangle & =& 0, \qquad \widehat{T}^m_{rs}|\chi^m_c\rangle \ =\ 0, \label{constbrstm}\\
\hspace{-1.9em} \texttt{for }|\chi^m_c\rangle  &=&\hspace{-0.5em}
\sum_{\{n\}_b=0}^{\infty}
\sum_{\{n\}_f=0}^{1}
\eta_{0}^{n_{\eta_{0}}}
 \prod_{i=1}^{k}
(\eta_{i}^{G})^{ n_{i}}
q_i^{+n_{q_i}}
p_i^{+n_{p_i}}
\prod_{j=1}^{k}f_j^{+n_{f_j}}\left|\chi(\hat{a}{}_i^+)_{c|n_{\eta_{0}}n_{i} n_{q_i}n_{p_i}n_{f_j}}\rangle\right.
 \hspace{-0.3em},
\label{extStaterm}
 \end{eqnarray}
with gauge independent parameter $|\chi_c^{m(n)}\rangle$ and  where in (\ref{extStaterm})  the vector $|\chi^m_c\rangle$ is written without definite spin and ghost  numbers values.
The BRST  operator, $Q^{\prime m}_c$,  for constrained massive LF has analogous structure to decomposition (\ref{Q'Qr})--(\ref{s'r}):
\begin{eqnarray}
Q^{\prime m}_c & = &  q^+_i\tilde{l}_i +  q_i\tilde{l}{}^+_i + \eta_0\tilde{l}_0 +iq_iq^+_i \mathcal{P}_0 + \eta^G_{i}\sigma^m_{i{}c} \ = \ Q^m_c + \eta^G_{i}\sigma^i_{m{}c} , \label{Q'rm}\\
\texttt{ where } && \sigma^i_{m{}c}\ =\ g^{i}_0 -\frac{1}{2}+ f_i^+f_i-  q_i^+ p_i  - q_i p_i^+, \ \mathrm{no}\ \mathrm{summation}\ \mathrm{in}\ i \label{s'rm}
\end{eqnarray}
with massive incomplete spin operator. These operators as well as the BRST-extended set of massive holonomic constraints $\widehat{L}^m_{ij}$, $\widehat{T}^m_{rs}$,
 \begin{eqnarray}
\widehat{L}^m_{ij} \ = \  \widehat{L}_{ij} - \frac{1}{2}f_if_j, \qquad  \widehat{T}^m_{rs} \ = \ \widehat{T}_{rs}  -f_r^+f_s \label{eontr}
\end{eqnarray}
are given on constrained Fock space $\mathcal{H}^f\otimes\mathcal{H}'\otimes H^{o_A}_{gh}$, with auxiliary Fock space  $\mathcal{H}'$ spanned by $f_1^+, ..., f_k^+$  creation operators.
The set of the operators $Q^m_c , \sigma^i_{m{}c}, \widehat{L}^m_{ij}, \widehat{T}^m_{rs}$ satisfies literally to the superalgebra of (\ref{eqQctot}), (\ref{algrest})    but for massive  HS fields. The spectral problem for the incomplete operator $Q^{\prime m}_c$ leads to the same as in (\ref{Qchir})--(\ref{QLambdair}) relations with only changing for middle set:
\begin{align}
 & \sigma^i_{m{}c}|\chi^{m(n)}_c\rangle=(s_i-\textstyle\frac{d-1}{2})|\chi^{m(n)}_c\rangle \ \mathrm{for} \ |\chi^{m(0)}_c\rangle \equiv  |\chi^{m}_c\rangle.
\label{Qsmr}
\end{align}
The resulting constrained BRST-BFV LF for massive HS field subject to $Y[s_1, ..., s_k]$ is described by the action
\begin{eqnarray}
\mathcal{S}^{m{}c}_{s[k]}(\chi^m_c) = \int d \eta_0 \; {}_{s[k]}\langle\chi_c^m
|Q^m_c| \chi^m_c \rangle_{s[k]}\quad   \label{Sclredm}
\end{eqnarray}
being invariant with respect to  $(\sum_is_i -1)$-th stage reducible gauge transformations
\begin{equation}\label{gtrnscm}
\delta| \chi^{m(l)}_c \rangle_{s[k]} =
Q^m_c| \chi^{m(l+1)}_c \rangle{s[k]}, \ l=0,1,..., \sum_is_i-1;\ \delta| \chi^{m(\sum_is_i)}_c \rangle_{s[k]}=0
\end{equation}
with  off-shell algebraic constraints imposed on whole set of $| \chi^{m(l)}_c \rangle_{s[k]}$:
\begin{eqnarray}
&& \widehat{L}^m_{ij}| \chi^{m(l)}_c \rangle_{s[k]}\ =\ \widehat{T}^m_{rs}| \chi^{m(l)}_c \rangle_{s[k]}\ =\ 0, \quad l=0,1,...,\sum_is_i. \label{offshellconm}
\end{eqnarray}
In turn, the metric-like  component LF in terms of  initial  massive tensor field $\Phi_{\mu^1[{s_1}],...,\mu^k[{s_k}]}$  may be obtained  from $Q^m_c$-BRST  complex resolution of  the LF (\ref{Sclredm})--(\ref{offshellconm}).

The result of dimensional reduction application to massless LF in tensor form  was considered in part for $k=2$ in \cite{resh1} and we will develop this issue in a separate work.

The relations (\ref{Sclredm})--(\ref{offshellconm}) for massive BRST constrained LF  present the basic results of this Section.

Let us turn to the interacting theory.

\section{Local cubic vertices for massless MAS fields}  \label{cubic}

It is well known that consistent interactions of a given gauge model (which need not be a free one) can be classified into three groups:  those that leave the gauge transformations unchanged; those that modify the gauge transformations while preserving their algebra; and those that modify both the gauge transformations and their algebra.

The Noether deformation procedure, considered for irreducible gauge theories in  \cite{BarnichHenneaux1} and then for first-stage reducible theories in \cite{BarnichHenneaux2}, was developed in \cite{RBP} for the construction of interacting LFs with cubic vertices for $p$ copies of MAS HS fields, with preservation of the Poincaré group irreps, in the deformed (non-Abelian) gauge theory with the incomplete BRST operator, for the case of  $3$-column Young tableaux. This procedure applies verbatim to the general case of fields  $\Phi^{(t)}_{\mu^1[{s_{1_t}}],...,\mu^k[{s_{k_t}}]}$  with $k_t\geq 1$, $t=1,...,p$. Here again, in order to obtain a deformed Lagrangian formulation, one should introduce $p$, $p\geq 3$, copies of the  free LFs with actions $\mathcal{S}_{(c)|s[k]_{t}}[\chi^{(t)}_{(c)}]$   equipped with vectors 
$|\chi^{(t)}_c\rangle_{{s}[k]_t}$, reducible gauge parameters
$|\Lambda^{(t)l}_c\rangle_{{s}[k]_t}\equiv |\chi^{(t)l+1}_c\rangle_{{s}[k]_t}$ for $t=1,...,p$ (so that the model is adapted to Yang–Mills-type interactions with gauge group  $SU(N)$, for $p=N^2-1$). 

\subsection{General structure}\label{gencubic}

We suggest, here the solutions for cubic vertex for interacting massless MAS fields when solving  the generating equations for partial case of coinciding the vertices $\big|  V^{(3)}_c\rangle=\big|  V^{(3)1}_c\rangle$ = ... = $\big|  V^{(3)L_t}_c\rangle$ defined in (\ref{S[3]}), (\ref{gt1})
\begin{eqnarray}\label{gencubBRSTc}
 &&  Q_c^{tot}
\big|  V^{(3)}_c\rangle_{(s[k])_{(i)_3}} =0, \  Q_c^{tot}\equiv \sum_{i=1}^p  Q_c^{(i)},  \qquad    \big( L^{(t)}_{ij},\, T^{(t)}_{rs}\big) \big|  V^{(3)}_c\rangle_{({s}[k])_{(i)_3}} \ =\ 0, \\
 && \sigma^{(t)i_t}_c\big|  V^{(3)}_c\rangle_{({s}[k])_{(i)_3}}\ =\  \Big(s_{i_t}+\frac{d-2}{2}\Big)\big|  V^{(3)}_c\rangle_{({s}[k])_{(i)_3}}, 
\label{gencubBRSTc1}
\end{eqnarray}
{for}  $(\varepsilon, gh_H)\big|  V^{(3)}_c\rangle_{({s}[k])_{(i)_3}} = (1+\sum_{j=1}^{3}\varepsilon_{\chi^{(i_j)}},  3)$. 
The equations determine a model of interacting MAS fields $\Phi^{(t)}_{\mu[{{s}{}^{(t)}_1}],..., \rho[{{s}{}^{(t)}_{k_t}}]}, t=1,...,p$, with the action 
\begin{eqnarray}\label{S[e]}
  && S_{[1]|(s[k])_p}[(\chi_c)_p] \  = \  \sum\nolimits_{t=1}^{p} \mathcal{S}_{c|s[k]_{t}}[\chi^{(t)}_c]   +  g S_{1|(s[k])_p}[(\chi_c)_p],\\
\label{S[3]}
\hspace{-0.7em}&\hspace{-0.7em}&\hspace{-0.7em}  S_{1|(s[k])_p}[(\chi_c)_p] =   \sum_{1\leq i_1<i_2<i_3\leq p} \hspace{-0.0em} \int \prod_{j=1}^3 d\eta^{(i_j)}_0  \Big( {}_{s[k]_{i_j}}\langle \chi^{(i_j)}_c
  \big|  V^{(3)}_c\rangle_{(s[k])_{(i)_3}}+h.c. \Big)  ,
  \end{eqnarray}
and reducible $l$-th level gauge transformations: $\delta^l_{[1]}  |\Lambda^{(t)l}_c\rangle_{s[k]_{t}}=(\delta^l_0+g\delta^l_1 )|\Lambda^{(t)l}_c\rangle_{s[k]_{t}} $ :
\begin{eqnarray}\label{gt1}
\hspace{-1.1em}&\hspace{-1.1em}&\hspace{-1.1em} \delta^l_1|\Lambda^{(t)l}_c\rangle_{s[k]_{t}}\hspace{-0.1em} = \hspace{-0.1em} - \hspace{-1.0em}\sum_{1\leq i_1<i_2\leq p}\int \hspace{-0.3em}\prod_{j=1}^2 d\eta^{(i_j)}_0 \hspace{-0.15em} \Big[\hspace{-0.1em} {}_{s[k]_{\{i_1}}\hspace{-0.25em}\langle \chi^{(\{i_1)}_c
  \big|  {}_{{s}[k]_{i_2\}}}\hspace{-0.25em}\langle \Lambda^{(i_2\})l+1}_c
  \big| {V}{}^{(3)l}_c\rangle_{({s}[k])_{(i)_2{}t}}\hspace{-0.1em}+h.c. \hspace{-0.15em}\Big]\hspace{-0.1em}  ,
  \end{eqnarray}
(with the notation  $|\Lambda^{(t)-1}_c\rangle \equiv |\chi^{(t)}_c\rangle$,  for $l=-1,0,...,L_t$) up to linear in deformation constant $g$
approximation. This deformation procedure modifies both the gauge symmetry and its algebra. 

The resolution of the system   (\ref{gencubBRSTc})  for $x$-local cubic vertices 
\begin{equation}\label{xdep}
  \big|  V^{(3)}_c\rangle_{(s[k])_{(i)_3}}  = \prod_{l=1}^3 \delta^{(d)}\big(x -  x_{i_l}\big) V^{(3)}_{c|(s[k])_{(i)_3}}(x)
  \prod_{l=1}^3 \eta^{(i_l)}_0 |0\rangle , \ \  |0\rangle\equiv \otimes_{t=1}^p |0\rangle^{t},
\end{equation} 
can be found in terms of the products of BRST-closed (i.e., $[Q_c^{tot}, L^{(i_j)}_{n_{i_j}}\} = [Q_c^{tot}, Z^{(i)_3}_{m_{i_1}n_{i_2}o_{i_3}} \} $ $=0$)  Grassmann-odd  $x$-local polynomials of the first degree in powers of momenta and of the first (for $L's$) and third (for $Z's$) degrees in oscillators 
\begin{eqnarray}\label{LrZLL}
      &&  L^{(i_j)}_{n_{i_j}} \ = \   (p^{(i_{j+1})}_{\mu}-p^{(i_{j+2})}_{\mu})\hat{a}^{(i_j)\mu+}_{n_{i_j}} - \imath \big(\mathcal{P}^{(i_{j+1})}_0- \mathcal{P}^{(i_{j+2})}_0
  \big)q_{n_{i_j}}^{(i_j)+} , \nonumber \\
  && Z^{(i)_3}_{m_{i_1}n_{i_2}o_{i_3}} \ = \  L^{(i_1i_2)+}_{m_{i_1}n_{i_2}}L^{(i_3)}_{o_{i_3}} + L^{(i_2i_3)+}_{n_{i_2}o_{i_3}}L^{(i_1)}_{m_{i_1}} + L^{(i_3i_1)+}_{o_{i_3}m_{i_1}}L^{(i_2)}_{n_{i_2}}. \label{LrZ1}
\end{eqnarray}
Here we have used   $p^{(i_j)}_{\mu} = -i\partial^{(i_j)}_{\mu}$,  relation  $[i_{j+3}\sim i_j]$ and the definitions for "mixed traces" (of indices from columns $m_{i_1}$, $n_{i_2}$ of different copies enumerated by $i_1$ and $i_2$) with antisymmetry property
\begin{eqnarray}
L^{(i_1i_2)+}_{m_{i_1}n_{i_2}} \ = \ \textstyle \hat{a}^{(i_1)\mu+}_{m_{i_1}}\hat{a}^{(i_2)+}_{\mu{}n_{i_2}} -\frac{1}{2}\big(
p^{(i_1)+}_{m_{i_1}}q_{n_{i_2}}^{(i_2)+} -p^{(i_2)+}_{n_{i_2}}q_{m_{i_1}}^{(i_1)+}\big); \quad  L^{(i_1i_2)+}_{m_{i_1}n_{i_2}} \ = \ - L^{(i_2 i_1)+}_{n_{i_2}m_{i_1}}
.\label{Lrr+1}
 \end{eqnarray}
As consequence the same antisymmetric property is true for  $Z^{(i)_3}$:
  \begin{equation}\label{antisymZ3}
 Z^{(i)_3}_{m_{i_1}n_{i_2}o_{i_3}} = - Z^{(i_2i_1i_3)}_{n_{i_2}m_{i_1}o_{i_3}} = - Z^{(i_1i_3 i_2)}_{m_{i_1}o_{i_3}n_{i_2}}
 \end{equation}
 It is easy to check that Grassmann-even   "mixed traces" are not BRST-closed quantities ($[Q_c^{tot}, L^{(i_1i_2)+}_{m_{i_1}n_{i_2}} \} \ne 0$)  and therefore can not be used to form the vertex as independent generating elements.
 In addition to operators (\ref{LrZ1}) the following recursively determined  fifth $Z^{(i)_5}$, and, in general, $(2e-1)$-th order $Z^{(i)_{2e+1}}$ polynomials in power of $a^+, p^+, q^+$ for natural $e$, such that $0<i_1<i_2<...<i_{2e+1}\leq p$:
 \begin{eqnarray}\label{LrZ5LL}
    \hspace{-0.7em}  &\hspace{-0.7em}& \hspace{-0.7em} Z^{(i)_5}_{m^1_{i_1}m^2_{i_2}...m^5_{i_5}}  = \ L^{(i_1i_2)+}_{m^1_{i_1}m^2_{i_2}}Z^{(i_3i_4i_5)}_{m^3_{i_3}m^4_{i_4}m^5_{i_5}} + cycl.perm.((i_1,m^1_{i_1}), \ldots (i_5,m^5_{i_5}))  ,  \\
 \hspace{-0.7em}  &\hspace{-0.7em}& \hspace{-0.7em} Z^{(i)_{2e+1}}_{m^1_{i_1}m^2_{i_2}...m^{2e+1}_{i_{2e+1}}}  = \   L^{(i_1i_2)+}_{m^1_{i_1}m^2_{i_2}}Z^{(i_3i_4...i_{2e+1})}_{m^3_{i_3}... m^{2e+1}_{2e+1}} + cycl.perm.((i_1,m^1_{i_1}), \ldots (i_{2e+1},m^{2e+1}_{2e+1})). \label{LrZe}
\end{eqnarray}
  are $Q_c^{tot}$-BRST-closed as well. They as the  operators $ L^{(i_j)}, Z^{(i)_3}$ do not contain BRST-exact summands of the respective form $Q_c^{tot} X^{(i_j)}$ or $Q_c^{tot} Y^{(i)_3}$ with any Grassmann-even first (third) order in oscillators local $X^{(i_j)}$ ($Y^{(i)_3}$) with ghost number equal to $(-1)$, thus cannot be simplified anymore by using field redefinitions. They satisfy as well the antisymmetric property (\ref{antisymZ3}). 
Hence, due to  nilpotency   of the first order in $p^\mu$ polynomials: $(L^{(i_j)}_{n_{i_j}})^2 = ( Z^{(i)_3}_{m_{i_1}n_{i_2}o_{i_3}})^2 = \ldots $ $ =(Z^{(i)_{2e+1}}_{m^1_{i_1}...m^{2e+1}_{i_{2e+1}}})^2 =0$)  have not sufficient BRST-closed independent polynomials to compose  parity  invariant vertex  $3$-vector for any MAS interacting fields as it was for MS HS fields \cite{2505.02190}, \cite{BRST-BV3} (without using Levi-Civita tensors)\footnote{the  polynomials $Z^{(i)_{2e+1}}_{m^1_{i_1}...m^{2e+1}_{i_{2e+1}}}$ are used to compose the $({2e+1})$-rtic vertex $\big|V^{(2e+1)}_{c|(s[k])_{(i)_{2e+1}}}\rangle$ in case of its necessity to make deformed theory to be closed in all orders in $g$.}.

The parity  invariant local cubic interaction
vertex for $p$ copies of irreducible massless MAS fields  without divergencies takes the form for $1 \leq i_1, i_2, i_3\leq p$:
\begin{eqnarray}
\label{vertexgen}  V^{(3)}_{c|(s[k])_{(i)_3}}(x) & = & V^{(3)}_{c|(s[k])_{(i)_3}}\Big(L^{(i_1)}_{n_{i_1}}, L^{(i_2)}_{n_{i_2}}, L^{(i_3)}_{n_{i_3}},
Z^{(i)_3}_{m_{i_1}n_{i_2}o_{i_3}}\Big| \widehat{L}{}^{(i_j)+}_{m_{i_j}n_{i_j}} ,\widehat{T}{}^{(i_j)+}_{r_{i_j}s_{i_j}}\Big)\,,
\end{eqnarray}
where $V^{(3)}_{c|(s[k])_{(i)_3}}$ for $(\varepsilon, gh_H)V^{(3)} = (\sum_{j=1}^{3}\varepsilon_{\chi^{(i_j)}},  0)$ is an arbitrary polynomial  of quantities
determined in (\ref{LrZLL})--(\ref{Lrr+1})  for traces $\widehat{L}{}^{(i_j)+}_{m_{i_j}n_{i_j}}$ and dual Young, $\widehat{T}{}^{(i_j)+}_{r_{i_j}s_{i_j}}$,  operators for $j=1,2,3$.

The latter are given according to (\ref{BRSTextTr}), (\ref{BRSTextYo}), (\ref{constbrstop}) as follows
\begin{eqnarray}\label{tracYoungoper}
\hspace{-1em}
\widehat{L}{}^{(i_j)+}_{m_{i_j}n_{i_j}}&=& l^{(i_j)+}_{m_{i_j}n_{i_j}} + \frac{1}{2} q^{(i_j)+}_{[m_{i_j}} p^{(i_j)+}_{n_{i_j}]} \,,\quad \widehat{T}{}^{(i_j)+}_{r_{i_j}s_{i_j}} = \ {t}^{(i_j)+}_{r_{i_j}s_{i_j}}+(q^{(i_j)+}_{s_{i_j}}p^{(i_j)}_{r_{i_j}}+q^{(i_j)}_{r_{i_j}}p^{(i_j)+}_{s_{i_j}}).
\end{eqnarray}
We will call the polynomials as the first  for $L^{(i_j)}_{n_{i_j}}$, the second  for $L^{(i_1i_2)+}_{m_{i_1}n_{i_2}}$, the third for $Z^{(i)_3}_{m_{i_1}n_{i_2}o_{i_3}}$,  the $({2e+1})$-th for $ Z^{(i)_{2e+1}}_{m^1_{i_1}...m^{2e+1}_{i_{2e+1}}} $ orders for  forms in power of oscillators.
The first order forms ($L^{(i_j)}$) relate to gravitational type of interactions, whereas the third order ones  ($Z^{(i)_3}$) to Yang--Mills type of interactions in accordance with \cite{Metsaev0512}.  The maximal order of derivatives in the vertex (\ref{vertexgen}) may be equal to $(\sum_{j=1}^3k_{i_j})$ in such a way, that with each column of number $m_{i_j}$ with height $s_{i_j}$ in $\Phi^{(i_j)}_{\mu^1[{s_{1_{i_j}}}],... ,\mu^{m}[{s_{m_{i_j}}}], .., \mu^k[{s_{k_{i_j}}}]}$ from any of interacting MAS fields maybe associated at most 1 degree of momenta $p^{(i_j)}_{\mu}$. 

Without imposing algebraic constraints the deformed LF describes a dynamics for many particles with less spins then for original free HS  fields $\Phi^{(t)}_{\mu^1[{s_{1_t}}],...,\mu^k[{s_{k_t}}]}$, $t=1,...,p$.

To have the cubic vertex for irreducible fields we should to extract from the vertex (\ref{vertexgen}) the traceless and Young-antisymmetric modes by means of use the second group operators $\widehat{L}{}^{(i_j)+}_{m_{i_j}n_{i_j}} ,\widehat{T}{}^{(i_j)+}_{r_{i_j}s_{i_j}}$.  This problem maybe solved  by 2 ways. First, in implicit form within  BRST-BV approach with incomplete BRST operator when the field and transformed (into ghost fields)  gauge parameters are combined together with theirs antifield vectors in unique generalized field-antifield vector, $|\chi^{(i)}_{g|c}\rangle_{s[k_i]}$, the deformed minimal BRST-BV action,  $S_{[1]c|(s[k])_p}[(\chi_{g|c})_p]$ = $S_{[1]c|(s[k])_p}[(\chi_{c})_p]\big|_{(\chi^{(i)}_{c}\to \chi^{(i)}_{g|c})}$ obtained from deformed classical action (\ref{S[e]}) (see for details \cite{2303.02870} to be applicable for the method with off-shell constraints) will completely select the traceless and Young-antisymmetric  parts from the vertex $\big|  V^{(3)}_c\rangle_{(s[k])_{(i)_3}}$ in the form $\big|  \overline{V}{}^{(3)}_c\rangle_{(s[k])_{(i)_3}}$  (\ref{L11V+})  without the rest vertices $\big| {V}{}^{(3)l}_c\rangle_{({s}[k])_{(i)_3}}$ for $l=0,1,...,L_t$.  Second, it may be done by constructing the projectors $\mathbf{P}^{(t|L)}_{i_t j_t}$  on traceless and $\mathbf{P}^{(t|T)}_{r_t q_t}$ Young-antisymmetric modes 
by the receipt suggested in \cite{2505.02190}  for mixed-symmetric  HS fields
\begin{eqnarray}\label{L11V+}
 \hspace{-0.5em}&\hspace{-0.5em}&\hspace{-0.5em}  \big|\overline{V}{}^{(3)}_{c}\rangle_{(s[k])_{(i)_3}} = \ \prod^{k_{i_j}}_{1\leq m_{i_j}<n_{i_j}<k_{i_j}}\mathbf{P}^{(i_j|L)}_{m_{i_j}n_{i_j}}\prod^{k_{i_j}}_{1\leq r_{i_j}<s_{i_j}<k_{i_j}}\mathbf{P}^{(i_j|T)}_{ r_{i_j}s_{i_j}}\big|  V^{(3)}_c\rangle_{(s[k])_{(i)_3}}.
 \end{eqnarray}
 Here, operators $\mathbf{P}^{(i_j|L)}_{m_{i_j}n_{i_j}}$  and $ \mathbf{P}^{(i_j|T)}_{r_{i_j}s_{i_j}}$ are respectively projectors for fixed  $i_j$ on  mixed traceless modes with respect to indices in columns $m_{i_j}$ and $n_{i_j}$ groups and Young-symmetry modes with respect to indices in $r_{i_j}$ and  $s_{i_j}$ groups of the same copy $i_j=1,...,p$.
 Explicitly, for $C(l,n)\equiv \prod_{e=0}^{l-1}(e+n)$
 \begin{eqnarray}
      &\hspace{-1.2em}& \hspace{-1.3em}  \mathbf{P}^{(i_j|L)}_{m_{i_j}n_{i_j}}   =  \sum_{k=0}^{\min([\hat{s}_{m_{i_j}}/2], [\hat{s}_{n_{i_j}}/2])}\hspace{-1.3em}(-1)^k\frac{4^k{}C(\hat{s}_{m_{i_j}}+\hat{s}_{n_{i_j}}-1-k,d)}{k!{}C(\hat{s}_{m_{i_j}}+\hat{s}_{n_{i_j}}-1,d)}(\widehat{L}{}^{(i_j)+}_{m_{i_j}n_{i_j}})^k (\widehat{L}{}^{(i_j)}_{m_{i_j}n_{i_j}})^k,  \label{L11V+w}\\
   &\hspace{-1.2em}& \hspace{-1.3em}  \mathbf{P}^{(i_j|T)}_{r_{i_j}s_{i_j}}   =  \sum_{k=0}^{ \hat{s}_{s_{i_j}}}(-1)^k\frac{(\hat{s}_{r_{i_j}}-\hat{s}_{s_{i_j}}+1-k)!}{k!{}(\hat{s}_{r_{i_j}}-\hat{s}_{s_{i_j}}+1)!}(\widehat{T}{}^{(i_j)+}_{r_{i_j}s_{i_j}})^k (\widehat{T}{}^{(i_j)}_{r_{i_j}s_{i_j}})^k . \nonumber 
\end{eqnarray}
Substituting the found local cubic vertex $\big| \overline{ V}{}^{(3)}_c\rangle $ (\ref{L11V+}) into  the action (\ref{S[e]}), (\ref{S[3]})  and sequence of gauge transformations  (\ref{gt1}),  leads to the same properties (with preserved p.d.f.) of the LF for cubically  interacting HS fields with given spins as ones for undeformed  model for $p$-samples of free fields. 

To get nontrivial filling of the vertex $V^{(3)}_{c|(s[k])_{(i)_3}}(x)$ (\ref{vertexgen}) by means of the forms additional to $L^{(i_j)}_{n_{i_j}},
Z^{(i)_3}_{m_{i_1}n_{i_2}o_{i_3}}$ may be realized with modified $Q_c^{tot}$-BRST-closed second order forms $\overline{L}{}^{(i_1i_2)+}_{m_{i_1}n_{i_2}}$  necessary to be by non-local:
\begin{equation}\label{Lrr+1mod}
\overline{L}{}^{(i_1i_2)+}_{m_{i_1}n_{i_2}} \ = \ L^{(i_1i_2)+}_{m_{i_1}n_{i_2}} - \imath\frac{ \big(\mathcal{P}^{(i_1)}_0- \mathcal{P}^{({i_2})}_0
  \big)}{(p^{(i_1)})^2-(p^{(i_2)})^2}  [Q_c^{tot}, L^{(i_1i_2)+}_{m_{i_1}n_{i_2}}\}.
  \end{equation}
  Indeed,  the graded Jacobi identity for double supercommutators $(-1)^{\varepsilon(A)\varepsilon(C)}[A,[B,C\}\}+cycl.perm.(A,B,C)=0$, for  $(A,B,C)=(Q_c^{tot},Q_c^{tot},L^{(i_1i_2)+}_{m_{i_1}n_{i_2}})$, as well as the nilpotency of BRST operator  lead to
  \begin{eqnarray}\label{Lrr+1modcheck}
&\hspace{-1.2em}& \hspace{-1.3em} [Q_c^{tot},\overline{L}{}^{(i_1i_2)+}_{m_{i_1}n_{i_2}}\} \ = \  \imath\frac{ \big(\mathcal{P}^{(i_1)}_0- \mathcal{P}^{({i_2})}_0
  \big)}{(p^{(i_1)})^2-(p^{(i_2)})^2} [Q_c^{tot}, [Q_c^{tot}, L^{(i_1i_2)+}_{m_{i_1}n_{i_2}}\}\} \\
  &\hspace{-1.2em}& \hspace{-1.3em} \phantom{[Q_c^{tot},\overline{L}{}^{(i_1i_2)+}_{m_{i_1}n_{i_2}}\}} \  = \  {\imath}\frac{ \big(\mathcal{P}^{(i_1)}_0- \mathcal{P}^{({i_2})}_0
+  \big)}{(p^{(i_1)})^2-(p^{(i_2)})^2} [(Q_c^{tot})^2, L^{(i_1i_2)+}_{m_{i_1}n_{i_2}}\} =0.\nonumber
  \end{eqnarray}
  As the result, the non-vanishing parity-even cubic vertex exists for any interacting MAS HS  triple of fields in non-local form\footnote{which may be useful for a non-local representation of LF in  \cite{Boulanger3}} for reducible HS fields
  \begin{eqnarray}
\label{vertexgen1} 
    &\hspace{-1.2em}& \hspace{-1.3em}  V^{(3)nl}_{c|(s[k])_{(i)_3}}(x) \ = \ V^{(3)nl}_{c|(s[k])_{(i)_3}}\Big(L^{(i_j)}_{n_{i_j}},
Z^{(i)_3}_{m_{i_1}n_{i_2}o_{i_3}}; \overline{L}{}^{(i_ji_{j+1})+}_{m_{i_j}n_{i_{j+1}}} \Big| \widehat{L}{}^{(i_j)+}_{m_{i_j}n_{i_j}} ,\widehat{T}{}^{(i_j)+}_{r_{i_j}s_{i_j}}, j=1,2,3\Big),
\end{eqnarray}
together with the vertices $\big|\overline{V}{}^{(3)nl}_{c|(s[k])_{(i)_3}}\rangle_{(s[k])_{(i)_3}}$ for irreducible MAS HS field derived according to receipt (\ref{L11V+}).
  
  Therefore, we have shown the non-trivial local cubic vertex exists for massless  interacting  MAS HS fields  subject to the Young tables $Y[1, ... , 1_{k_t}]$, ... , $Y[2, ... 2, 1_{k_t}]$ of heights of columns equal at least to 1 and at most to 2. These  fields, due to Young-antisymmetry conditions correspond to sets of totally-symmetric higher 
  integer spin  $s_k=k_t$  fields $\Phi^{(t)}_{\{\mu^1,\mu^2,...,\mu^{k_t}\}}(x)$ for $t=1,...,p$ and for mixed-symmetric HS fields with 2 group indices.   
  
  Non-local  non-trivial cubic vertex may be realized for larger class of  massless  interacting  MAS HS fields subject to $Y[\hat{s}_{1_t}, ... , \hat{s}_{k_t}]$ as compared to local cubic vertex. 
  The structure for  local (\ref{vertexgen}), (\ref{L11V+}) and non-local cubic vertices (\ref{vertexgen1}) present the main results of the subsection.
  
\subsection{Totally antisymmetric interacting HS fields}\label{cubictotmas}

In case of totally antisymmetric interacting massless fields $\Phi^{(t)}_{\mu[{\hat{s}{}^{(t)}}]}$, $t=1,...,p$, there are no any holonomic constraints and the LF within approaches with complete and incomplete BRST operators coincide.  The local vertex  $V^{(3)}_{c|(s[1])_{(i)_3}} \equiv V^{(3)}_{(i)_3}(s_{i_1},s_{i_2},s_{i_3};k_{(i)_3})$  may be found in accordance with \cite{BarnichHenneaux12} explicitly with omitting $\hat{\ }$ over $s$ and for the simplified notations of the forms: $L^{(i_j)}_{1_{i_j}} \equiv L^{(i_j)}$, $Z^{(i)_3}_{1_{i_1}1_{i_2}1_{i_3}}\equiv Z^{(i)_3}$, for local not BRST-closed $L^{(i_ji_{j+1})+}_{1_{i_j}1_{i_{j+1}}} \equiv L^{(i_ji_{j+1})+} $ and non-local BRST-closed $\overline{L}{}^{(i_ji_{j+1})+}$ as
\begin{eqnarray}\label{totsymV} && \hspace{-1em} V_{(i)_3}^{(3)}\big(s_{i_1},s_{i_2},s_{i_3};k_{(i)_3}\big)\ 
=  \  \sum_{j=1}^3 L^{(i_j)}  + \sum_{j=1}^3 L^{(i_j)}L^{(i_{j+1})} 
+ \prod_{e=1}^3 L^{(i_e)} \\
&&\phantom{\hspace{-1em} V_{(i)_3}^{(3)}\big(s_{i_1},s_{i_2},s_{i_3};k_{(i)_3}\big)
}
+ Z^{(i)_3}\Big[1+  \sum_{j=1}^3 L^{(i_j)} +  \sum_{j=1}^3 L^{(i_j)}L^{(i_{j+1})} \Big] \nonumber
.
\end{eqnarray}
Here the power in derivatives is bounded by the interval $ 1 \leq k_{(i)_3} \leq 3$ and minimal in complete agreement with restriction for local cubic vertex for massless totally-symmetric and mixed-symmetric higher integer spin fields \cite{Metsaev0512}, \cite{BRST-BV3}, because of the summands with $ k_{(i)_3}> 3$ lead after integration by parts in the action to the  multiplier:  $\Phi^{(i_1)}_{\mu[{{s}_{i_1}}]}\partial^{\mu_{i_1}}\partial^{\mu_{i_2}}\equiv 0$, when the boundary terms  vanish. 

The list of admissible local cubic vertices exhausted by triples of  fields: 
\begin{itemize}
  \item $V_{(i)_3}^{(3)}\big(0,0,1;1\big)$, $V_{(i)_3}^{(3)}\big(0,1,1;2\big)$, $V_{(i)_3}^{(3)}\big(1,1,1;3\big)$ with only forms $L^{(i_j)}$;
  \item $V_{(i)_3}^{(3)}\big(1,1,1;1\big)$ with only form $Z^{(i)_3}$;
  \item $V_{(i)_3}^{(3)}\big(1,1,2;2\big)$, $V_{(i)_3}^{(3)}\big(1,2,2;3\big)$ with forms $L^{(i_j)}$, $Z^{(i)_3}$.
\end{itemize}
The above  local nilpotent BRST-closed forms  do not permit to construct 
local nontrivial parity-even  cubic vertices with any field $\Phi^{(t)}_{\mu[s_t]}(x)$ for $\sum_{j=1}^3{s_{i_j} \equiv  \mathbf{s}^{(i)} \geq 6}$ and $s_{i_j}>2$ with modifying the gauge transformations. 

Let us introduce two new sets of operators. First, it is  the polynomials of the first order in momenta $p_\mu$ and $s_{i_1}$-th order in oscillators
\begin{eqnarray}\label{WrZLL1}
  && W^{(i_j)}_{\mu[s_{i_j}+1]} \ = \  p^{(i_{j})}_{[\mu_{0}}\prod_{e_1=1}^{s_{i_j}}\hat{a}^{(i_j)+}_{\mu_{e_1}]}, \ \ \  j=1, ... , 3. 
\end{eqnarray}

Second set contains the forms being by the  third order in momenta $p_\mu$ and $(d-3)$-th order in oscillators forms to pass over the parity-even-invariance condition for the vertices:
\begin{eqnarray}\label{WrZLL}
 \hspace{-1em} &\hspace{-1em}& W^{(i)_3}_{} \ = \  \varepsilon^{\mu_1\mu_2...\mu_d}\prod_{e_1=1}^{s_{i_1}}\hat{a}^{(i_1)+}_{\mu_{e_1}}\prod_{e_2=1}^{s_{i_2}}\hat{a}^{(i_2)+}_{\mu_{e_2}}\prod_{e_3=1}^{s_{i_3}}\hat{a}^{(i_3)+}_{\mu_{e_3}}\prod_{j=1}^{3} p^{(i_{j})}_{\mu_{d-j+3}}, \ \sum_{j=1}^3s_{i_j}=d-3
\end{eqnarray}
with Levi-Civita tensor $\varepsilon^{\mu_1\mu_2...\mu_d}$. 

Polynomials $W^{(i_j)}_{\mu[s_{i_1}+1]} $ and forms $W^{(i)_3}$  are $Q_c^{tot}$-BRST-closed, because of presence exactly 2 degrees of commuting momenta when commuting with BRST operator for former and 4 degree  of momenta -- for the latter: 
\begin{equation}\label{wbrst}
[Q_c^{tot}, W^{(i)_3}\} \sim \varepsilon^{\mu_1\mu_2...\mu_d} \Big(\sum_{l=1}^3 q^{(i_l)+}p^{(i_{l})}_{\mu_{d-l+6}}\prod_{j=1}^{3} p^{(i_{j})}_{\mu_{d-j+3}} \prod\hat{a}^{+} \Big)= 0.
\end{equation}
Hence, if the values of antisymmetric spins $s_{i_1}$, $s_{i_2}$, $s_{i_3}$  for the fields  satisfy to the condition (\ref{WrZLL}) then the  local nontrivial parity-even and parity-odd  cubic vertices with  field $\Phi^{(i_j)}_{\mu[s_{i_j}]}(x)$, $j=1,2,3$ maybe constructed by the rule
\begin{eqnarray}\label{totsymVev}\hspace{-1em} &\hspace{-1em}& V_{(i)_3}^{(3)\texttt{ev}}\big(s_{i_1},s_{i_2},s_{i_3}; 3\big) 
=   W^{(i_1)}_{\mu[k]\nu[s_{i_1}-k+1] }W^{(i_2)\mu[k]\rho[s_{i_2}-k+1]}W^{(i_3)}_{\rho[s_{i_2}-k+1]}{}^{\nu[s_{i_3}-s_{r}+k]}, \\ 
\label{totsymVodd}\hspace{-1em} &\hspace{-1em}& V_{(i)_3}^{(3)\texttt{odd}}\big(s_{i_1},s_{i_2},s_{i_3}; 3\big) 
=    W^{(i)_3}_{}.
\end{eqnarray}
(for $2k =s_{i_1}+s_{i_2} -s_{i_3}+1$ and $k\leq \min(s_{i_1},s_{i_2})$). The latter vertex determines the topological  term in the deformed action.
 
At the same time it is possible to get non-trivial non-local cubic vertices $V^{(3)nl}_{c|(s[1])_{(i)_3}}(x)$ from (\ref{vertexgen1}) with $\overline{L}{}^{(i_ji_{j+1})+}$ (\ref{Lrr+1mod}) for any heights of $s^{s_{i_1},s_{i_2},s_{i_3}}$ not greater than $[(d-2)/2]$. Due to nilpotency  the term with $\frac{\big(\mathcal{P}^{(i_j)}_0- \mathcal{P}^{(i_{j+1})}_0
  \big)}{(p^{(i_j)})^2-(p^{(i_{j+1})})^2}$ in $\overline{L}{}^{(i_ji_{j+1})+}$  is contained in the product $(\overline{L}{}^{(i_ji_{j+1})+})^k$, $k=1,2,...$ in the first degree only, as well as the momenta $p^{(i_{j})}_{\mu}$ from $[Q_c^{tot}, L^{(i_ji_{j+1})+}\}$: 
    \begin{equation}\label{ant}
   (\overline{L}{}^{(i_ji_{j+1})+})^k = ({L}{}^{(i_ji_{j+1})+})^k -\imath(k-1)({L}{}^{(i_ji_{j+1})+})^{k-1}\frac{ \big(\mathcal{P}^{(i_j)}_0- \mathcal{P}^{({i_{j+1}})}_0
  \big)}{(p^{(i_j)})^2-(p^{(i_{j+1})})^2}[Q_c^{tot}, L^{(i_ji_{j+1})+}\} .
  \end{equation}
The non-locality in $V^{(3)nl}_{c|(s[1])_{(i)_3}}(x)$: $\sim$ $\prod_{j=1}^2\frac{ \big(\mathcal{P}^{(i_j)}_0- \mathcal{P}^{({i_{j+1}})}_0
  \big)}{(p^{(i_j)})^2-(p^{(i_{j+1})})^2}$ can not be more than fourth-order in $p_\mu$.
 The nontrivial  non-local cubic vertex may be obtained for interacting   massless fields $\Phi^{(t)}_{\mu[{\hat{s}_{t}}]}$, when $\hat{s}_{t}>1$ as follows
\begin{eqnarray}\label{totsymVnl} && \hspace{-1em} V_{(i)_3}^{(3)nl}\big(s_{i_1},s_{i_2},s_{i_3}; k_{(i)_3}\big)
=  \prod_{j=1}^3(\overline{L}{}^{(i_ji_{j+1})+})^{\sigma_{i_{j+2}}}  + \sum_{j=1}^3 L^{(i_j)}\prod_{p=1}^3(\overline{L}{}^{(i_{j+p}i_{j+p+1})+})^{\tau_{i_{j+p+2}}} \\
&& + \sum_{j=1}^3 L^{(i_j)}L^{(i_{j+1})}\prod_{p=1}^3(\overline{L}{}^{(i_{j+p}i_{j+p+1})+})^{\kappa_{i_{j+p+2}}} + \prod_{e=1}^3 L^{(i_e)}\prod_{p=1}^3(\overline{L}{}^{(i_{j+p}i_{j+p+1})+})^{\rho_{i_{j+p+2}}}\nonumber 
\\
&& + Z^{(i)_3}\Big[
\prod_{j=1}^3(\overline{L}^{(i_ji_{j+1})+})^{\sigma_{i_{j+2}}-1} + \sum_{j=1}^3 L^{(i_j)}\prod_{p=1}^3(\overline{L}{}^{(i_{j+p}i_{j+p+1})+})^{\tau_{i_{j+p+2}}-1} \nonumber \\
&& + \sum_{j=1}^3 L^{(i_j)}L^{(i_{j+1})}\prod_{p=1}^3(\overline{L}{}^{(i_{j+p}i_{j+p+1})+})^{\kappa_{i_{j+p+2}}-1}  \Big]\,, \nonumber \\
&\hspace{-1.1em}& \texttt{for}  \ \sigma_{i_{j}}=\frac{1}{2}\mathbf{s}^{(i)}-s_{i_j}; \  \tau_{i_{j+\alpha}}=\frac{1}{2}(\mathbf{s}^{(i)}-1)-s_{i_{j+\alpha}}, \alpha=1,2,\  \tau_{i_{j}}=\frac{1}{2}(\mathbf{s}^{(i)}+1)-s_{i_j}\,;\label{restr000} \\
&\hspace{-1.1em}& \kappa_{i_{j+1}}=\frac{1}{2}(\mathbf{s}^{(i)}-1)-s_{i_{j+1}}, \kappa_{i_{j+\beta}}=\frac{1}{2}(\mathbf{s}^{(i)}+1)-s_{i_{j+\beta}}, \beta=2,3;  \ \rho_{i_{j}}=\frac{1}{2}(\mathbf{s}^{(i)}-1)-s_{i_j}\nonumber
.
\end{eqnarray}
  The vertex $V_{(i)_3}^{(3)nl}\big(s_{i_1},s_{i_2},s_{i_3};k_{(i)_3}\big)$ for fixed $(i)_3$ composes the family from 7 vertices parameterized by presence of differential forms  being by the number of powers of derivatives $p^{(i_{j})}_{\mu}$ entering in the vertices. For $p=3$ the vertices take the form $V^{(3)}_{111}(s_{1},s_{2},s_{3};k_{1}) \equiv V^{(3)}(s_{1},s_{2},s_{3};k)$.

Note, when considering the local cubic vertices that does not modify the gauge transformations of any levels with parameters $\xi^{(t)l}_{\mu[s_t-l-1]}$
\begin{equation}\label{gaugantsymm}
 \delta \Phi^{(t)}_{\mu[s_t]}(x) = \partial_{[\mu_{s_t}}\xi^{(t)0}_{\mu[s_t-1]]}(x), \ \delta \xi^{(t)l-1}_{\mu[s_t-l]}(x) = \partial_{[\mu_{s_t-l}}\xi^{(t)l}_{\mu[s_t-l-1]]}(x), \ l=1,...,s_t-1.
\end{equation}
 there exists  non-trivial local vertex $V_{(q,r,t)}^{(3)\texttt{ev}}\big(s_{q},s_{r},s_{t}; 3\big)$ (\ref{totsymVev})  constructed from the  field strengths $F^{(t)}_{\mu[s_{t+1}]}(x)$ for $\Phi^{(t)}_{\mu[s_t]}(x)$ 
 \begin{eqnarray}
F^{(t)}_{\mu[s_{t+1}]}(x) &=&
\frac{1}{s_t!}\Bigl[
\partial_{\mu_1}\Phi^{(t)}_{\mu_2\ldots\mu_{s_t+1}} + \sum_{P(\mu_1,...,\mu_{s_t+1})}(-1)^{\epsilon(P(\mu_1,...,\mu_{s_t+1}))}\partial_{\mu_2}\Phi^{(t)}_{\mu_1\ldots\mu_{s_t+1}}
\Bigr]
\label{strength1}
\end{eqnarray}
with $({s_t+1})!$ terms in square brackets in sum with respect to all permutations of indices $\mu_1,...,\mu_{s_t+1}$ with account for its parity $\epsilon(P(\mu_1,...,\mu_{s_t+1}))$. Indeed,   the vertex in the deformed Lagrangian with completely antisymmetric constants $f^{qrt}=-f^{qpr}=-f^{prq}$, $p,q,r=1,...,p$,  
 \begin{eqnarray}\label{L1int}
 \mathcal{L}_{}(\Phi^{(t)}_{\mu[s_t]}(x)) &=& -\frac{1}{({s_t+1})!}\;
\sum_{t=1}^{p} F^{(t)}_{\mu[s_{t+1}]}F^{(t)\mu[s_{t+1}]} 
\\
& +& g f^{pqr}F^{(q)}_{\mu[k]\nu[s_q-k+1] }F^{(r)\mu[k]\rho[s_{r}-k+1]}F^{(t)}_{\rho[s_{r}-k+1]}{}^{\nu[s_{t}-s_{r}+k]}\nonumber
\end{eqnarray}
with a sum over repeated indices $p,q,r$ solves the problem. Natural parameter $k$ should be restricted by the condition:
\begin{equation}\label{kcond}
 \big(s_q-k+1 =s_{t}-s_{r}+k\big) \Leftrightarrow 2k =s_q+s_r -s_t+1 \ \texttt{and} \  k \leq \min(s_q,s_{r}) +1, \  \forall q,s,r.
\end{equation}
A solution for $k$ exists, for instance, for equal odd heights $s_q=s_r=s_t = 2l-1$. In $d=4$ Minkowski space a vertex of this kind is possible for the vector  (Yang-Mills) field  $\Phi^{(t)}_\mu$ with values in some representation of algebra $su(N)$, for $p=N^2-1$.    
 A more detailed study of consistent local interactions for exterior-form gauge fields has been given in \cite{BarnichHenneaux12}. The local cubic vertices presented in this section are in good agreement with those results.

Note that cubic interaction vertices for massless and massive fields may be obtained both via the dimensional reduction procedure from massless HS fields and in an independent way. A detailed classification of the vertices requires their analysis within the light-cone formalism — for example, along the lines of that performed for mixed-symmetric HS fields in  \cite{Metsaev0512} — together with a simultaneous consistency check against the local BRST cohomology analysis of 
\cite{BarnichHenneaux12},  on the basis of the BRST–BV machinery \cite{2303.02870}.
 Thus, the derivation of the subset of cubic vertices for massless (ir)reducible MAS fields within the Lagrangian formulation with the incomplete BRST operator constitutes the basic result of this Section.

\section{Conclusions}

We developed   BRST approaches with complete  and incomplete BRST operators to solve a problem of constructing local LFs for
mixed-antisymmetric HS field $\Phi_{\mu^1[{\hat{s}_1}],...,\mu^k[{\hat{s}_k}]}$ of generalized integer spin $\mathbf{s} = (k,k,...,k,k-1,k-1,...,k-1,...,1,...,1)$
in a  Minkowski space $\mathbb{R}^{1,d-1}$ whose symmetry of indices is subject to $k$-column Young tableaux $Y[\hat{s}_1, ... , \hat{s}_k]$. It is established the respective LFs  for free HS field have equivalent dynamics and are   qualified as Abelian gauge theories of  stages  reducibility depending on sum of spin components  equal respectively to $(\sum_i\hat{s}_i+\frac{1}{2}k(k-1)-1)$ given in (\ref{dx0})--(\ref{Scl})  and  $(\sum_i\hat{s}_i-1)$ given in (\ref{Sclred})--(\ref{gaugecvectf})  with fewer auxiliary fields for the model with algebraic constraints. LF with incomplete $Q_c$ BRST operator in case without traceless and mixed-antisymmetric constraints describes the gauge theory on the same configuration space of fields, but for reducible Poincare group representation  including with initial field $\Phi_{\mu^1[{\hat{s}_1}],...,\mu^k[{\hat{s}_k}]}$ the fields with  lesser (in lexicographical sense) values of spin.
The results  are applicable both for massless  and massive particles. One should stress the obtained here local LFs in antisymmetric basis, first, are related with local LFs (in symmetric basis) for mixed-symmetric HS fields subject to $l$-rows Young tableaux $Y({s}_1, ... , {s}_l)$ (\cite{1110.5044}), but can not be simply reformulated  from the latter and deserves to be a new independent tool to describe Lagrangian dynamics of Poincare group irreducible representations. Second, as compared to the gauge-invariant LF with non-local action for free  massless tensor fields transforming in arbitrary irreducible representations of general linear group in $d$-dimensional constant curvature space-times  \cite{Boulanger3}, it does not require to reformulate LF in local form necessary 
to make conventional calculations  within classical and quantum diagrammatic Feynman techniques (as it was done in the Standard model and its modifications).

For  unconstrained LF it was realized the additive conversion procedure of the initial
operator second-class constraint system
into a system of the first-class operators. To get new scalar  oscillator representation for the additional parts of constraints $o'_I$, it is found in the first time
 the  auxiliary representations in the form of Verma and Fock modules for
 the second-class operator constraints  
 subalgebra. It was shown that this subalgebra (in HS symmetry superalgebra of all operators $o_I$) is isomorphic due to Howe duality to orthogonal $so(k,k)$
 algebra.  
  To illustrate general prescription we derived known (from \cite{resh1}) reducible gauge-invariant  LF in terms of MAS massless irreducible field $\Phi_{\mu^1[{{s}_1}],\mu^2[{{s}_2}]}$ with two group indices from LF with incomplete BRST operator. 
  The obtained result may be used to get the interesting result for  HS fields subject to Young tableau $Y[\hat{s}_1,\hat{s}_2,1,...,1]$.
 
We enlarged the results of  deformation procedure, considered  
 for  constructing interacting LF with possible local cubic vertices for $p$-copies of  HS fields with preservation the  Poincare group irreducible representations for  deformed (non-Abelian) gauge theory  
 within approach
 with incomplete BRST operator in case $3$-column Young tableaux in \cite{RBP}  to the general case for $p\geq 3$ (to adapt the model for Yang-Mills type interactions with gauge group $SU(N)$ when $p=N^2-1$). 
   The resolution of the generating equations for the $3$-vector local  cubic vertex $\big|  V^{(3)}_c\rangle_{(s[k])_{(i)_3}}$   for massless fields  permitted one to present it as the sum of homogeneous products of $Q_c^{tot}$-BRST closed Grassmann-odd first,  and third orders forms  in power of oscillators and linear in momenta  $p^\mu$ firstly given in  (\ref{LrZ1}). For higher order local vertices it is found the homogeneous BRST closed Grassmann-odd linear in  $p^\mu$ form $Z^{(i)_{2e+1}}_{m^1_{i_1}...m^{2e+1}_{i_{2e+1}}}$ (\ref{LrZe}).
   The cubic vertex can  contain the  maximal degrees in powers of momenta $p_\mu$ at most equal to sum of all columns, and at most one momenta composing gradient with indices from given group $\mu[{{s}{}^{(t)}}]$: $\Phi^{(i_1)}_{\mu[{{s}{}^{(t)}}]}\partial^{\mu_{i_1}}$. Due to "mixed trace" Grassmann-even second order in oscillators operators $L^{(i_1i_2)+}_{m_{i_1}n_{i_2}}$ (\ref{Lrr+1})  are not  $Q_c^{tot}$-BRST closed we met the obstacle to construct 
   local parity invariant  cubic vertices with nontrivial deformation of the gauge transformations and gauge algebra for  any mixed-antisymmetric HS fields as compared for mixed-symmetric HS fields \cite{2505.02190}, \cite{BRST-BV3}. The  only for the fields with  heights of columns in Young tableaux not greater than $2$ it is able to construct local vertices with these forms (without using Levi-Civita tensors).
   
   It is shown the non-local non-trivial BRST-closed parity-even cubic vertex 
     $V^{(3)nl}_{c|(s[k])_{(i)_3}}$ (\ref{vertexgen1})  exists for any interacting mixed-antisymmetric  HS  triples of fields by means of  modified "mixed trace" non-local second order  BRST-closed forms (\ref{Lrr+1mod}) $\overline{L}{}^{(i_1i_2)+}_{m_{i_1}n_{i_2}}$. 
   
   The  nontrivial local cubic vertices $\big|\overline{V}{}^{(3)}_{c}\rangle_{(s[k])_{(i)_3}}$ for  irreducible HS fields satisfying to traceless and Young-antisymmetric constraints (\ref{gencubBRSTc})  are derived in (\ref{L11V+}) by means of system of projectors on traceless and Young-antisymmetric modes. 
   
   For the simplest case of local cubic vertices for $p$ copies of massless totally antisymmetric HS fields (without using Levi-Civita tensors) with modifying gauge transformations and their algebra we have found the admissible local cubic vertices (\ref{totsymV}) exhausted by 6 triples, including  2 vertices  with one derivative, 2 vertices  with two derivatives,  and 2 vertices  with three derivatives. The nontrivial non-local cubic vertex (\ref{totsymVnl}) exists with filling its  $3$-vector by  the powers of modified  forms $\overline{L}{}^{(i_1i_2)+}_{m_{i_1}n_{i_2}}$ for any triples of totally antisymmetric fields. The local cubic vertices (of Yang--Mills type in the first order in coupling constant $g$) of the 3-rd order in $p^\mu$ that does not modify the gauge transformations constructed from the field strengths exist (\ref{L1int}) for the case when sum of all heights: $(s_{i_1}+s_{i_2}+s_{i_3})$ is odd natural (\ref{kcond}) is derived from the new linear in momenta BRST-closed polynomials $ W^{(i_j)}_{\mu[s_{i_j}+1]} \ne Q^{tot}_c Y^{(i_j)}(a^{i_j+})$ (\ref{WrZLL1}). These polynomials as well as the new third order in momenta forms $ W^{(i)_3}_{}$ (\ref{WrZLL})  generating parity-odd local vertices (with topological terms in the deformed action)     maybe generalized for the case of MAS HS fields.
   
   The local cubic vertices  for totally antisymmetric HS fields  are in good agreement with the general results from BV-cohomological analisys  by Henneaux and Knaepen \cite{BarnichHenneaux12}.
   
   Consider presented research as the initial steps in constructing cubic interactions for the fields in question it is interesting to find and classify all local cubic vertices  both for mixed-antisymmetric HS fields and for mixed-symmetric with mixed-antisymmetric HS fields.  One of the efficient way is to develop
theirs study in terms of physical degrees of freedom within light-cone formalism \cite{Metsaev0512}, \cite{Metsaev:2007rn}, \cite{Metsaev2022}. 

 The procedure permits to  study also an important problem of consistent interactions of massless and massive  mixed-antisymmetric HS fields.  
  We stress, that the structure of cubic vertices for massless MAS fields is significantly differed from one for massless mixed-symmetric HS fields.  
  The study of the interacting HS fields within approach with complete BRST operator \cite{2105.12030}, containing in itself all the constraints describing irreducible HS fields, promises to reveal new features in the structure of cubic vertices.
  
  The 
interesting BRST example of higher-spin gravity with supersymmetry in any dimension was recently discussed
in \cite{Vasiliev2025}, while the  application of the BRST approach to the study spinning black
holes was realized in \cite{SkvortsovTsulaia1, SkvortsovTsulaia2}. The problem of constructing the BRST-BV (Batalin-Vilkovisky) minimal,  quantum  and effective actions for the interacting LF  may be  considered  following to
extension of BRST-BV approach \cite{2303.02870} with complete, $Q$,  and  with incomplete, $Q_c$, BRST operator according to~\cite{2010.15741} in order to get an effective action in terms of initial basic HS field with minimal set of appropriate Nakanishi-Lautrup and ghost-antighost fields (as it was done for totally-antisymmetric  HS fields in AdS spaces for integer \cite{Kuzenko} and half-integer \cite{BarvBK} spins)\footnote{See, as well the work \cite{Ohta1} with using antisymmetric reducible ghost tensors for Lagrangian quantizing a model of Unimodular Gravity.}
in order to perturbatively evaluate  average expectation values of the quantities composed from mixed-antisymmetric   HS fields.
In fact, the BRST-BV approaches for MAS fields in question is immediately  obtained from BRST approaches by extending the field contents for the general vectors $|\chi\rangle$ (\ref{extState}) or $|\chi_c\rangle$ (\ref{extStater}) by the minimal ghost fields of all levels for positive values of Lagrangian ghost number $gh_L$, as it is described in \cite{2303.02870}, as well as by theirs respective antifields for negative values of $gh_L$. The only requirement, that there are no any gauge transformations for an BRST-BV action, which now satisfies to the (controlling a consistency of interactions) master-equation.

Finalizing, we will plan also to study the problem of LF construction for (ir)reducible  MAS HS fields with half-integer generalized spin on Minkowski space-times.




\appendix
\section{Verma module for superalgebra $\mathcal{A}(Y[k],\mathbb{R}^{1,d-1})$}\label{ap1}
\renewcommand{\theequation}{\Alph{section}.\arabic{equation}}
\setcounter{equation}{0}

In this appendix, we describe the method of auxiliary
representation  construction (known for mathematicians as Verma
module \cite{Dixmier}) for the Lie algebra   with
second-class constraints $\{o'_a, {o'}^+_a \} = \{l'_{ij},
t^{\prime}_{rs}, l^{\prime +}_{ij}, t^{\prime +}_{rs}\}$ and Cartan
subalgebra elements $g_0^{\prime i}$.

Following to Poincare--Birkhoff--Witt  theorem, we start to
construct Verma module, based on Cartan  decomposition of the subalgebra
\begin{equation}\label{Cartandecomp}
   \{o'_a, {o'}^+_a ; g_0^{\prime i}\} \ =\  \{l^{\prime +}_{ij}, t^{\prime +}_{rs}\} \oplus \{g_0^{\prime i}\} \oplus \{l'_{ij},
t^{\prime}_{rs}\} \equiv \mathcal{E}^-_{k(k-1)}\oplus
H_k \oplus\mathcal{E}^+_{k(k-1)}.
\end{equation}
Then, we
consider the highest-weight representation of the algebra  with the highest-weight vector
$|0\rangle_V$, which should be
annihilated by the positive roots from  $\mathcal{E}^+_{k(k-1)} \equiv  (l'_{ij},
t^{\prime}_{rs})$, and being a proper one for
the Cartan elements $g_0^i$,
\begin{align}\label{hwrep}
& l'_{ij}|0\rangle_V =t^{\prime}_{rs}|0\rangle_V =0, && g_0^i |0\rangle_V =
h^i|0\rangle_V.
\end{align}
The general vector of the  Verma module  $V(so(k,k)) := \mathcal{U}(\mathcal{E}^-_{k(k-1)})\otimes |0\rangle_V$,  
written concisely as $|\vec{N}\rangle_V$ is determined  as:
\begin{eqnarray}
 |\vec{N}\rangle_V \ \stackrel{def}{=}\  |\vec{n}_{ij}, \vec{p}_{rs} \rangle_V \ = \ \prod_{i=1}^{k-1}\Bigr[\prod_{j=i+1}^{k}
\bigl(l^{\prime +}_{ij}\bigr){}^{n_{ij}}\Bigl]\prod_{r=1}^{k-1}\Bigr[\prod_{s=r+1}^{k}
\bigl(t^{\prime +}_{rs}\bigr){}^{p_{rs}}\Bigl] |0\rangle_V
  \label{module},
\end{eqnarray}
(for $\vec{n}_{ij} = (n_{12},...; n_{1k}; n_{23},...,n_{2k};...; n_{k-1{} k } \big)$ and $\vec{p}_{rs} = (p_{12},...; p_{1k}; p_{23},...,p_{2k};...; p_{k-1{} k } \big)$). Note,
that  all $\sum_{j>i}n_{ij}\leq d$ for any fixed number  $i$ and $\sum_{s>r}p_{rs}\leq d$ for any fixed  $r$.

In order to find the result of the action of all the  elements $\{o'_a, {o'}^+_a \}$ on the vector $|\vec{N}\rangle_V$ we should  to use a general formula
\begin{eqnarray}\label{ABnprod}
A B^n & = & \sum_{k = 0}^n
\binom{n}{k} B^{n - k}{ad}_{B}^{k} A \mbox{, \;with \;}{ad}_{B}A \equiv [A,B] ,
\end{eqnarray}
for  any $A, B$ from the set $\{o'_a, {o'}^+_a \}$.

It is easy to get the result of the action of negative root vectors, i.e.
$(l^{'+}_{i'j'}, t^{'+}_{r's'})$ and Cartan generators,
$g_{0}^{\prime i} $ on $|\vec{N}\rangle_V$
\begin{eqnarray}
\label{t'+lm} && t^{\prime+}_{r's'}  |\vec{N}\rangle_V  =  \left|\vec{N} + \delta_{r's',rs} \rangle_V \right. -
\sum_{k'=1}^{r'-1}p_{k'r'}\left|\vec{N} - \delta_{k'r',rs}+ \delta_{k's',rs} \rangle_V \right.  \\
  &&\ \ - \sum_{k'=1}^{r'-1}n_{k'r'}\left|\vec{N}-
  \delta_{k'r',ij}  + \delta_{k's',ij}\rangle_V\right. + \sum_{k'=r'+1}^{s'-1}n_{r'k'}\left|\vec{N}-
  \delta_{r'k',ij}  + \delta_{k's',ij}\rangle_V\right. \nonumber \\
  &&\ \ - \sum_{k'=s'+1}^{k}n_{r'k'}\left|\vec{N}-
  \delta_{r'k',ij}  + \delta_{s'k',ij}\rangle_V
 \right.\,,\nonumber
 \\
  && l_{i'j'}^{+\prime} |\vec{N} \rangle_V \ = \  | \vec{n}_{ij}+\delta_{i'j',i j}, \vec{p}_{rs}
  \rangle_V, \label{tl12+}\\
&& g_{0{}i}^{\prime } |\vec{N}\rangle_V \ = \ \left( \sum_{l<i}n_{li} + \sum_{l>i} n_{il}  - \sum_{s>i}p_{is}+\sum_{r<i}p_{ri} + h^i\right)
 \left|\vec{N}\rangle_V
 \right.\,.\label{g'0i}
\end{eqnarray}
Second, the relation (\ref{ABnprod}) permits to find both the
identities,
\begin{equation}\label{identllm}
l^{\prime }_{i'j'} \left|{\vec{0}}_{ij}, {\vec{p}}_{rs}\rangle_V
\right. = 0
\end{equation}
and the equation
 in acting of the positive root vectors $ t^{'}_{l'm'}$  on the vector $|\vec{0}_{ij},
\vec{p}_{rs}\rangle_V$ (due to non-commutativity of the negative
root vectors $t^{\prime +}_{rs}$ among each other) in the form,
\begin{eqnarray}
 \label{t'recurr}
  t^{\prime}_{l'm'}
|\vec{0}_{ij},\vec{p}_{rs}\rangle_V &=&
\left|C^{l'm'}_{\vec{p}_{rs}}\rangle_V
 \right.-
\sum_{n'=1}^{l'-1}p_{n'm'}\left| \vec{0}_{ij},
\vec{p}_{rs}-\delta_{n'm',rs}+\delta_{n'l',rs}\rangle_V
 \right. \\
  && + \sum_{k'=l'+1}^{m'-1}p_{l'k'}\Bigr[\prod_{r'< l', s'>r'}\prod_{r'=l', m'> s'>r'} \bigl(t^{\prime
+}_{r's'}\bigr){}^{p_{r's'}-\delta_{l'k',r's'}}\Bigl]
t'_{k'm'}\nonumber\\
&& \times\prod_{q'= l', t'\geq m'}\prod_{q'> l', t'>q'}
\bigl(t^{\prime +}_{q't'}\bigr){}^{p_{q't'}} \left|0\rangle_V
 \right.,\nonumber
\end{eqnarray}
where the vector $ \left|C^{l'm'}_{\vec{p}_{rs}}\rangle_V
 \right.$, $l'<m'$, is determined as follows,
\begin{eqnarray}\label{Clmin}
 \left|C^{l'm'}_{\vec{p}_{rs}}\rangle_V
 \right. &=& p_{l'm'}\Big(h^{l'}-h^{m'}-\sum_{k'=m'+1}^{k}(p_{l'k'}- p_{m'k'})- \sum_{k'=l'+1}^{m'-1}p_{k'm'}-p_{l'm'}+1\Big)
 \\
&& \times \left|
\vec{0}_{ij}, \vec{p}_{rs}-\delta_{l'm',rs}\rangle_V
 \right.  + \sum_{k'=m'+1}^{k}p_{l'k'}\Bigl\{\left| \vec{0}_{ij},
\vec{p}_{rs}-\delta_{l'k',rs}+\delta_{m'k',rs}\rangle_V
 \right. \nonumber\\
  && - \sum_{n'=l'+1}^{m'-1} p_{n'm'}\left|\vec{0}_{ij},
\vec{p}_{rs}-\delta_{l'k',rs}-\delta_{n'm',rs}+\delta_{n'k',rs}\rangle_V
 \right. \Bigr\} .\nonumber
 \end{eqnarray}
Recurrent relation (\ref{t'recurr}) maybe easily resolved  so
the solution has the form, (for $k'_{-1}\equiv 1$)
\begin{eqnarray}
 \label{t'fin}
  t^{\prime}_{l'm'}
|\vec{0}_{ij},\vec{p}_{rs}\rangle_V
&=&\left|C^{l'm'}_{\vec{p}_{rs}}\rangle_V\right.- 
\sum_{n'=1}^{l'-1}p_{n'm'}\left| \vec{0}_{ij},
\vec{p}_{rs}-\delta_{n'm',rs}+\delta_{n'l',rs}\rangle_V
 \right. \\
  &&+  \sum_{p=1}^{m'-l'-1}\bigg\{\sum_{k'_1=k'_0+1}^{m'-1}\ldots
\sum_{k'_p=k'_{p-1}+1}^{m'-1}\prod_{j=1}^{p}p_{k'_{j-1}k'_{j}}\bigg(
 \left|C^{k'_{p}m'}_{\vec{p}_{rs}-\sum_{j=1}^{p}\delta_{k'_{j-1}k'_j,rs}}\rangle_V
 \right. \nonumber
   \\
 &&-
\sum_{n'=k'_{p-1}}^{k'_p-1}p_{n'm'}\left| \vec{0}_{ij},
\vec{p}_{rs}-\sum_{j=1}^{p}\delta_{k'_{j-1}k'_j,rs}-\delta_{n'm',rs}+\delta_{n'k'_p,rs}\rangle_V
 \right.\bigg)\bigg\}, \   k'_{0}\equiv l' \nonumber.
 \end{eqnarray}
Therefore, the final result for the action of $t^{\prime}_{l'm'}$
on a vector $|\vec{N}\rangle_V$ maybe written as follows,
 \begin{eqnarray}
\label{t'lm}
 \hspace{0.5em} t^{\prime}_{l'm'}
|\vec{N}\rangle_V &=&
-\sum_{k'=1}^{l'-1}n_{k'm'}\left|\vec{N}-
  \delta_{k'm',ij}  + \delta_{k'l',ij}\rangle_V\right. \\
  &&\hspace{-0.5em}+ \sum_{k'=l'+1}^{m'-1}n_{k'm'}\left|\vec{N}-
  \delta_{k'm',ij}  + \delta_{l'k',ij}\rangle_V - \hspace{-0.5em}\sum_{k'=m'+1}^{k}n_{m'k'}\left|\vec{N}-
  \delta_{m'k',ij}  + \delta_{l'k',ij}\rangle_V
 \right.\right.
  \nonumber\\
  &&\hspace{-0.5em} +\left|C^{l'm'}_{\vec{N}}\rangle_V\right.- 
\sum_{n'=1}^{l'-1}p_{n'm'}\left| \vec{n}_{ij},
\vec{p}_{rs}-\delta_{n'm',rs}+\delta_{n'l',rs}\rangle_V
 \right. \nonumber
  \\
 &&\hspace{-0.5em}+   \sum_{p=1}^{m'-l'-1}\bigg\{\sum_{k'_1=k'_0+1}^{m'-1}\ldots
\sum_{k'_p=k'_{p-1}+1}^{m'-1}\prod_{j=1}^{p}p_{k'_{j-1}k'_{j}}\bigg(
 \left|C^{k'_{p}m'}_{ \vec{n}_{ij}, \vec{p}_{rs}-\sum_{j=1}^{p}\delta_{k'_{j-1}k'_j,rs}}\rangle_V
 \right.\nonumber
 \end{eqnarray}
\vspace{-3ex}
 \begin{eqnarray}
  && -
\sum_{n'= k'_{p-1}}^{k'_p-1}p_{n'm'}\left| \vec{n}_{ij},
\vec{p}_{rs}-\sum_{j=1}^{p}\delta_{k'_{j-1}k'_j,rs}-\delta_{n'm',rs}+\delta_{n'k'_p,rs}\rangle_V
 \right.\bigg)\bigg\}.
  \nonumber
 \end{eqnarray}
Now, it is easy to get the rest  formulae for the positive root
vectors $l'_{l'm'}$, for $l' < m'$,

Preliminary, we have defining expression
 \begin{eqnarray}
\label{l'lmprem} \hspace{-0.8em} &\hspace{-0.8em} &\hspace{-0.8em} l^{\prime }_{l'm'}
\bigl|\vec{N}\bigr\rangle_V =  -
\frac{1}{4}\Big(\sum\limits_{k'=1}^{l'-1}{n}_{k'l'}\bigg\{\prod_{i=1}^{k'}\Bigr[\prod_{j=i+1}^{l'}
\bigl(l^{\prime +}_{ij}\bigr){}^{n_{ij}-\delta_{k'l',ij}}\Bigl] t^{\prime }_{k'm'}\prod_{i=1}^{k'}\Bigr[\prod_{j=l'+1}^{k}
\bigl(l^{\prime +}_{ij}\bigr){}^{n_{ij}}\Bigl]\bigg\}
\\
&\hspace{-0.5em} & \hspace{3.0em} \times
\prod_{i=k'+1}^{k-1}\Bigr[\prod_{j=i+1}^{k}
\bigl(l^{\prime +}_{ij}\bigr){}^{n_{ij}}\Bigl]\bigl|\vec{0}_{ij},\vec{p}_{rs}\bigr\rangle_V \nonumber\\
&&\hspace{2.0em}
-\prod_{i=1}^{l'-1}\Bigr[\prod_{j=i+1}^{k}
\bigl(l^{\prime +}_{ij}\bigr){}^{n_{ij}}\Bigl]\sum\limits_{k'=l'+1}^{m'-1}\bigg\{{n}_{l'k'} \prod_{j=l'+1}^{m'-1}
\bigl(l^{\prime +}_{l'j}\bigr){}^{n_{l'j}-\delta_{l'k',ij}} t^{\prime }_{k'm'} \prod_{j=m'}^{k}
\bigl(l^{\prime +}_{l'j}\bigr){}^{n_{l'j}}\bigg\}
\nonumber\\
&\hspace{-0.5em} & \hspace{3.0em} \times \prod_{i=l'+1}^{k-1}\Bigr[\prod_{j=i+1}^{k}
\bigl(l^{\prime +}_{ij}\bigr){}^{n_{ij}}\Bigl]\bigl|\vec{0}_{ij},\vec{p}_{rs} \bigr\rangle_V\Big)\nonumber\\
&\hspace{-0.5em} & \hspace{1.0em} +
\frac{1}{4}\Big\{\sum\limits_{k'=1}^{l'-1}\bigg\{{n}_{k'm'}\prod_{i=1}^{k'}\Bigr[\prod_{j=i+1}^{k}
\bigl(l^{\prime +}_{ij}\bigr){}^{n_{ij}-\delta_{k'm',ij}}\Bigl]\bigg\} t^{\prime }_{k'l'}\prod_{i=1}^{k'}\Bigr[\prod_{j=l'+1}^{k}
\bigl(l^{\prime +}_{ij}\bigr){}^{n_{ij}}\Bigl]\Big\}\nonumber\\
&\hspace{-0.5em} & \hspace{3.0em} \times\prod_{i=k'+1}^{k-1}\Bigr[\prod_{j=i+1}^{k}
\bigl(l^{\prime +}_{ij}\bigr){}^{n_{ij}}\Bigl]\bigl|\vec{0}_{ij},\vec{p}_{rs}\bigr\rangle_V \nonumber
\end{eqnarray}
\vspace{-3ex}
 \begin{eqnarray}
&&\hspace{2.0em} -
\frac{1}{4}{n}_{l'm'}\prod_{i=1}^{l'}\Bigr[\prod_{j=i+1}^{m'}
\bigl(l^{\prime +}_{ij}\bigr){}^{n_{ij}-\delta_{l'm',ij}}\Bigl]\Big\{{n}_{l'm'}-1 + g_{0l'}+g_{0m'}\Big\}\prod_{i=l'}^{k-1}\Bigr[\prod_{j=m'+1}^{k}
\bigl(l^{\prime +}_{ij}\bigr){}^{n_{ij}}\Bigl]  \nonumber\\
&&\hspace{2.0em}\times \bigl|\vec{0}_{ij},\vec{p}_{rs}\bigr\rangle_V
+\frac{1}{4}\prod_{i=1}^{l'-1}\Bigr[\prod_{j=i+1}^{k}
\bigl(l^{\prime +}_{ij}\bigr){}^{n_{ij}}\prod_{j=l'+1}^{m'}
\bigl(l^{\prime +}_{l'j}\bigr){}^{n_{l'j}}\Bigl]\sum\limits_{k'=m'+1}^{k}\bigg\{{n}_{l'k'} \nonumber
\\
&\hspace{-0.5em} & \hspace{3.0em} \times \prod_{j=m'+1}^{k}
\bigl(l^{\prime +}_{l'j}\bigr){}^{n_{l'j}-\delta_{l'k',l'j}} t^{\prime +}_{m'k'} \bigg\}
\prod_{i=l'+1}^{k-1}\Bigr[\prod_{j=i+1}^{k}
\bigl(l^{\prime +}_{ij}\bigr){}^{n_{ij}}\Bigl]\bigl|\vec{0}_{ij},\vec{p}_{rs} \bigr\rangle_V\Big)\nonumber\\&&
 \hspace{2.0em}+
\frac{1}{4}\Big(\sum\limits_{k'=l'+1}^{m'-1}{n}_{k'm'} \bigg\{\prod_{i=1}^{k'}\Bigr[\prod_{j=i+1}^{k}
\bigl(l^{\prime +}_{ij}\bigr){}^{n_{ij}-\delta_{k'm',ij}}\Bigl]\bigg\}
t^{\prime +}_{l'k'}   \prod_{i=m'}^{k-1}\Bigr[\prod_{j=i+1}^{k}
\bigl(l^{\prime +}_{ij}\bigr){}^{n_{ij}}\Bigl]  \bigl|\vec{0}_{ij}, \vec{p}_{rs} \bigr\rangle_V\nonumber\\
&&\hspace{2.0em}
-\sum\limits_{k'=m'+1}^{k}{n}_{m'k'}  \bigg\{\prod_{i=1}^{m'}\Bigr[\prod_{j=i+1}^{k}
\bigl(l^{\prime +}_{ij}\bigr){}^{n_{ij}-\delta_{m'k',ij}}\Bigl]\bigg\} t^{\prime +}_{l'k'} \prod_{i=m'+1}^{k-1}\Bigr[\prod_{j=i+1}^{k}
\bigl(l^{\prime +}_{ij}\bigr){}^{n_{ij}}\Bigl]\bigl|\vec{0}_{ij},\vec{p}_{rs}\bigr\rangle_V\Big).
 \nonumber
 \end{eqnarray}
As the result, (for $k'_0\equiv k'$)
   \begin{eqnarray}
\label{l'lm} \hspace{0.5em}l^{\prime }_{l'm'}
\bigl|\vec{N}\bigr\rangle_V &\hspace{-0.3em} =\hspace{-0.3em}&  -
\frac{1}{4}\sum\limits_{k'=1}^{l'-1}{n}_{k'l'}
 \bigg[\sum_{p'=k'+1}^{m'-1}n_{p'm'}\left|\vec{N}-
  \delta_{p'm',ij}-\delta_{k'l',ij}  + \delta_{k'p',ij}\rangle_V\right. \\
  && - \sum_{p'=m'+1}^{k}n_{m'p'}\left|\vec{N}-
  \delta_{m'p',ij} -\delta_{k'l',ij} + \delta_{k'p',ij}\rangle_V
 \right.
  \nonumber
  \end{eqnarray}
\vspace{-3ex}
 \begin{eqnarray}
  && +\left|C^{k'm'}_{\vec{N}-\delta_{k'l',ij}}\rangle_V\right.- 
\sum_{n'=1}^{k'-1}p_{n'm'}\left| \vec{n}_{ij}-\delta_{k'l',ij},
\vec{p}_{rs}-\delta_{n'm',rs}+\delta_{n'k',rs}\rangle_V
 \right. \nonumber\\
  && +   \sum_{p=1}^{m'-k'-1}\bigg\{\sum_{k'_1=k'_0+1}^{m'-1}\ldots \sum_{k'_p=k'_{p-1}+1}^{m'-1}\prod_{j=1}^{p}p_{k'_{j-1}k'_{j}}\bigg(
 \left|C^{k'_{p}m'}_{ \vec{n}_{ij}-\delta_{k'l',ij}, \vec{p}_{rs}-\sum_{j=1}^{p}\delta_{k'_{j-1}k'_j,rs}}\rangle_V
 \right.\nonumber\\ && -
\sum_{n'=k'_{p-1}}^{k'_p-1}p_{n'm'}\left| \vec{n}_{ij}-\delta_{k'l',ij},
\vec{p}_{rs}-\sum_{j=1}^{p}\delta_{k'_{j-1}k'_j,rs}-\delta_{n'm',rs}+\delta_{n'k'_p,rs}\rangle_V
 \right. \bigg)\bigg\}
\bigg]\nonumber\\
  && +\frac{1}{4}\sum\limits_{k'=l'+1}^{m'-1}{n}_{l'k'}
    \bigg[-\sum_{p'=l'}^{k'-1}n_{p'm'}\left|\vec{N}-
  \delta_{p'm',ij}-\delta_{l'k',ij}  + \delta_{p'k',ij}\rangle_V\right.+ \hspace{-0.5em}\sum_{p'=k'+1}^{m'-1}\hspace{-0.5em}n_{p'm'}\times \nonumber\\
  &&  \left|\vec{N}-
  \delta_{p'm',ij}-\delta_{l'k',ij}  + \delta_{k'p',ij}\rangle_V\right. - \hspace{-0.7em}\sum_{p'=m'+1}^{k}\hspace{-0.7em}n_{m'p'}\left|\vec{N}-
  \delta_{m'p',ij}-\delta_{l'k',ij}  + \delta_{k'p',ij}\rangle_V\right. \nonumber\\
&&   +\left|C^{k'm'}_{\vec{N}-\delta_{l'k',ij}}\rangle_V\right.- 
\sum_{n'=1}^{k'-1}p_{n'm'}\left| \vec{n}_{ij}-\delta_{l'k',ij},
\vec{p}_{rs}-\delta_{n'm',rs}+\delta_{n'k',rs}\rangle_V
 \right. \nonumber\\
   && +   \sum_{p=1}^{m'-k'-1}\bigg\{\sum_{k'_1=k'_0+1}^{m'-1}\ldots \sum_{k'_p=k'_{p-1}+1}^{m'-1}\prod_{j=1}^{p}p_{k'_{j-1}k'_{j}}
\bigg(\left|C^{k'_{p}m'}_{ \vec{n}_{ij}-\delta_{l'k',ij}, \vec{p}_{rs}-\sum_{j=1}^{p}\delta_{k'_{j-1}k'_j,rs}}\rangle_V
 \right. \nonumber\\
  && -
\sum_{n'=k'_{p-1}}^{k'_p-1}p_{n'm'}\left| \vec{n}_{ij}-\delta_{l'k',ij},
\vec{p}_{rs}-\sum_{j=1}^{p}\delta_{k'_{j-1}k'_j,rs}-\delta_{n'm',rs}+\delta_{n'k'_p,rs}\rangle_V
 \right.\bigg)
\bigg\}\bigg]\nonumber\\
  && + \frac{1}{4}\sum\limits_{k'=1}^{l'-1}{n}_{k'm'}
 \bigg[ \sum_{p'=k'+1}^{l'-1}n_{p'l'}\left|\vec{N}-
  \delta_{p'l',ij}-\delta_{k'm',ij}  + \delta_{k'p',ij}\rangle_V
  \right.
  \nonumber \\ 
    && - \sum_{p'=l'+1}^{k}n_{l'p'} \left|\vec{N}-
  \delta_{l'p',ij} -\delta_{k'm',ij} + \delta_{k'p',ij}\rangle_V
 \right.\nonumber
 \\
  &&   +\left|C^{k'l'}_{\vec{N}-\delta_{k'm',ij}}\rangle_V\right.- 
\sum_{n'=1}^{k'-1}p_{n'm'}\left| \vec{n}_{ij}-\delta_{k'm',ij},
\vec{p}_{rs}-\delta_{n'm',rs}+\delta_{n'k',rs}\rangle_V
 \right. \nonumber
 \\
  && +   \sum_{p=1}^{l'-k'-1}\bigg\{\sum_{k'_1=k'+1}^{l'-1}\ldots \sum_{k'_p=k'_{p-1}+1}^{l'-1}\prod_{j=1}^{p}p_{k'_{j-1}k'_{j}}
 \bigg(\left|C^{k'_{p}l'}_{ \vec{n}_{ij}-\delta_{k'm',ij}, \vec{p}_{rs}-\sum_{j=1}^{p}\delta_{k'_{j-1}k'_j,rs}}\rangle_V
 \right.
\nonumber\\
  && -
\sum_{n'=k'_{p-1}}^{k'_p-1}p_{n'_pl'}\left| \vec{n}_{ij}-\delta_{k'm',ij},
\vec{p}_{rs}-\sum_{j=1}^{p}\delta_{k'_{j-1}k'_j,rs}-\delta_{n'l',rs}+\delta_{n'k'_p,rs}\rangle_V
 \right.
 \bigg)
\bigg\}\bigg]
 \nonumber
\\
&& -
\frac{1}{4}{n}_{l'm'}\Big\{{n}_{l'm'}-1 +  \sum_{k'=m'+1}^{k} (n_{l'k'}+n_{m'k'})+\sum_{k'=l'+1}^{m'-1} n_{k'm'}   - \sum_{s>l'}p_{l's}+\sum_{r<l'}p_{rl'}
\nonumber\\
&&- \sum_{s>m'}p_{m's}+\sum_{r<m'}p_{rm'}+ h^{l'}+h^{m'}\Big\}\bigl|\vec{N}- \delta_{l'm',ij}\bigr\rangle_V  \nonumber
\end{eqnarray}
\vspace{-3ex}
 \begin{eqnarray}
&& +
\frac{1}{4}\hspace{-0.3em}\sum\limits_{k'=m'+1}^{k}{n}_{l'k'} \bigg\{\hspace{-0.2em}
  \left|\vec{N} \hspace{-0.1em} -\delta_{l'k',ij} \hspace{-0.1em} + \delta_{m'k',rs} \rangle_V \right. - \hspace{-0.2em}
\sum_{p'=1}^{m'-1}p_{p'm'}\hspace{-0.1em}\left|\vec{N}-\delta_{l'k',ij} - \delta_{p'm',rs}+ \delta_{p'k',rs} \rangle_V \right. \nonumber\\
&& - \hspace{-0.7em}\sum_{p'=l'+1}^{m'-1}\hspace{-0.5em}n_{p'm'}\hspace{-0.2em}\left|\hspace{-0.2em}\vec{N}-\hspace{-0.1em}\delta_{l'k',ij}-\hspace{-0.1em}
  \delta_{p'm',ij}  + \hspace{-0.1em}\delta_{p'k',ij}\rangle_V +\hspace{-0.7em}\sum_{p'=m'+1}^{k'-1}\hspace{-0.5em}n_{m'p'}\hspace{-0.1em}\left|\hspace{-0.1em}\vec{N}-\delta_{l'k',ij}-\hspace{-0.1em}
  \delta_{m'p',ij} \hspace{-0.1em} + \delta_{p'k',ij}\rangle_V
 \right.\right. \nonumber\\
%
&&  - \sum_{p'=k'+1}^{k}n_{m'p'}\left|\vec{N}-\delta_{l'k',ij}-
  \delta_{m'p',ij}  + \delta_{k'p',ij}\rangle_V
 \right. \bigg\}\nonumber
\\
&& +
\frac{1}{4}\hspace{-0.2em}\sum\limits_{k'=l'+1}^{m'-1}\hspace{-0.5em}{n}_{k'm'} \bigg\{
  \left|\vec{N}  -\delta_{k'm',ij}  + \delta_{l'k',rs} \rangle_V \right. -\hspace{-0.4em}
\sum_{p'=1}^{l'-1}\hspace{-0.2em}p_{p'l'}\left|\vec{N}-\delta_{k'm',ij} - \delta_{p'l',rs}+ \delta_{p'k',rs} \rangle_V \right. \bigg\}\nonumber \\
&& -\frac{1}{4}\hspace{-0.2em}\sum\limits_{k'=m'+1}^{k}\hspace{-0.5em}{n}_{m'k'}  \bigg\{
  \left|\vec{N}  -\delta_{m'k',ij}  + \delta_{l'k',rs} \rangle_V \right. -\hspace{-0.5em}
\sum_{p'=1}^{l'-1}\hspace{-0.2em}p_{p'l'}\left|\vec{N}-\delta_{m'k',ij} - \delta_{p'l',rs}+ \delta_{p'k',rs} \rangle_V \right. \bigg\}\nonumber.
\end{eqnarray}

The formulae (\ref{t'+lm})-- (\ref{l'lm}) completely solve the problem of   Verma module construction for  the algebra $so(k,k)$ with second-class operator constraints.

\subsection{Note on additional parts construction for  massive HS fields}\label{addalgebram}

To solve the same problem, but for construction of auxiliary representation
 for  HS symmetry massive superalgebra
$\mathcal{A}_m(Y[k]),\mathbb{R}^{1,d-1})$  we may to enlarge the
Cartan decomposition (\ref{Cartandecomp})  up to one for
$\mathcal{A}_m(Y[k],\mathbb{R}^{1,d-1})$.  Then we could make all
the same steps again with only the fact, that the Cartan
subalgebra would now contain the element $l_0'$ whereas the
highest weight vector $|0\rangle_V$ and basis vector
$|\vec{N}^m\rangle_V$ of $\mathcal{A}_m(Y[k],\mathbb{R}^{1,d-1})$ in
addition to definitions (\ref{hwrep})--(\ref{module}) determines as
follows,
\begin{align}\label{masshwrep}
& l'_i|0\rangle_V =0 && l'_0 |0\rangle_V = m^{\prime 2}|0\rangle_V,\\
& |\vec{N}^m\rangle_V \sim
\prod_{i}^k\textstyle\bigl(\frac{l^{\prime
+}_{i}}{m_i}\bigr){}^{n^0_{i}}|\vec{N}\rangle_V \equiv |n^0_{1},n^0_{2},...,n^0_k; \vec{N}\rangle_V, \ \ \mathrm{for} \ \ n^0_i=0,1
\end{align}
for some positive parameters $m_i \in \mathbb{R}$ of dimension of mass, so
that central charge $m^2$ in the initial algebra
$\mathcal{A}_m(Y[k],\mathbb{R}^{1,d-1})$ will vanish in the
deformed (converted) algebra $\mathcal{A}_{mC}(Y[k],\mathbb{R}^{1,d-1})$ because
of the additive composition law
\begin{align}\label{vancentcharge}
    & m^2 \to M^2 = m^2+{m'}^2 =0, &&  \tilde{l}_0  \to L_0 = \tilde{l}_0 + l'_0 = \tilde{l}_0
+m^{\prime 2},
\end{align}
for the central elements $m^2, {m'}^2$ and Casimir operators $\tilde{l}_0,
l'_0$ respectively of the original algebra of $o_I$ and algebra of
additional parts $o'_I$. The parameters $m_i$  may be used to provide some specific properties of Lagrangian formulation.

\section{Oscillator realization  of the
additional parts $o'_I$} \label{oscrealsl2kdet}
\setcounter{equation}{0}

 Following general  result
of \cite{Burdik} and making use of the mapping between basis $\{|\vec{N}\rangle_V\}$ (\ref{module})
 of the constructed
 Verma module for $so(k,k)$
 and the one in new Fock space
$\mathcal{H}'$, we have
\begin{equation}\label{map}
    \left| \vec{n}_{ij}, \vec{p}_{rs}\rangle_V \right.
    \leftrightarrow \left| \vec{n}_{ij}, \vec{p}_{rs}\rangle
    \right.,\qquad
\left| \vec{n}_{ij}, \vec{p}_{rs}\rangle
    \right. = \prod_{i=1}^{k-1}\prod_{j= i+1}^k\bigl(b^{+}_{ij}\bigr){}^{
 n_{ij}}\prod_{r=1}^{k-1}\prod_{s= r+1}^k\bigl(d^{+}_{rs}\bigr){}^{p_{rs}}|0\rangle\,.
\end{equation}
Here the vector $\left| \vec{n}_{ij},
\vec{p}_{rs}\rangle\right.$, first, has the same structure as the
vector $|\vec{N}\rangle_V$ in the equation (\ref{module}), for $n_{ij},
{p}_{rs} \in \mathbb{N}_0$ and, second, appears by the basis
vectors of a Fock space $\mathcal{H}'$ generated by new
 bosonic, $b^{+}_{ij}, d^+_{rs}, b_{ij},
d_{rs}$, $i,j,r,s =1,\ldots, k; i<j; r<s$, creation and
annihilation operators with the only nonvanishing commutation
relations
\begin{equation}\label{commrelationsf}
 [b_{i_1j_1}, b^+_{i_2j_2}] =
 \delta_{i_1i_2}\delta_{j_1j_2}\,, \   \qquad [d_{r_1s_1}, d^+_{r_2s_2}]
 =\delta_{r_1r_2}\delta_{s_1s_2}.
\end{equation}
Thus, we can represent the action of the elements $o'_I$ on
$|\vec{N}\rangle_V$ given by the equations (\ref{t'+lm})--
(\ref{g'0i}), (\ref{t'lm}) -- (\ref{l'lm}) as polynomials in the
creation operators of the Fock space $\mathcal{H}'$, therefore realizing Fock module. The only
requirement on the number of pairs of the above bosonic operators
that it must coincides with one for pairs of second-class
constraints, i.e. with the numbers of negative (or positive) root
vectors in Cartan decomposition of $so(k,k)$.

Finally, the oscillator realization of the elements $o'_I$
 may be uniquely presented as follows, for Cartan elements and negative
root vectors,
\begin{eqnarray}
g_0^{\prime i}& = &  \sum_{l<i}b^+_{li}b_{li} + \sum_{l>i} b^+_{il}b_{il}  - \sum_{s>i}d^+_{is}d_{is}+\sum_{s<i}d^+_{si}d_{si}+ h^i
 \,,\label{g'0iFa} \\
  t^{\prime+}_{lm}   & = & d^+_{lm} - \sum_{n=1}^{l-1}d^+_{nm}d_{nl}-\sum_{n=1}^{l-1} b^+_{nm}b_{nl}+
  \sum_{n=l+1}^{m-1} b^+_{nm}b_{ln}-\sum_{n=m+1}^{k} b^+_{mn}b_{ln}
  \,,
 \label{t'+lma}
 \\
   l^{\prime+}_{ij} & = & b_{ij}^+\,,
 \label{l'+ijFa}
\end{eqnarray}
 for the elements $l^{\prime }_{lm}$ of upper-triangular
subalgebra $\mathcal{E}^+_{k(k-1)}$  for  $k_{-1}\equiv 1$
 \begin{eqnarray}
l^{\prime }_{lm}&=&-
 \frac{1}{4}\sum\limits_{n=1}^{l-1} \Bigl[\sum_{p=n+1}^{m-1}
 b^+_{np}b_{pm }- \sum_{p=m+1}^{k}
 b^+_{np}b_{mp }+ C^{nm}(d^+,d)- \sum_{n'=1}^{n-1}d^+_{n'n}d_{n' m}
 \label{l'lmbosea}\\
 \hspace{-1em}&\hspace{-1em}+&\hspace{-1em}
\sum_{p=1}^{m-n-1}\Big(\sum_{k_1=n+1}^{m-1}\ldots \sum_{k_p=k_{p-1}+1}^{m-1}\Big\{
 C^{k_{p}m}(d^+,d)- \sum_{n'=k_{p-1}}^{k_p-1}d^+_{n'k_p}d_{n' m} \Big\}\prod_{j=1}^{p}d_{k_{j-1}k_{j}}\Big)\Bigr]b_{nl}
 \nonumber\\
\hspace{-1.5em}&\hspace{-1em}+& \hspace{-0.5em}
\frac{1}{4}\textstyle\sum\limits_{n=l+1}^{m-1} \Bigl[-\sum\limits_{p=l}^{n-1}
 b^+_{pn}b_{pm }+ \sum\limits_{p=n+1}^{m-1}
 b^+_{np}b_{pm }-\sum\limits_{p=m+1}^{k}
 b^+_{np}b_{mp }  + C^{nm}(d^+,d)- \displaystyle\sum_{n'=1}^{n-1}d^+_{n'n}d_{n' m}\nonumber\\
 \hspace{-1.5em}&\hspace{-1em}+&\hspace{-0.5em}
  \sum_{p=1}^{m-n-1}\Big(\sum_{k_1=n+1}^{m-1}\ldots \sum_{k_p=k_{p-1}+1}^{m-1}\Big\{
 C^{k_{p}m}(d^+,d)- \sum_{n'=k'_{p-1}}^{k_p-1}d^+_{n'k_p}d_{n' m} \Big\}\prod_{j=1}^{p}d_{k_{j-1}k_{j}}\Big)\Bigr]b_{ln}
 \nonumber\\
\hspace{-1.5em}&\hspace{-1em}+& \hspace{-0.5em} \frac{1}{4}\textstyle\sum\limits_{n=1}^{l-1} \Bigl[\sum\limits_{p=n+1}^{l-1}
 b^+_{np}b_{pl }- \sum\limits_{p=l+1}^{k}
 b^+_{np}b_{lp }+C^{nl}(d^+,d)- \displaystyle\sum_{n'=1}^{n-1}d^+_{n'n}d_{n' l}
\nonumber\\
\hspace{-1.5em}&\hspace{-1em}+& \hspace{-0.5em}
\sum_{p=1}^{l-n-1} \Big(\sum_{k_1=n+1}^{l-1}\ldots \sum_{k_p=k_{p-1}+1}^{l-1}\Big\{
 C^{k_{p}l}(d^+,d)- \sum_{n'=k'_{p-1}}^{k_p-1}d^+_{n'k_p}d_{n' l} \Big\}\prod_{j=1}^{p}d_{k_{j-1}k_{j}} \Big)\Bigr]b_{nm}
 \nonumber
 \end{eqnarray}
\vspace{-3ex}
 \begin{eqnarray}
\hspace{-1.5em}&\hspace{-1em}-& \hspace{-0.5em}\frac{1}{4}\Bigl(b^+_{lm}b_{lm}+\sum_{n=m+1}^k (b^+_{ln}b_{ln}+b^+_{mn}b_{mn}) +  \sum_{n=
l+1}^{m-1}b^+_{nm} b_{nm}  - \sum_{s>l}d^+_{ls}d_{ls} -
\sum_{s>m}d^+_{ms}d_{ms} \nonumber\\
\hspace{-1em}&\hspace{-1em}+& \sum_{r<l}d^+_{rl}d_{rl} +\sum_{r<m}d^+_{rm}d_{rm} + h^{l}+
h^{m}\Bigr)b_{lm} \nonumber\\
 \hspace{-1.5em}&\hspace{-1.2em}+& \hspace{-0.2em} \frac{1}{4}\hspace{-0.1em}\textstyle\sum\limits_{n=m+1}^{k} \Bigl[ d^+_{mn}-\hspace{-0.2em}
\sum\limits_{n'=1}^{m-1} d^+_{n'n}d_{n'm} -\hspace{-0.2em}
\sum\limits_{n'=l+1}^{m-1}b^+_{n'n}b_{n'm}+\sum\limits_{n'=m+1}^{n-1}b^+_{n'n}b_{mn'} - \hspace{-0.2em}\sum\limits_{n'=n+1}^{k}b^+_{n n'}b_{mn'}   \Bigr]{b}_{ln}\nonumber\\
\hspace{-1.5em}&\hspace{-1em}+& \hspace{-0.7em}  \frac{1}{4}\textstyle\sum\limits_{n=l+1}^{m-1}
\Bigl[ d^+_{ln} - \sum\limits_{n'=1}^{l-1}d^+_{n'n}d_{n'l}
\Bigr]{b}_{nm} - \displaystyle\frac{1}{4} \textstyle\sum\limits_{n=m+1}^{k}
\Bigl[ d^+_{ln} - \sum\limits_{n'=1}^{l-1}d^+_{n'n}d_{n'l}
\Bigr]{b}_{mn}\nonumber,
 \end{eqnarray}
and for the "mixed antisymmetry" elements $t^{\prime }_{lm}$,
 \begin{eqnarray}
t^{\prime }_{lm} &=&C^{lm}(d^+,d)- \sum_{n'=1}^{l-1}d^+_{n'l}d_{n'm} +
\sum_{p=1}^{m-l-1}\bigg[\sum_{k_1=k_0+1}^{m-1}\ldots \sum_{k_p=k_{p-1}+1}^{m-1}
 \Big\{C^{k_{p}m}(d^+,d)\label{t'lmFa} \\
 && - \sum_{n'=k'_{p-1}}^{k_p-1}d^+_{n'k_p}d_{n'm} \Big\}\prod_{j=1}^{p}d_{k_{j-1}k_{j}}\bigg] -\sum_{n=1}^{l-1}b^+_{nl}b_{nm}
 + \sum_{n=l+1}^{m-1}b^+_{ln}
b_{nm}-\sum_{n=m+1}^{k}b^+_{ln}
b_{mn}
 \,,    \nonumber
\end{eqnarray}
(for $k_0\equiv l$) where the  vector $\left|C^{lm}_{\vec{p}_{rs}}\rangle_V
 \right.$, $l<m$  (\ref{Clmin}) is transformed to
 the operator $C^{lm}(d^+,d)$ 
\begin{eqnarray}
 \label{Clm}\hspace{-1.7em}
C^{lm}(d^+,d)&\equiv &
\Bigl(h^{l}-h^{m}-\sum_{n=m+1}^{k}\bigl(d^+_{ln}d_{ln} -d^+_{mn}d_{mn}\bigl)-
\sum_{n=l+1}^{m-1}d^+_{nm}d_{nm}-d^+_{lm}d_{lm}\Bigr)d_{lm} \\
 &&  + \sum_{n=m+1}^{k}\Bigl\{d^+_{mn}  - \sum_{n'=l+1}^{m-1} d^+_{n'n} d_{n'm}\Bigr\}d_{ln}.\nonumber
  \end{eqnarray}
In the above expressions  quantities $h^i, i=1,...,k$ are the
arbitrary dimensionless constants whose values will be determined
later.
Thus, we have obtained the expressions of the additional parts
$o'_I(B,B^+)$ ) for the operator
subalgebra $so(k,k)$  given in the table~\ref{table in}.

We introduce new inner product in $\mathcal{H}'$ to restore the proper Hermitian conjugation properties for $\{o'_a, {o'}^+_a \}$ as follows,
\begin{eqnarray}
\label{nsproduct}
\langle{\Psi}|\Phi\rangle_{new} & =  & \langle{\Psi}|K'|\Phi\rangle\texttt{ for any } |\Phi\rangle, |\Psi\rangle \in \mathcal{H}',
\end{eqnarray}
with some unknown Hermitian (in standard sense) operator $K'$ which should be found from the system of $k^2$ equations (for $i<j, i,j=1,...,k $),
\begin{eqnarray}
\label{systemK}
\langle{\Psi}|K'c_{ij}'|\Phi\rangle\ =\  \langle{\Phi}|K'c_{ij}^{+\prime}|\Psi\rangle^*,\quad c \in \{t\,,l\},  \quad \langle{\Psi}|K'g_{0}^{i\prime}|\Phi\rangle\ =\  \langle{\Phi}|K'g_{0}^{i\prime}|\Psi\rangle^*.
\end{eqnarray}
The solution to the equations (\ref{systemK}) may be presented in the form,
\begin{eqnarray}
\label{explicit K}
 K'&=&Z^+Z, \quad
Z \ =\ \sum_{\vec{n}_{ij}=\vec{0}}^{\infty}\sum_{\vec{p}_{rs} = \vec{0}}^{\infty}
|\vec{n}_{ij}, \vec{p}_{rs} \rangle_V\langle
0|\prod_{(i,j)=(1,2)}^{(k-1,k)} \frac{{b}_{ij}^{n_{ij}}}{(n_{ij})!}\prod_{(r,s)=(1,2)}^{(k-1,k)} \frac{{d}_{rs}^{p_{rs}}}{(p_{rs})!},\\
Z^+&=& \sum_{\vec{n}_{ij}=\vec{0}}^{\infty}\sum_{\vec{p}_{rs} = \vec{0}}^{\infty}
\prod_{(i,j)=(1,2)}^{(k-1,k)} \frac{({b}^+_{ij})^{n_{ij}}}{(n_{ij})!}\prod_{(r,s)=(1,2)}^{(k-1,k)} \frac{({d}^+_{rs})^{p_{rs}}}{(p_{rs})!}|0\rangle_V\langle\vec{n}_{ij}, \vec{p}_{rs} |. \nonumber
\end{eqnarray}

 One can show
by direct calculation that the following relation holds true:
\begin{equation}
{}_V\left\langle\vec{n}'_{lm}, \vec{p}'_{rs}\right.
\left|\vec{n}_{lm},\vec{p}_{rs}\rangle_V\right.\sim
\prod_{l=1}^k\delta^{\textstyle\sum_{i<l} n_{il}+ \sum_{i>l}n_{li}-\sum_{r>s}
p_{sr}+\sum_{r<s}
p_{rs}}_{\textstyle\sum_{i<l} n'_{il}+ \sum_{i>l}n'_{li}-\sum_{r>s}
p'_{sr}+\sum_{r<s}
p'_{rs}}.
\end{equation}
 For practical calculations for  low values of sets of $k$
numbers
\begin{equation} (\sum_{i>1}n_{i1}-\sum_{r>1}
p_{1r}, \sum_{i>2}n_{i2} + n_{12}-\sum_{r>2} p_{2r} + p_{12},...
, \sum_{i<k}n_{ik} + \sum_{r<k} p_{rk}),
\end{equation}
with $p_{rt}, n_{ij}$ being the numbers of formal ``particles''
associated with $d^+_{rt}, b_{ij}^+$ for $i\leq j, r<t$ (where
$d^+_{rt}$ reduces the spin number $s_r$ by one unit and increases
the spin number $s_t$ by one unit simultaneously), the operator
$K'$ reads with use of normalization condition
${}_V\langle0|0\rangle_V = 1$
\begin{eqnarray}\label{Ka}
K' &=&  |0\rangle\langle0| +
\sum_{r<s}\big(h^r-h^s\big)d^+_{rs}|0\rangle\langle0|d_{rs}
 -\frac{1}{4} \sum_{i < j}\big(h^i+h^j\big)
b_{ij}^+|0\rangle\langle 0|b_{ij}\\
\phantom{K'}&& +\frac{1}{4} \sum_{i<j}b_{ij}^+|0\rangle\langle0|\Bigl(
 \sum_{l<i}b_{lj}d_{li} (h^i-h^l) + \sum_{l < j}(h^j-h^l)b_{il}d_{lj}
 \Bigr)\nonumber\\
\phantom{K'}&& + \frac{1}{4}
\sum_{i<j}\Bigl(\sum_{l<i}b_{lj}^+d^+_{li}
|0\rangle\langle0|(h^i-h^l)
  + \sum_{l<j} b_{il}^+d^+_{lj}
|0\rangle\langle0|(h^j-h^l)
 \Bigr) b_{ij} + \ldots
\nonumber
\end{eqnarray}

\section{Resolution of the holonomic constraints  for $k=2$}\label{holconstrres}

The resolution of the systems (\ref{n0>=s1})--(\ref{n11<s1}) is based on two auxiliary  Lemmas.

     \noindent
\textbf{Lemma 1}: The general non-vanishing solution of the system of linear homogeneous $s_2$  equations (\ref{n0>=s1}) with $(s_2+1)$ unknown  vectors $\big|\varphi^{s_1}_{00k}\rangle \in \mathcal{H}^f$ for $k=0,...,s_2-1$, reads:
\begin{eqnarray}\label{gensolgphos1-100kv}
&& \big|\varphi^{s_1}_{00k}\rangle_{[k,s_2-k]} = \frac{t^{k}_{12}}{k!} \big|\varphi^{s_1}_{000}\rangle_{(0,s_2)},\quad k = 0,\ldots, s_2,\label{gensolgphos1-100kv2}\\
&& \big|\varphi^{s_1}_{00k}\rangle_{[k,s_2-k]}= \frac{\imath^{s_2}}{k!(s_2-k)!}\varphi^{s_1}_{00k|[[\mu^1]_{k}, [\mu^2]_{s_2-k}]} \prod_{i=1}^{k}\hat{a}{}^{\mu^1_i +}_1\prod_{j=1}^{s_2-k}\hat{a}{}^{\mu^2_j +}_2 |0\rangle\nonumber
\end{eqnarray}
 with arbitrary antisymmetric  vector $\big|\varphi^{s_1}_{000}\rangle_{(0,s_2)}$  (tensor $\varphi^{s_1}_{000| \mu^2[{s_2}]} \in Y[0,s_2]$\footnote{Equivalently, we will  use the Young antisymmetry  operators $t_{12}$ or $Y$ (\ref{Eq-3b}) to turn the respective vector $\big|\varphi^{s_1}_{000}\rangle_{(0,s_2)}$ or tensor $\varphi^{s_1}_{000| \mu^2[{s_2}]}$, to  satisfy the correct notations $\big|t^{s_2}_{12}\varphi^{s_1}_{000}\rangle_{(s_2,0)}$ ($(Y^{s_2}\varphi^{s_1})_{000| \mu^1[{s_2}]}$ $\in Y[s_2,0]$)}).
 \vspace{1ex}

 To proof the Lemma we need  to use the decomposition of any vector $ \big|\varphi^{s_1}_{00k}\rangle_{(k,s_2-k)}$ (tensor $\varphi^{s_1}_{00k|\mu^1[{k}], \mu^2[{s_2-k}]}$) in sum of Young antisymmetry irreducible vectors (tensors) following to \cite{Hamermesh}:
 \begin{eqnarray}\label{decompYir0}
\hspace{-0.8em} &\hspace{-0.8em}& \hspace{-0.8em}\big|\varphi^{s_1}_{00k}\rangle_{(k,s_2-k)} =  \sum_{i=0}^k  \big|\varphi^{s_1|i}_{00k}\rangle_{(k,s_2-k)} , \  \big|\varphi^{s_1|i}_{00k}\rangle_{(k,s_2-k)} \in Y[i, s_2-i]   \\
\hspace{-0.8em}   &\hspace{-0.8em}& \hspace{-0.8em}  \big|\varphi^{s_1|i}_{00k}\rangle_{(k,s_2-k)}\equiv \big|\varphi^{s_1|i}_{00k}\rangle_{[k-i+\{i,\,i\}+s_2-k-i]}: \label{decompYir01}\\
  \hspace{-0.8em}   &\hspace{-0.8em}& \hspace{-0.8em}  \phantom{\big|\varphi^{s_1|i}_{00k}\rangle_{(k,s_2-k)}} (Y^{s_2-k-i}\varphi^{s_1|i})_{00k|\mu^1[{k}], \mu^2[{s_2-k}]} \in Y[s_2-k-i, i]\nonumber
 \end{eqnarray}
for $k=0,...,s_2 $ with  $i$ pairs of symmetrized indices  in $\varphi^{s_1|i}_{00k|\mu^1[{k}], \mu^2[{s_2-k}]}$ and rest $(s_2-2i)$ indices to be antisymmetrized.
Decomposing the vectors $\big|\varphi^{s_1}_{00k}\rangle_{(k,s_2-k)}$ by the receipt above, we get to the validity of the Lemma~1.

For the rest systems    in (\ref{n0<s1}), (\ref{n10<s1}), (\ref{n11<s1}) $l=s_1-1$ we need

     \noindent
\textbf{Lemma 2}: The general non-vanishing solutions of the systems  of linear homogeneous  $s_2$  equations (\ref{n0<s1}) with $(\min(s_2,s_1-1)+1)$ unknowns  $\big|\varphi^{s_1-1}_{00k}\rangle \in \mathcal{H}^f$ and $(2s_2-1)$ equations (\ref{n10<s1}), (\ref{n11<s1}) with $2s_2$  unknowns $\big|\varphi^{s_1-1}_{10k}\rangle_{(k-1,s_2-k)}$, $\big|\varphi^{s_1-1}_{11k}\rangle_{[0,\,s_2-1]}$  for $k=1,...,s_2$, read in dependence on value of $r=s_1-s_2$:
\begin{eqnarray}\label{gsols1-1k}
   && \big|\varphi^{s_1-1|i}_{00k}\rangle_{[k+1-i+\{i,\,i\}+ s_2-i-k]} =   \frac{t^k_{12}}{k!}\big|\varphi^{s_1-1|i}_{000}\rangle_{[1-i+\{i,\,i\}+ s_2-i]},   i=\left\{\begin{array}{cc}
     0,1 &\texttt{ for } r>0 \\
           1 & \texttt{ for } r=0
             \end{array}\right.
   , \\
   &&  \big|\varphi^{s_1-1|i}_{00k}\rangle_{[k-i+\{i,\,i\}+ s_2-i-k]}=0, \ i=\left\{\begin{array}{cc}
     2,...,k &\texttt{ for } r>0 \\
      0,2,...,k & \texttt{ for } r=0
             \end{array}\right.,\ k = 0,..., \min(s_2,s_1-1), \nonumber \\
   && \big|\varphi^{s_1-1|0}_{11k}\rangle_{[k-1, s_2-k]} =   \frac{t^k_{12}}{k!}\big|\varphi^{s_1-1}_{111}\rangle_{[0,\,s_2-1]}, \  \big|\varphi^{s_1-1|i}_{11k}\rangle_{[k-1-i+\{i,\,i\}+ s_2-i-k]}=0,  \label{vps1-1111}\\
   && \big|\varphi^{s_1-1|0}_{10k}\rangle_{[k-1, s_2-k]} =   -\frac{t^k_{12}}{k!}\big|\varphi^{s_1-1}_{111}\rangle_{[0,\,s_2-1]} , \  \big|\varphi^{s_1-1|i}_{10k}\rangle_{[k-1-i+\{i,\,i\}+ s_2-i-k]}=0,   \label{vps1-1101}
\end{eqnarray}
for $i=1,...,k$  and with  arbitrary $(s_1-1)$-level  gauge parameters
 \begin{eqnarray}\label{s1-1gp}
  &&\big|\varphi^{s_1-1}_{111}\rangle_{[0,\,s_2-1]}, \ \big|\varphi^{s_1-1}_{000}\rangle_{(1,s_2)}=\sum\limits_{i=0}^1\big|\varphi^{s_1-1|i}_{000}\rangle_{[1-i+\{i,\,i\}+s_2-i]}, \\
  && \qquad\texttt{ for } \big|\varphi^{s_1-1|0}_{000}\rangle_{[1,s_2]} =0 \texttt{ when  } r=0.\nonumber
  \end{eqnarray}

 \vspace{1.5ex}
\noindent
To prove, first,   we note that the solution for homogeneous subsystem (\ref{n10<s1}) follows from the Lemma 1 in the form (\ref{vps1-1111}). Then
the homogeneous system (\ref{n10>=s1}) for $n=1$ transforms as
\begin{equation}
\label{gphos1-11100kvm}     t_{12}\big|\varphi^{s_1-1}_{10k}\rangle_{(k-1,s_2-k)}\theta_{k,0}-  (k+1) \big|\varphi^{s_1-1}_{10k+1}\rangle_{(k,s_2-k-1)} - \frac{t^k_{12}}{k!}\big|\varphi^{s_1-1}_{111}\rangle_{[0,\,s_2-1]} = 0.
\end{equation}
For $k=0$ from (\ref{gphos1-11100kvm}) it follows that $\big|\varphi^{s_1-1}_{101}\rangle=-\big|\varphi^{s_1-1}_{111}\rangle$.  The last term in (\ref{gphos1-11100kvm}) contains $\forall k $ only antisymmetric vector (tensor). Repeating, for $k=1$ the arguments from the proof of  Lemma 1, starting from $k=1$ and continuing the resolution with use of  the decomposition  (\ref{decompYir0}) of any vectors $\big|\varphi^{s_1-1}_{10k}\rangle$ on Young irreducible ones we reach the representation (\ref{vps1-1101}).

Applying  for the vectors$ \big|\varphi^{s_1-1}_{00k}\rangle_{(k+1,s_2-k)}$  in the  system (\ref{n10<s1}) decomposition (\ref{decompYir0}) we have for $k=0$ that $ \big|\varphi^{s_1-1|i}_{001}\rangle_{[2-i+\{i,\,i\}+s_2-i-1]}$ vanish  for $i=0,1,2$ so that from homogeneous  equation (\ref{n0>=s1}) it follows
\begin{eqnarray}\nonumber
\hspace{-0.7em}   &\hspace{-0.7em} &\hspace{-0.7em} \big|\varphi^{s_1-1|i}_{001}\rangle_{[2-i+\{i,\,i\}+ s_2-i-1]} =   t_{12}\big|\varphi^{s_1-1|i}_{000}\rangle_{[1-i+\{i,\,i\}+ s_2-i]},  i=0,1,  \big|\varphi^{s_1-1|2}_{001}\rangle_{[\{2,\,2\}+ s_2-3]}=0,
\end{eqnarray}
 Continuing, we have for $k<s_2-1$:
   \begin{gather}\label{gsols1-1kpr}
              \big|\varphi^{s_1-1|i}_{00k}\rangle_{[k+1-i+\{i,\,i\}+ s_2-i-k]} =   \frac{t^k_{12}}{k!}\big|\varphi^{s_1-1|i}_{000}\rangle_{[1-i+\{i,\,i\}+ s_2-i]}, \ i=0,1;\\
              \big|\varphi^{s_1-1|i}_{00k}\rangle_{[k-i+\{i,\,i\}+ s_2-i-k]}=0, \ i=2,...,k.
            \end{gather}
      For $k=s_2-1$ there exists two cases $r=s_1-s_2>0$ and  $r=0$. For the former variant we have the validity of the representation (\ref{gsols1-1kpr}). For
the latter   one the last homogeneous equation in (\ref{n0>=s1}) looks as
\begin{equation}\label{gphos1-1100kvl}
    t_{12}\big|\varphi^{s_1-1}_{00s_1-1}\rangle_{(s_1,1)} = 0,
\end{equation}
so that $\big|\varphi^{s_1-1|0}_{00s_1-1}\rangle_{[s_1,1]}$ vanishes  $\big|\varphi^{s_1-1|0}_{00s_1-1}\rangle_{[s_1,1]}=0$ and  $\big|\varphi^{s_1-1}_{00s_1-1}\rangle_{(s_1,1)} \in Y[s_1,1]$. Hence, from  (\ref{gsols1-1kpr})
\begin{equation}\label{gsols1-1kpr1}
  \big|\varphi^{s_1-1|i}_{00s_1-1}\rangle_{(s_1-i+\{i,i\}+1-i)} =   \frac{t^{s_1}_{12}}{(s_1-1)!}\big|\varphi^{s_1-1|i}_{000}\rangle_{[1-i+\{i,\,i\}+ s_1-i]}
  \ \Rightarrow \ \big|\varphi^{s_1-1|0}_{000}\rangle_{[1,\, s_1]}=0,
\end{equation}
it follows the validity of representation (\ref{gsols1-1k}) for $r=0$ as well.

\vspace{1ex}
The compatibility of the gauge transformations
\begin{eqnarray}
   \label{s1-0rgt1v} \delta\left( \begin{array}{l}
                       \big|\varphi^{s_1-1}_{0;00k}\rangle_{(k,s_2-k)} \\
                        \big|\varphi^{s_1-1}_{00k}\rangle_{(k+1,s_2-k)}\\
                        \big|\varphi^{s_1-1}_{10k}\rangle_{(k-1,s_2-k)}\\
                        \big|\varphi^{s_1-1}_{11k}\rangle_{(k-1,s_2-k)}
                                             \end{array}\right) =  \left(  \begin{array}{c}
                       \partial^2  \\
                        - \big[(s_1-k)l_1^+ +l_2^+ t_{12}\theta_{s_2,k}\big] \\
                         l_1\\
                        k l_2 t_{12}^{-1}
                                              \end{array}\right)\frac{t^{k}_{12}}{k!}\big| \varphi^{s_1}_{000}\rangle_{[0,s_2]},
          \end{eqnarray}
          where $k=0,...,s_2$ for the first, $k=0,...,\min(s_2,s_1-1)$ for the second, $k=1,...,s_2$  for the third and fourth  equations
 for vanishing  parameter  $\big|\varphi^{s_1-1|0}_{000}\rangle_{[1,s_2]}$ when $r=0$ follows
from  (easily verified for the component tensor language according to (\ref{gensolgphos1-100kv2})):
\begin{equation}\label{trgtrs1-1as}
\delta  \big|\varphi^{s_1-1|0}_{000}\rangle_{[1,s_2]}= -  \big[s_1l_1^+ +l_2^+ t_{12}\big] \big|\varphi^{s_1}_{000}\rangle_{[0,s_2]}\equiv 0
\end{equation}

 Equivalently, for the final  gauge parameter we have the gauge transformations and Young antisymmetry conditions:
 \begin{eqnarray}
\label{gtrl=s1-1ap}  && \delta \sum_i\big|\varphi^{s_1-1|i}(a^+)\rangle_{[1-i+\{i,\,i\}+s_2-i]} = -\Big(s_1 l_1^+ +   l_2^+t^{}_{12}  \Big) \big|  \varphi^{s_1}(a^+)\rangle_{[0,s_2]},   \\
\label{gtrl=s1-2ap}   && \  \left\{\begin{array}{l}  t_{12}^{s_2} \big|\varphi^{s_1-1}(a^+)\rangle_{(1,s_2)} \ \in\  Y[s_2+1,0],\  s_1>s_2; \\
t_{12}^{s_1-1} \big|\varphi^{s_1-1}(a^+)\rangle_{(1,s_2)} \in Y[s_1,1],\ \ \ \  s_1=s_2\end{array}\right..
\end{eqnarray}

\acknowledgments
The author wish to acknowledge the  Organizing Committees of the International Workshop-School QFTHEP'270 in Moscow and International Conference AQFT'2025 in Dubna, where the significant parts of the work was done and many issues were discussed. I am thankful to referees for careful criticism, to I.~Buchbinder, to M.~Vasiliev for useful comments, to  N.~Ohta for correspondence,  to V. Krykhtin, Yu. Zinoviev, V. Tolstoy   for stimulating discussions, to Yu. Bogdanova, E. Lemeshko for the participation at the beginning of the work  and to E. Boos and D. Kazakov for wonderful scientific  spirit.


\paragraph{Note added.} After the paper has been written the paper \cite{Metsaevnew} was appeared with BRST--BV action related with Lagrangian formulations  (for reducible field representation) in "unconstrained" form without use of complete BRST operator and incomplete BRST operator for (constrained)
totally symmetric higher-spin fields on AdS space.


\end{document}